\documentstyle[aps,prb,floats,preprint,eqsecnum,tighten,epsfig]{revtex}
\begin{document}
\widetext
\thispagestyle{empty}

\title{Real-time non-equilibrium  dynamics of quantum glassy systems}

\author{Leticia F. Cugliandolo$^*$\cite{add1} and 
Gustavo Lozano$^{**}$\cite{add2}}
\address{ $^*$
\it Laboratoire de Physique Th\'eorique de l'\'Ecole Normale 
Sup\'erieure,
\cite{add3} 
\\
24 rue Lhomond, 75231 Paris Cedex 05, France and \\
 Laboratoire de Physique Th\'eorique  et Hautes Energies, Jussieu\\
 5 \`eme \'etage,  Tour 24, 4 Place Jussieu, 75005 Paris France
\\
$^{**}$
\it Division de Physique Th\'eorique,
\cite{add4} Institut de Physique 
Nucl\'eaire,
Universit\'e de Paris-Sud Orsay, 91406 Orsay Cedex, France}

\date\today
\maketitle
\widetext

\begin{abstract}

We develop a systematic analytic approach to 
aging effects in quantum disordered systems 
in contact with an environment. Within the closed-time
path-integral formalism 
we include dissipation by coupling the 
system to a set of independent harmonic oscillators that mimic
a quantum thermal bath. After integrating over the bath variables and 
averaging over disorder we obtain an effective action that 
determines the real-time dynamics of the system. The classical limit 
yields the Martin-Siggia-Rose generating functional associated 
to a colored noise. 
We apply this
general formalism to a prototype model 
related to the $p$ spin-glass.  We show that the 
model has a dynamic phase transition separating the paramagnetic
from the spin-glass phase and that quantum fluctuations
depress the transition temperature until a quantum critical 
point is reached. We show that the dynamics in the 
paramagnetic phase is stationary but presents an interesting crossover
from a region controlled by the classical critical point to another one 
controlled by the quantum critical point. 
The most characteristic property of the dynamics in a glassy phase,
namely aging, survives the quantum fluctuations. In the 
sub-critical  region the quantum fluctuation-dissipation theorem is 
modified in a way that is consistent with the notion of effective temperatures 
introduced for the classical case. We discuss these results in
connection with recent experiments in dipolar quantum spin-glasses and
the relevance of the effective 
temperatures with respect to the understanding 
of the low temperature dynamics. 
\end{abstract}
\vspace{.5cm}
{{\bf \footnotesize LPTENS 97/55, LPTHE 98/19}}
\pacs{PACS Numbers~: 75.10.Nr, 64.60.Cn, 64.70.Pf, 11.17.+y}

\newpage 

\vspace{0.5cm}
\section{Introduction}
\vspace{0.5cm}
\widetext
In the last few years,  
great progress has been made towards the understanding of
the out of equilibrium dynamics of {\it classical} systems with slow 
dynamics and aging effects. 
Experiments in several glassy materials such as  polymer glasses,\cite{Struick}
spin-glasses,\cite{spin-glasses} orientational glasses,\cite{levelut}
 simple liquids like  glycerol\cite{Nagel} and gels, \cite{Bonn} 
 have shown that
all these systems share an extremely slow
dynamics at low temperatures.
Although some aspects of the  dynamic evolution 
as the precise scaling laws or the dependence on cooling rates
vary from system to system,
all the examples mentioned above are characterized by 
the existence of a non-equilibrium low temperature phase with
{\it aging effects}.  This means that measurements show a  strong dependence 
upon the time elapsed since its preparation by a thermal quench or
annealing  from a high temperature phase. Needless to say, the 
standard equilibrium approach is not suited to describe such a 
phenomenology. 

In order to describe the spin-glass case, 
a number of theoretical methods have been 
exploited, ranging from scaling arguments based on 
coarsening ideas;\cite{Fihu}
phenomenological models founded on particular assumptions on the 
structure of phase space;\cite{Phasespace,rusos,Bo,Kula} 
analytical solutions to models with infinite-range
interactions;\cite{Cuku,Cuku2,Frme,Cukule} 
and numerical simulations of finite and infinite 
dimensional 
models.\cite{Ri} All of these approaches have
succeeded in capturing some of the effects observed experimentally.

Importantly enough, the solution of simplified models with infinite 
range interactions
has been fruitful for several reasons: \newline
(i) It has shown that indeed  a large family of such models do capture 
aging effects that resemble, at least qualitatively, the experimental 
observations.
\newline 
(ii) It has provided a general framework to analyze the non-equilibrium dynamics 
 with new predictions, like scaling laws for the time-delayed correlation 
function and asymptotic violations of the fluctuation-dissipation
theorem, that are not restricted to the infinite 
dimensional case. Some of these 
have  been confirmed by  numerical studies of more realistic 
models\cite{numerics}
and will be probed experimentally soon.\cite{Bonn2}
\newline
(iii) It has allowed to extend the {\it analytical} connection\cite{Kith} 
between mean-field theories of disordered systems and 
the mode-coupling theory of super-cooled liquids\cite{Go} to the 
low temperature glassy phase.\cite{Frhe,Bocukume} 
This is most welcome since many of the similarities observed in 
the behavior of real glasses of very different origin can now be understood  
theoretically. \newline
(iv) Problems of a more general nature such as the dynamics of manifolds
in a random quenched environment, relevant to the description of  e.g.
the pinning of the flux lattice by disorder in dirty high $T_c$ 
superconductors,\cite{Gile} can be studied in a similar way.\cite{Cukule}
The dependence
on the internal dimension of the manifold can be included in the 
formalism and the analysis yields, among other results, dynamic
scaling laws for the displacements with aging effects.

\vspace{.5cm}
The studies we have mentioned so far are mostly concerned with
the analysis of {\it classical} systems. This treatment is
in general justified since, in many cases, the critical
temperature at which the transition to the ``glassy phase"
occurs is sufficiently high as to make quantum effects
irrelevant. Nevertheless, in many cases of practical interest,
the critical temperature can be lowered by
tuning an external parameter. Experimental realizations
of this situation are given  by 
insulating magnetic materials\cite{Wuetal} such as 
LiHo$_x$Y$_{1-x}$F$_4$ and 
 randomly mixed hydrogen-bonded ferro-antiferroelectric crystals
of the Rb$_{1-x}$ (NH$_4$)$_{x}$H$_2$PO$_4$ type (known as
RADP).\cite{ferroelectrics}

Theoretical studies addressing 
the effects of quantum fluctuations in disordered media
have been considered over the last twenty years.
A variety of techniques including replica 
theory,\cite{quantumreplicas,Giledou,Niri}
 renormalization group\cite{Fisher}  and Monte Carlo 
simulations\cite{SKpara,Rogr,MC}
have been employed
in order to understand the {\it static} properties
and critical behavior of quantum systems
such as quantum rotors\cite{Sachdev}, 
transverse field Ising models in finite dimensions\cite{Fisher,Igri}
and mean-field systems like the Sherrington-Kirkpatrick model
in a transverse field.\cite{SKpara,Rogr,MC}

From the study of {\it classical}  glassy systems described in the first
paragraph, one knows that at low temperatures their dynamics becomes so 
slow that the systems
are not able to reach thermal equilibrium at any experimentally accessible 
time. The same occurs regarding numerical simulations of large systems. 
It is then plausible that even at very low temperatures, say $T=0$,
with only quantum fluctuations driving the dynamics, 
the materials and/or models will need an unaffordable long time to reach 
equilibrium and that  the relevant evolution will be characterized 
by non-equilibrium effects. Indeed, it is well-known that certain glasses
in the limit of very low temperatures have aging effects\cite{librolowT,Fayer}
and hence the approaches based on 
the assumption of equilibration have a restricted domain of application.
Instead, one should necessarily start from a quantum {\it dynamic} description 
in order to obtain sensible information about the
systems.  
To the best of our knowledge this problem has not been theoretically
tackled, from a microscopic point of view, yet.

\vspace{.5cm}

The question then arises as to how do quantum fluctuations
modify the {\it real time} dynamics of glassy systems. 
In this paper, we treat the case of a disordered model 
in contact with a thermal 
quantum bath. 

Our first aim is to present a general framework that allows us to
study the real time dynamics of a system with quantum {\it and}
thermal fluctuations starting from an arbitrary initial condition.
In the case of classical models this is most easily done within the 
Langevin approach. One generally starts by postulating that the 
dynamics of the microscopic variables is given by a 
stochastic Langevin equation. The noise and friction terms account for 
the bath-system interaction. A path-integral formulation of the generating 
functional, known as  the Martin-Siggia-Rose formalism\cite{MSR} (MSR), 
yields an 
elegant and simple formulation which is particularly well suited for 
problems with disorder. However,
the generalization of this procedure to the quantum case is far from being 
straightforward. The Langevin equation does not seem to be a good 
starting point since, being phenomenological,
it cannot be quantized in an obvious way.\cite{Langevin,Gardiner} In other 
words, it is not clear how to make the Langevin approach compatible with 
the intrinsic microscopic dynamics of a quantum system.

The approach we use here results from the combination of different techniques
which have been intensively discussed in the literature and successfully
applied to a wide range of phenomena. Essentially, it consists of the 
Schwinger and Keldysh\cite{closedpath} closed-time path-integral formalism 
applied to a system in contact with a thermal bath, combined with the 
integration of the bath variables leading to the  Feynman and Vernon
influence functional.\cite{Feve} A subsequent integration over the 
disorder gives the quantum analogue of the 
MSR generating functional for disordered systems.

Though the formalism allows us to consider general cooling
procedures or/and various temperature variations, we shall not
study involved thermal histories here but concentrate instead 
on a quench from a configuration associated to the disordered phase.

\vspace{.5cm}

For the sake of definiteness we focus
on a system whose classical counterpart, the 
so-called ``$p$-spin spherical spin-glass model'', has been studied in 
great detail. This model is described by a potential energy
\begin{equation}
V[{\bbox \sigma}, J] =
-
\sum_{i_1 < \dots < i_p}^N 
J_{i_i\dots i_p} \, \sigma_{i_1} \dots \sigma_{i_p} 
\end{equation}
with $J_{i_1,\dots,i_p}$ quenched independent Gaussian
variables and $p\geq 2$ a parameter. 

Let us attempt a short review of the behavior of the 
{\it classical} $p$-spin models.

In its Ising version it was first introduced by Derrida in relation 
with the random energy model.\cite{pspinIsing} Several studies of 
its static  properties, in particular using replicas,  followed.
\cite{pspinIsing}
Kirkpatrick and Thirumalai\cite{Kith} studied the equilibrium dynamics, 
\`a la Sompolinsky,\cite{Sozi,So} and remarked that the 
dynamic equations, in the high temperature phase, are 
those appearing as $F_{p-1}$ models in the mode-coupling theory of super-cooled 
liquids developed to describe the  glass transition.\cite{Go}  

The spherical version, where a time-dependent 
Lagrange multiplier is introduced to enforce the spherical constraint
$\sum_{i=1}^N \sigma_i^2 =N$, 
was first studied by Jones, Kosterlitz and 
Thouless\cite{p2} for $p=2$. Its 
statics is fully solvable either by using replicas -- 
a replica symmetric ansatz is exact -- or through a direct calculation.
Its dynamics\cite{Cide,Cude} is also fully solvable and it 
renders explicit the fact that the model is formally 
related\cite{Cide,Bocukume} to
the $O(N)$ model in the large $N$ limit and dimension $d=3$. In the 
classical limit, it is then a toy model for ferromagnetic domain-growth. 
In the quantum case one feels tempted to 
relate it to coarsening in quantum systems. 

The analysis of the spherical model for $p\geq 3$ is due to Crisanti 
and Sommers,\cite{Crso} who also studied the equilibrium
dynamics \`a la Sompolinsky in 
collaboration with Horner.\cite{Crhoso} 
In the classical limit and at the static level,
this model is the prototype spin-glass solved 
by a Parisi one-step replica symmetric ansatz. 
Still at the classical level and 
in the large $N$ limit, one derives exact evolution 
equations that exhibit a first order -- discontinuous -- dynamic 
transition at $T_d$.
Above $T_d$, as in the Ising case, one recovers the mode-coupling equation 
for model $F_{p-1}$.\cite{Crhoso} Below $T_d$ 
it is well-known that the dynamic approach used in Ref.[\cite{Crhoso}] 
is incorrect.\cite{Hojayo,Biyo,Mepavi} 
The dynamics in the low-temperature phase was solved in 
Ref.~[\raisebox{-.22cm}{\Large \cite{Cuku}}] 
showing that the dynamics of  mean-field spin-glasses 
strikingly resembles that of real systems. In the spin-glass phase
the evolution is intrinsically out of equilibrium and the systems age in
similar way to the one observed in experiments. 

In order to define the model at the quantum level, we need to 
specify the dynamics of the variables ${\bbox \sigma}$. For simplicity, we 
consider a kinetic term of the form
\begin{equation}
{\sc K} = \frac{m}{2} \sum_{i=1}^N {\dot {\bbox \sigma}}^2
\; 
\end{equation}
where the dot denotes a time-derivative.
Calling $z$ the Lagrange multiplier enforcing the 
spherical constraint, the model we here consider 
is then described by the action
\begin{equation}
S_{\sc s}[{\bbox \sigma}, J] = 
\int dt \, 
\left[ \frac{m}{2} {\dot {\bbox \sigma}}^2 - 
\frac{z}{2} \left( {\bbox \sigma}^2 -N \right) 
+\sum_{i_1 < \dots < i_p}^N 
J_{i_i\dots i_p} \, \sigma_{i_1} \dots \sigma_{i_p} 
\right] 
\; .
\end{equation} 

Our main goals are to 
detect a dynamic transition from a disordered phase 
to an ordered glassy phase and to characterize 
the dynamics in these phases. With this purpose, 
we pay particular attention 
to the departure of correlation and response functions 
from stationarity and, we search for one of the hallmarks 
of a glassy phase, namely, violations of the fluctuation-dissipation
theorem. We also 
intend to discuss our results in view of previous and possible experimental 
tests. 

\vspace{.5cm}

The paper is organized as follows. In Section \ref{CTP} we discuss
the closed-time path-integral formalism of Keldysh and Schwinger
and the coupling of the quantum system to a quantum heat bath. We obtain 
a generating functional with an effective action that takes into account 
the effect of the coupling to the bath. In three Appendices 
we give details of the Keldysh Schwinger formalism (Appendix 
\ref{appKeldysh}), we recall the quantum FDT (Appendix  \ref{appFDT}) and 
we discuss the classical limit (Appendix  \ref{classicallimit}).
Section \ref{meanfieldequations} 
is devoted to the derivation 
of the dynamic equations through a saddle-point approximation of 
the generating functional. The method is similar to the one used in the 
classical case when manipulating the Martin-Siggia-Rose generating functional.
The general scenario for the real-time dynamics of this system is 
discussed in Section \ref{scenario}. It is inspired by the 
behavior of the classical counterpart but some generalizations are
necessary. In Section \ref{equilibriumdynamics} the equations and 
their solution in the paramagnetic phase are studied with special emphasis
set on the analysis of the out of phase susceptibility close to the 
transition line. In Section
\ref{Nonequilibrium} the dynamic equations in the 
glassy phase are studied. 
The non-equilibrium dynamics is shown to survive quantum fluctuations
in a finite region of the phase diagram. Further details of the approach to solve the equations are given in Appendix \ref{integrals}.
The modification of FDT needed to construct an ansatz to solve 
these equations and its relation to effective temperatures 
are discussed in Section \ref{generalFDT}.
Finally, in Section \ref{conclusions} we present a summary of our
results, our conclusions and 
some ideas as to how to continue investigating along the lines
here proposed.
A short account of some of these results has appeared 
in Ref.~[\raisebox{-.22cm}{\Large \cite{letter}}].

\vspace{0.5cm}
\section{Closed-time path-integral formalism}
\label{CTP}
\vspace{0.5cm}

The closed-time path-integral (CTP) formalism of
Keldysh and Schwinger\cite{closedpath} for treating the non-equilibrium
dynamics of a quantum system
has been reviewed several times in the literature.
\cite{chinos,alemanes,alemanes2,Hupazh,Huetal,deVega}
In order to set the notation 
we briefly present it in Section~\ref{generalframework}.
We then consider the situation in which the system under study is
coupled to a quantum thermal bath and derive the effective action following
Feynman and Vernon \cite{Feve} (Section~\ref{Couplingtoaheatbath}). 
In Section~\ref{CTPfordisorderedsystems} we finally address the 
case of a disordered system and we obtain the effective action 
from which the dynamic equations are derived in 
Section~\ref{meanfieldequations}.

\subsection{General framework}
\label{generalframework}

The basic ingredient of this formulation is the ``in-in'' generating
functional from which,  after derivation with respect to the
external sources $\xi_i^+$, $\xi_j^-$, the different Green
functions can be obtained. Let us consider a model with degrees of freedom described by a field
${\bbox \phi}(t)=(\phi_1(t),\dots,\phi_N(t))$
in the Heisenberg picture. Denoting $\hat \rho(t_o)$ the density
matrix at the initial time $t_o$ (which from now on we set to zero),
the generating functional is defined as
\begin{eqnarray}
{\cal Z}[{\bbox \xi}^+, {\bbox \xi}^-] 
&=&
\mbox{Tr} \left[T^* \exp\left(- \frac{i}{\hbar} 
\int_{0}^\infty dt \, 
{\bbox \xi}^-(t) {\bbox \phi}(t)\right) 
T \exp\left(  
 \frac{i}{\hbar} \int_{0}^\infty dt \, 
{\bbox \xi}^+(t) {\bbox \phi}(t) \right) 
 \hat\rho(0) 
\right]
\; 
\label{generating}
\end{eqnarray}
where ${\bbox \xi}^+(t)$ and ${\bbox \xi}^-(t)$ are the 
$N$-vector external sources and  a scalar product between $N$ vectors is 
assumed.  The symbols
$T$ and $T^*$ are the time and anti-time  ordering operators:
\begin{eqnarray}
T( A(t_1) B(t_2))& =& \theta(t_1-t_2) A(t_1) B(t_2)+ 
                      \theta(t_2-t_1) B(t_2) A(t_1)
\; ,
\\
T^*( A(t_1) B(t_2))& =& \theta(t_2-t_1) A(t_1) B(t_2)+ 
                      \theta(t_1-t_2) B(t_2) A(t_1)
\; .
\end{eqnarray} 
Defining ensemble averages in the usual way
\begin{equation}
 \langle A(t) \rangle  = \frac{ \mbox{Tr}( A(t) \hat \rho(0) )}
 { \mbox{Tr} \hat \rho(0) }
\; ,
\end{equation}
one obtains
\begin{eqnarray}
\left.
\frac{\delta^2 {\ln} {\cal Z}}{\delta \xi_i^+(t) \delta \xi_j^+(t')}
\right|_{{\bbox \xi}=0}
& = &-\frac{1}{\hbar^2} \langle T(\phi_i(t) \phi_j(t')) \rangle
\end{eqnarray}
which is directly related to the Feynman causal propagator. In the
same way,
\begin{eqnarray}
C_{ij}(t,t')&\equiv&\frac{1}{2} \langle  \phi_i(t) \phi_j(t')
 + \phi_j(t') \phi_i(t) \rangle  \\
 &=&
\left.
\frac{\hbar^2}{2} 
 \left[
\frac{\delta^2}{\delta \xi_i^+(t) \delta \xi_j^-(t')}
+
\frac{\delta^2}{\delta \xi_j^+(t') \delta \xi_i^-(t) }
\right]
{\ln} {\cal Z}
\right|_{{\bbox \xi}=0}
\; 
\end{eqnarray}
where $C_{ij}(t,t')$ is the {\it symmetrized correlation function}
($C_{ij}(t,t')=C_{ji}(t',t)$).

The response function $R_{ij}(t,t')$ is defined as the variation of the
averaged field $\langle \phi_i\rangle$ 
with respect to the ``strength''
$f_j$ of a perturbation that modifies the potential energy
according to $V \to V - {\bbox f} {\bbox \phi}$:
\begin{eqnarray}
R_{ij}(t,t')&\equiv&
\left. \frac{\delta \langle \phi_i(t) \rangle}
{\delta f_j(t')} \right|_{f=0}
\label{respdef}
\; .
\end{eqnarray}
In linear-response theory, it can be expressed in terms of the 
averaged commutator:
\begin{eqnarray}
R_{ij}(t,t')
&=& \frac{i}{\hbar} \theta(t-t')
\langle [\phi_i(t),\phi_j(t')] 
\rangle  \\
&=&
\left.
\frac{\hbar}{i} 
 \left[
\frac{\delta^2}{\delta \xi_i^+(t) \delta \xi_j^+(t')}
+
\frac{\delta^2}{\delta \xi_j^+(t) \delta \xi_i^-(t')}
\right]
{\ln} {\cal Z}
\right|_{{\bbox \xi}=0}
\; .
\end{eqnarray}

One of the main advantages of this formalism is that the generating
functional (\ref{generating}) admits a path-integral representation
which is a slight modification of the usual ``in-out'' Feynman path
integral. This  will allow us to make formal manipulations
similar to the ones performed in the study
of the dynamics of classical systems via the Martin-Siggia-Rose\cite{MSR} (MSR) 
formalism. In a short-hand notation,
the generating functional reads
\begin{eqnarray}
{\cal Z}[{\bbox \xi}^+, {\bbox \xi}^-]  &=&\int 
{\cal D}{\bbox \phi}^+
{\cal D}{\bbox \phi}^-
\exp\left[ \frac{i}{\hbar} \, \left( S[{\bbox \phi^+}]
-S[{ \bbox \phi^-}]
+ \int dt \, {\bbox \xi}^+(t) {\bbox \phi}^+(t)
- \int dt \, {\bbox \xi}^-(t) {\bbox \phi}^-(t)
\right) \right]
\nonumber\\
& & 
\;\;\;\;\;\;\;\;\;\; 
\times 
\langle {\bbox \phi}^+  | \hat \rho(0) |  {\bbox \phi}^-  \rangle
\label{aaab}
\end{eqnarray}
where $S[{\bbox \phi}]$ is the action of the system and
$\langle {\bbox \phi}^+  | \hat \rho(0) |  {\bbox \phi }^-  \rangle$
stands for the matrix element of the density matrix which has support
only at $t=0$ (see Appendix \ref{appKeldysh} for a proof of this equation). 
Hereafter, 
we omit the limits of the time-integrals 
that, unless otherwise stated, go from $t_o=0$ to $\infty$.

In the following it will be useful to write the correlation and response 
in terms of the fields ${\bbox \phi}^+$, ${\bbox \phi}^-$:
\begin{eqnarray}
C_{ij}(t,t') 
&=&
\frac12 \; 
\langle \phi_i^+(t) \phi_j^-(t')  + \phi_j^+(t') \phi_i^-(t)  \rangle
\; ,
\label{defcorr}
\\
R_{ij}(t,t')
&=&
\frac{i}{\hbar} \, 
\langle \phi_i^+(t) \, \left( \phi_j^+(t') - \phi_j^-(t') 
\right) \rangle
\label{defresp}
\; .
\end{eqnarray}

Notice that as a result of having two external sources (one related to 
the evolution with time ordering and the other with anti-time ordering)
a doubling of degree of freedom is necessary when writing the
path-integral. As we explicitly show in Appendix \ref{classicallimit}, 
the doubling of 
degrees of freedom 
characteristic of this description is
intimately linked to the introduction of Lagrange multipliers in the MSR
description of the classical model.

The time-integration in Eq.~(\ref{aaab})
 can be interpreted as being closed, going 
forward from $t_0=0$ to $t=\infty$ and then 
backwards from $t=\infty$ to $t_0=0$. 
This motivates the name of the method.

\subsection{Coupling to a heat bath}
\label{Couplingtoaheatbath}

Up to now, the discussion has been completely general and it applies
to systems with an arbitrary number of degrees of freedom. We 
now consider the coupling of the system of interest 
to a thermal quantum bath, assumed  to be in equilibrium.
A convenient way to deal with the interaction between the system and 
the bath in the path-integral formalism is due 
to Feynman and Vernon. \cite{Feve}
In this approach one starts by describing the system plus bath  
 by ${\bbox \phi}=({\bbox \sigma},{\bbox v}^a)$
with ${\bbox \sigma}=(\sigma_1,\dots,\sigma_N)$ denoting the 
variables of the system 
and ${\bbox v}^a=(v_1^a,\dots,v_N^a)$, $a=1,\dots,N_b$, denoting the 
variables of the bath. The total action is
\begin{equation}
S[{\bbox\phi}]= S_s[{\bbox \sigma}] + S_b[{\bbox v}^a] + 
S_{sb}[{\bbox{\sigma,v}^a}]
\end{equation}
where $S_s$ is the action characterizing the system (and eventually
depending upon disorder), $S_b$ is the action for the bath, and
$S_{sb}$ contains the system-bath interaction terms.
For definiteness we assume  that our system interacts linearly
with the bath which we model as a set of independent harmonic oscillators,
\begin{equation}
S[{\bbox \phi}] =S_{ s}[{\bbox \sigma}] + \frac12  \sum_{a=1}^{N_b} \, M_a \, 
\left( \dot {\bbox v}_a^2 - \omega_a^2 \, {\bbox v}_a^2 \right) 
- \sum_{a=1}^{N_b} C_a {\bbox v}_a {\bbox \sigma}
\; .
\end{equation}
$M_a$ are their masses, $\omega_a$ their frequencies and $C_a$ the coupling
constants between the oscillators and the system.
Of course, more general couplings between bath and system are 
possible but this is the simplest choice that allows us to go a bit
farther analytically. 
If we further assume that the initial density 
matrix $\hat \rho(0)$ is factorizable as
\begin{equation}
\hat \rho(0) = \hat{ \rho_s}(0) \times \hat{\rho_b} (0)
\; ,
\end{equation}
where $\hat{\rho_b} (0)$ is the  Boltzman
distribution for the bath variables at equilibrium at a temperature
$k_B T=1/\beta$, one can
integrate out the bath variables,  obtaining in this way
an effective thermal action 
$S_{T}[{\bbox \sigma}^+,{\bbox \sigma}^-]$, that enters  the 
{\it Feynman-Vernon
influence functional}:
\begin{eqnarray}
 S_{T} [ {\bbox \sigma}^+ , {\bbox \sigma}^-] &=& 
- \int dt \int dt' \;
\left( {\bbox \sigma}^+(t) - 
{\bbox \sigma}^-(t) \right) \, \eta(t-t') \,  
\left( {\bbox \sigma}^+(t') + {\bbox \sigma}^-(t') \right) 
\nonumber\\
& & 
+i\int dt \int dt' \; \left( {\bbox \sigma}^+(t) - 
{\bbox \sigma}^-(t) \right) \, \nu(t-t') \,  
\left( {\bbox \sigma}^+(t') - {\bbox \sigma}^-(t') \right) 
\; .
\label{Sthermal}
\end{eqnarray}
The noise  and dissipative kernels $\nu$ and $\eta$ are given by 
\cite{Feve,Cale,Hupazh}
\begin{eqnarray}
\nu(t-t') &=&  \int_0^\infty d\omega I(\omega) \, 
\coth\left( \frac12 \beta \hbar \omega \right) \; 
\cos(\omega (t-t'))
\; ,
\label{nu}
\\
\eta(t-t') &=&  
-\theta(t-t') \, 
\int_0^\infty d\omega \; I(\omega) \, \sin(\omega (t-t')) 
\; .
\label{eta}
\end{eqnarray}
In these equations, $I(\omega)$ is the spectral 
density of the bath:
\begin{equation}
I(\omega) \equiv \sum_{a=1}^{N_b} 
\delta(\omega-\omega_a) \, \frac{C_a^2}{2 M_a \omega_a}
\; .
\end{equation}
The system-bath interaction manifests via the appearance of 
two quadratic and non-local terms (\ref{Sthermal}) 
in the effective action. As it can be most
easily seen in the classical limit presented in Appendix \ref{classicallimit}
they can be related to the noise and dissipative terms
of the associated Langevin formulation of the classical dynamics. 

In addition, it is easy to see that the kernels $\nu$ and $\eta$ are 
related to the correlation
and response functions of the bath. In fact $4\eta(t-t')$ is the 
response of the bath 
while $-2\hbar\nu(t-t')$ is the correlation of the bath. They are related in a way 
dictated by the fluctuation-dissipation theorem (FDT): 
\begin{equation}
4 \eta(\omega)
=
\frac1{\hbar} \, \lim_{\epsilon\to 0^+} 
\int \frac{d\omega'}{\pi} \; \frac{1}{\omega-\omega'+i \epsilon} \;
\tanh \left( \frac{\beta \hbar \omega'}{2}\right) \, 
2 \hbar \nu(\omega')
\; .
\end{equation}
(See Appendix \ref{appFDT} for the derivation and 
properties of the quantum FDT.)
This equation is the quantum counterpart of the 
fluctuation-dissipation relation of classical relaxational dynamics,
that fixes the dissipative coefficient in terms of the
strength of the noise correlation of the  Langevin equation. 
It expresses the fact that the bath is assumed to be 
in equilibrium at the initial time $t_o=0$ and that it stays 
in equilibrium at all subsequent 
times $t > 0$.  

It is important to notice that the assumption of thermal 
equilibrium of the bath  {\it does not 
imply} that the system -- that is in contact with  it --
is in equilibrium at the initial time nor that it 
will reach equilibrium at a given equilibration time $t_{\sc eq}$. 
It is known that ``mean-field'' classical disordered models with relaxational 
dynamics -- the $p$ spin-glass model is an example --- do not reach equilibrium at 
any time if the limit of $N$ 
going to infinity is taken at the outset of the calculation.\cite{Cuku} 
It is one of the main aims of 
this paper to show that this result carries through to the quantum case. 
As a consequence of the out of equilibrium behavior, we shall 
find that the FDT {\it does not hold} between response and correlation
of the system in a sub-critical region of the phase diagram.

The integrals in the $\eta$ and  $\nu$ definitions are, in general,
ill-defined (divergent) and might need to be
regularized. A way to regularize  them\cite{Hupazh} is to observe that 
one expects $I(\omega)$ to decrease to zero for large $\omega$ and
introduce an explicit  cut-off:
\begin{equation}
I(\omega) \sim 0 \;\;\;\;\;\;\;\;
{\mbox{for}} \;\;\;\; \omega > \Lambda
\; .
\end{equation}
By modifying $I(\omega)$ in this way, we regularize the kernels
 $\eta$ and  $\nu$ at the same time preserving the
 FDT relation for the bath variables; in other words, we do not  break
 the 
assumption of equilibrium of the bath.

Different environments are characterized  by different behaviors 
of $I(\omega)$  such that for $\omega < \Lambda$:
\begin{eqnarray}
I(\omega) &\sim& \omega^a \;\;\;\;\;\; \mbox{and} \;\;
\left\{ 
\begin{array}{l}
a = 1 \;\;\;\; \mbox{Ohmic} \; ,
\nonumber\\
a < 1 \;\;\;\; \mbox{Subohmic} \; ,
\nonumber\\
a > 1 \;\;\;\; \mbox{Superohmic} \; .
\end{array}
\right.
\end{eqnarray}
A typical example is \cite{Leggetreview,Gardiner}
\begin{equation}
I(\omega) = \frac{M \gamma_o}{\pi} \; \omega \left( \frac{\omega}{\Lambda} 
\right)^{a-1} \,
 \exp\left(-\frac{|\omega|}{\Lambda}\right)
\; 
\end{equation}  
where
$M\gamma_o$ is a constant that plays the r\^ole of a friction coefficient.

Throughout this paper, we concentrate 
in the Ohmic case. 
The kernel $\eta$ is
\begin{equation}
\eta(t-t') =  \theta(t-t') \; \frac{M\gamma_o}{\pi} \;
\frac{d}{d(t-t')} \left( \frac{\Lambda}{1 + (\Lambda (t-t'))^2} \right)
\; .
\end{equation}
In the limit where the cut-off tends to infinity it becomes
\begin{eqnarray}
\lim_{\Lambda\to\infty} \eta(t-t') &=&   
\theta(t-t') \;  M \gamma_o \delta'(t-t') 
\; .
\end{eqnarray}
Of particular importance is the zero temperature limit of the kernel $\nu$:
\begin{equation}
\lim_{T\to 0} \nu(t-t') 
=
\frac{M\gamma_o}{\pi} \, \Lambda^2 \, 
\frac{1-(\Lambda(t-t'))^2}{(1+(\Lambda(t-t'))^2)^2}
\; .
\end{equation}
One sees that the quantum fluctuations yield a non-trivial kernel even in the 
absence of thermal fluctuations. In the zero temperature limit 
the characteristic time goes to infinity and  
the  long-time decay of $\nu$ becomes power-law:  
\begin{equation}
\lim_{t-t'\to\infty} \lim_{T\to 0} \nu(t-t') 
=
- 
\frac{M\gamma_o}{\pi} \, (t-t')^{-2}
\; .
\end{equation}
In the classical limit, for any temperature, $\nu$ becomes
\begin{eqnarray}
\lim_{\hbar\to 0} 2 \hbar \nu(t-t') &=&
\frac{4M\gamma_o k_B T}{\pi} \; \frac{\Lambda}{1+\Lambda^2 (t-t')^2}
\; .
\end{eqnarray}
Note that for a finite cut-off $\Lambda$,  the classical limit of the kernel 
$\nu$ 
also has a rather slow, power-law decay at large time differences.
If, next, the cut-off is sent to infinity one recovers a delta function 
\begin{eqnarray}
\lim_{\Lambda\to\infty} \lim_{\hbar\to 0} 2 \hbar \nu(t-t')
&=&
4M\gamma_o k_B T  \, \delta(t-t')
\end{eqnarray}
characteristic of e.g. a white thermal noise. 
The classical limit of these terms is further discussed in Appendix
\ref{classicallimit} where we
show that the dynamic equations of 
the corresponding 
classical model correspond to  a friction coefficient
$\Gamma_o^{-1} \equiv 2 M \gamma_o$.

The coupling to the quantum thermal bath introduces 
three time-scales into the problem. First, 
the inverse cut-off $1/\Lambda$ characterizes the memory
of the bath. Second, the coupling strength 
$M \gamma_o$ yields the inverse
relaxational characteristic time. Third, 
$(\beta\hbar)^{-1}$ accounts for the relative importance
of quantum to thermal effects.

\subsection{CTP for disordered systems}
\label{CTPfordisorderedsystems}

The CTP formalism results to be well suited for problems with 
disorder.\cite{chinos}
Let us suppose that the system under consideration is described by 
the action,
\begin{equation}
S_s [{\bbox{\sigma}}, J] = S_{\sc o}[\bbox{\sigma}] - 
\int dt \, V[\bbox{\sigma}, J]
\end{equation}
where $S_{\sc o}$ is the disordered independent part of the action and
$V[{\bbox \sigma},J]$ is the potential energy that
depends upon random couplings collectively denoted by $J$.   As usual in 
disordered systems, we concentrate on 
quantities that are averaged over the disorder distribution $P[J]$.

We wish to consider the real-time dynamics of the system 
starting from a {\it random initial condition}
at time $t_o=0$ when it is set in contact with the environment (that 
is itself in equilibrium at a constant temperature $T$).  
Given that the initial condition is chosen to be 
random, it is not correlated with the 
disorder. Therefore, the density operator $\hat \rho(0)$ does 
not depend upon disorder and the generating functional without sources 
\begin{equation}
{\cal Z}[{\bbox \xi}^+=0,{\bbox \xi}^-=0,J] 
= \mbox{Tr} \, [ \, \hat\rho(0) \, ]
\end{equation}
{\it is also independent of disorder}. 

This property  is equivalent to the independence 
of the classical Martin-Siggia-Rose (MSR) generating functional \cite{MSR} 
without sources 
upon disorder.\cite{cirano} 
As in the classical case, it allows us to write dynamic equations for 
random initial conditions 
without having to compute the average over disorder of 
$\ln {\cal Z}[{\bbox \xi}^+,{\bbox \xi}^-,J]$ 
and hence without 
resorting to the use of replicas. 
We are then interested in the averaged generating functional:
\begin{equation}
\overline{ {\cal Z} [ \bbox{\xi}^+, \bbox{\xi}^-] }= \int d J P[J] \; 
{\cal Z} [ \bbox{\xi}^+, \bbox{\xi}^-, J]
\; ,
\end{equation}
from which any averaged operator can be computed as
\begin{eqnarray}
\overline{ \langle {\bbox \sigma}(t) \rangle }
\equiv
\left. 
\overline{
\frac{ \delta \ln {\cal Z}[ {\bbox \xi}^+,{\bbox \xi}^-,J ] }{\delta {\bbox 
\xi}^+(t)} 
}
\right|_{
{\bbox \xi} =0
}
=
\left.
\frac{1}{{\cal Z}[0,0,J]} \;
\frac{\delta \overline{{\cal Z}[{\bbox \xi}^+,{\bbox \xi}^-,J]} }{\delta {\bbox 
\xi}^+(t)} 
\right|_{
{\bbox \xi} =0
}
\; ,
\end{eqnarray}
with all sources set to zero and ${\bbox \sigma}^+={\bbox \sigma}^-={\bbox 
\sigma}$. Here and in what follows, the overline represents an average 
over the disorder.

In many cases of interest, the integration over the disorder can be
performed explicitly, 
\begin{equation}
\int dJ P[J] \; \exp \left[-\frac{i}{h}  \int dt 
\left( \, V[{\bbox \sigma}^+, J] -
V[\bbox{\sigma} ^-, J] \, \right)
\right]=
\exp \left[- \frac{i}{h} V_{\sc d} [\bbox{ \sigma}^+, \bbox{\sigma}^-] \right]
\; 
\end{equation}
and it introduces a non-linear  interaction term
that is, in general, non-local both in time and in the ``space'' indices $i,j$. 

A rather general way of introducing randomness is to 
consider a model with a Gaussian random potential energy term 
$V[{\bbox \sigma}, J]$
depending upon the variables ${\bbox \sigma}=(\sigma_1,\dots,\sigma_N)$ and 
correlated as 
\begin{equation}
\overline {V[{\bbox \sigma}, J] V[{\bbox \sigma}', J]}  = 
- \frac{N}{2} {\cal V}
\left( \frac{ ( {\bbox \sigma} - {\bbox \sigma}')^2}{N} \right) 
\label{randompotential}
\; .
\end{equation}
The variables 
${\bbox \sigma}$ may represent spins leading to a  
spin-glass model or a position in an $N$-dimensional 
space becoming the problem of the dynamics and/or diffusion of a particle 
in a random potential. 
Since we are interested in following the real time 
dynamics of such systems, ${\bbox \sigma}$ will be a time-dependent 
$N$-vector.

For the sake of definiteness, we shall solve 
 a particular realization 
of the random potential that in the classical case 
defines the so-called $p$ spin-glass: \cite{pspinIsing}
\begin{equation}
V[{\bbox \sigma},J] =  - \sum_{i_1 < \dots < i_p}^N 
J_{i_i\dots i_p} \, \sigma_{i_1} \dots \sigma_{i_p} 
\;. 
\end{equation}
$p$ is a parameter, $p \geq 2$.
The interaction strengths are quenched independent random variables
with a Gaussian distribution
\begin{equation}
P[J] 
= 
\sqrt{ \frac{N^{p-1}}{{\tilde J}^2\pi p!}} \; \exp \left( - 
\frac{N^{p-1}}{{\tilde J}^2 p!}
\sum_{i_1 \neq \dots \neq i_p}  (J_{i_i\dots i_p})^2  \right) 
\;\;\;\;\;\;\;\;
\Rightarrow
\;\;\;\;\;\;\;\;
\overline{ 
(J_{i_i\dots i_p})^2 }
=
\frac{{\tilde J}^2 p!}{2 N^{p-1}}
\; .
\end{equation}
This corresponds to a random potential correlation ${\cal V}(x) = 
-1/2 (1-x/2)^p$ and it leads to an effective,  non-local in time, interaction
\begin{eqnarray}
V_{\sc d}[ {\bbox \sigma}^+,{\bbox \sigma}^-] &=& 
\frac{{\tilde J}^2 Ni}{4 \hbar} \int dt \int dt' 
\left[ \;
\left( \frac1N {\bbox \sigma}^+(t) {\bbox \sigma}^+(t') \right)^p +  \left( 
\frac1N 
{\bbox \sigma}^-(t) {\bbox \sigma}^-(t') \right)^p
\right.
\nonumber\\ 
& &
\left. 
\;\;\;\;\;\;\;\;\;\;\;\;\;\;\;\;\;\;\;\;\;\;\;\;\;\;\;\;\;\;\;\;\;
- 
\left( \frac1N {\bbox \sigma}^+(t) {\bbox \sigma}^-(t') \right)^p -  \left( 
\frac1N 
{\bbox \sigma}^-(t) {\bbox \sigma}^+(t') \right)^p
\;\; \right]
\; .
\label{actionV}
\end{eqnarray}

In its spherical version,\cite{Crso}
$\sigma_i, i=1,\dots,N$ are continuous dynamic variables 
$-\sqrt{N} < \sigma_i < \sqrt{N}$, $\forall i$,  that satisfy the global spherical 
constraint 
\begin{equation}
\frac1N \, {\bbox \sigma}^2(t) = \frac1N \, \sum_{i=1}^N \sigma_i^2(t) = 1
\; ,
\end{equation}
at each instant. We enforce this constraint by 
introducing a time-dependent Lagrange multiplier $z(t)$.

The disordered-independent part of the action is
\begin{equation}
S_{\sc o}[{\bbox \sigma}, J] = 
\int dt \, 
\left[ \frac{m}{2} {\dot {\bbox \sigma}}^2 - 
\frac{z}{2} \left( {\bbox \sigma}^2 -N \right) 
\right] 
\label{actiono}
\; .
\end{equation}
As explained in the Introduction, for simplicity 
we have chosen a kinetic term that dictates a microscopic dynamics 
with second derivatives in time. Other choices, for example leading to 
first time-derivatives, are also possible. The derivation of the dynamic 
equations that we explain in Section~\ref{meanfieldequations}
applies with only minor modifications.

\vspace{.2cm}
\section{Mean-field equations}
\label{meanfieldequations}
\vspace{.2cm}

We now have all the elements to derive the equations which describe
the dynamic evolution of a disordered quantum system. According to the 
discussion in Section \ref{CTP}, the system is described by the generating 
functional
\begin{equation}
\overline{{\cal Z}[{\bbox \xi}^+, {\bbox \xi}^-, J] } =
\int 
{\cal D}{\bbox \sigma}^+
{\cal D}{\bbox \sigma}^-
\exp\left[ \frac{i}{\hbar} \, \left( S_{\sc eff}[{\bbox \sigma^+}, 
{\bbox \sigma^-}] + \int dt \, {\bbox \xi}^+(t) {\bbox \sigma^+}(t) -
\int dt \, {\bbox \xi}^-(t) {\bbox \sigma^-(t)}
\right) \right]
\end{equation}
with
\begin{equation}
S_{\sc eff}[{\bbox \sigma}^+, {\bbox \sigma}^-]
= S_{\sc o}[{\bbox \sigma}^+] - S_{\sc o}[{\bbox \sigma}^-]
+ S_{\sc t}[{\bbox \sigma}^+, {\bbox \sigma}^-] 
- V_{\sc d}[{\bbox \sigma}^+, {\bbox \sigma}^-]
\; .
\label{action1}
\end{equation}
We remind that $S_{\sc o}$ is the disorder independent part
of the action given by Eq.~(\ref{actiono}), $S_{\sc t}$  accounts for the
the system-bath interaction, Eq.~(\ref{Sthermal}), and $V_{\sc d}$ is the effective
potential which arises as a result of the integration over
the disorder and is given by Eq.~(\ref{actionV}). 

The quantum dynamic equations of motion
follow from similar  steps  to those usually used \cite{Sozi} 
to obtain the classical equations of motion in the Martin-Siggia-Rose formalism. 
We introduce macroscopic order parameters and derive, through a 
saddle-point point approximation of the KS generating functional 
(that becomes exact when $N\to\infty$),  
the dynamic equations of motion. Since we are interested in considering the 
dynamics of the system for {\it large but  finite} 
times with respect to $N$,
out of equilibrium effects are expected.
We shall therefore make no 
assumption about the time-dependence of the order-parameters.

\subsection{Dynamic order parameters}

The quadratic terms in the action can be condensed into one term 
by introducing the operator
\begin{eqnarray}
{\cal O}p(t,t') 
&=&
\left(
\begin{array}{rl}
\mbox{Op}^{++}(t,t') \;& \; \mbox{Op}^{+-}(t,t') 
\nonumber\\
\mbox{Op}^{-+}(t,t') \; & \; \mbox{Op}^{--}(t,t') 
\end{array}
\right)
=
\{ \mbox{Op}^{\alpha\beta} (t,t')\}
\; ,
\nonumber\\
\;
\nonumber\\
\mbox{Op}^{++}(t,t') &=&
(m \partial^2_t + z^+(t)) \, \delta(t-t') - 2 i \nu(t-t') 
\nonumber\\
\mbox{Op}^{+-}(t,t') &=&
2 \eta(t-t') + 2 i \nu(t-t') 
\nonumber\\
\mbox{Op}^{-+}(t,t') &=&
- 2 \eta(t-t') + 2 i \nu(t-t') 
\nonumber\\
\mbox{Op}^{--}(t,t') &=&
-(m \partial^2_t + z^-(t)) \, \delta(t-t') - 2 i \nu(t-t') 
\; 
\end{eqnarray}
in such a way that
\begin{equation}
S_{\sc eff}[{\bbox \sigma}^+, {\bbox \sigma}^-]
= 
-\frac12 
\int dt  \int dt'
{\bbox \sigma}^\alpha(t) \,
\mbox{Op}^{\alpha\beta}(t,t') {\bbox \sigma}^\beta(t') 
- V_{\sc d}[{\bbox \sigma}^+, {\bbox \sigma}^-]
\end{equation}
where Greek indices label $\alpha=+,-$ and 
the sum convention is assumed.

Introducing the identity
\begin{eqnarray}
1 &=& \int \prod^{\alpha\beta} DQ^{\alpha\beta} \;
     \delta\left( \frac{1}{N}{\bbox \sigma}^\alpha(t) {\bbox \sigma}^\beta(t') - 
Q^{\alpha\beta}(t,t') \right)
\nonumber\\
&\propto& 
\int \prod_{\alpha\beta} DQ^{\alpha\beta} \; D\lambda^{\alpha\beta} 
\exp\left( \;  - \frac{i}{2\hbar} \, \lambda^{\alpha\beta} \, 
\left( {\bbox \sigma}^\alpha(t) {\bbox \sigma}^\beta(t')  - 
N Q^{\alpha\beta}(t,t') \right)
\; 
\right)
\; 
\end{eqnarray}
into the generating functional, 
the full action can be rewritten as
\begin{eqnarray}
S_{\sc eff}[{\bbox \sigma}^+,{\bbox \sigma}^-]
&=& 
- \frac{1}{2}\int dt  \int dt' \; 
{\bbox \sigma^\alpha}(t) \,
\left( \, \mbox{Op}^{\alpha\beta}(t,t') 
+  \lambda^{\alpha\beta}(t,t') \right)
{\bbox \sigma}^\beta(t')  
\nonumber\\
& & 
+ \frac{N}{2} \int dt \int dt' \lambda^{\alpha\beta}(t,t') \, 
Q^{\alpha\beta}(t,t')
+ \frac{N}{2}  \int dt \left( z^+(t) - z^-(t) \right) 
\nonumber\\
& & 
+ \frac{i {\tilde J}^2 N}{4 \hbar} \int dt \int dt' 
\left[ \,
\left(Q^{++}(t,t') \right)^p +  \left( Q^{--}(t,t') \right)^p
\right.
\nonumber\\
& & 
\;\;\;\;\;\;\;\;\;\;\;\;\;\;\;\;\;\;\;\;
-
\left.
\left( Q^{+-}(t,t') \right)^p -  \left( Q^{-+}(t,t') \right)^p
\, \right]
\; .
\end{eqnarray}
The stationary-point values of the 
order parameters $Q^{\alpha\beta}(t,t')$ with $\alpha,\beta=+,-$ are related to 
the ``physical'' correlations and responses defined in Eqs.~(\ref{defcorr})  
and (\ref{defresp}) as follows
\begin{eqnarray}
N Q^{++}(t,t') = \overline{\langle {\bbox \sigma}^{+}(t)  {\bbox \sigma}^{+}(t') 
\rangle}
&=&
N \left( C(t,t') - \frac{i\hbar}{2} \; (R(t,t') + R(t',t)) \right)
\; ,
\label{Q++}
\\
N Q^{+-}(t,t') = \overline{ \langle {\bbox \sigma}^{+}(t)  {\bbox 
\sigma}^{-}(t') \rangle}
&=&
N \left( C(t,t') + \frac{i\hbar}{2} \; (R(t,t') - R(t',t)) \right)
\; ,
\label{Q+-}
\\
N Q^{-+}(t,t') = \overline{\langle {\bbox \sigma}^{-}(t)  {\bbox \sigma}^{+}(t') 
\rangle}
&=&
N \left( C(t,t') - \frac{i\hbar}{2} \; (R(t,t') - R(t',t)) \right)
\; ,
\label{Q-+}
\\
N Q^{--}(t,t') = \overline{\langle {\bbox \sigma}^{-}(t)  {\bbox \sigma}^{-}(t') 
\rangle}
&=&
N \left( C(t,t') + \frac{i\hbar}{2} \; (R(t,t') + R(t',t)) \right)
\; ,
\label{Q--}
\end{eqnarray}
with 
\begin{eqnarray}
N C(t,t') &\equiv& \frac{1}{2} 
\sum_{i=1}^N \overline{\langle \sigma_i^+(t) \sigma_i^-(t') + \sigma_i^-(t) \sigma_i^+(t') \rangle}
\; ,
\nonumber\\
N R(t,t') &\equiv& \frac{i}{\hbar} \sum_{i=1}^N \overline{\langle \sigma_i^+(t) 
\left( \sigma_i^+(t') -\sigma_i^-(t') \right) \rangle }
\; .
\end{eqnarray} 
It is easy to check that these functions  satisfy the identity
\footnote{At the 
classical level and in the Martin-Siggia-Rose language,
this identity reduces to the 
condition $\langle i \hat {\bbox s}(t)  i \hat {\bbox s}(t')  \rangle = 0$ 
for all pairs of times $t,t'$. In the classical case this is a 
requirement to be imposed 
 upon the saddle-point two-point functions in order to 
preserve causality. In the quantum case, this 
condition emerges as a property of the Green 
functions.}
\begin{equation}
Q^{++}(t,t') +Q^{--}(t,t') -Q^{+-}(t,t')-Q^{-+}(t,t') =0
\label{identityQo}
\; .
\end{equation}

The functional  integration over $\sigma_i^+(t)$ and $\sigma_i^-(t)$ is now 
quadratic and can be performed.
This amounts to replace the quadratic term in $i/\hbar S_{\sc eff}$ by
\begin{eqnarray}
-\frac{N}{2}
\int dt  \int dt'
\, \mbox{Tr} \log 
\left( \, \frac{i}{\hbar} \mbox{Op}^{\alpha\beta}(t,t') + 
\frac{i}{\hbar} \lambda^{\alpha\beta}(t,t') \right)
\; .
\end{eqnarray}

\subsection{Saddle-point evaluation}

At this stage, all terms in the action depend upon the 
``macroscopic'' quantities 
$\lambda^{\alpha\beta}, Q^{\alpha\beta}$ and $z^\alpha$ and are proportional to 
$N$. 
Since it is easier to write the equations in matrix notation, we encode 
$\lambda^{\alpha\beta}$ and 
$Q^{\alpha\beta}$ in two matrices
\begin{eqnarray}
{\cal L} = \left( 
\begin{array}{c}
\lambda^{++} \;\; \lambda^{+-}
\nonumber\\
\lambda^{-+} \;\; \lambda^{--}
\end{array}
\right)
& \;\;\;\;\;\;\;\;\;\;\;\;\;\;\;\;\;\;\;\;\;\;
{\cal Q} = \left( 
\begin{array}{c}
Q^{++} \;\; Q^{+-}
\nonumber\\
Q^{-+} \;\; Q^{--}
\end{array}
\right)
\; 
\end{eqnarray}
and we define
\begin{eqnarray}
F[{\cal Q}](t,t') 
&\equiv&
\left(
\begin{array}{rl}
(Q^{++}(t,t'))^{p-1} 
\; &  \;
-(Q^{+-}(t,t'))^{p-1}
\\
-(Q^{-+}(t,t'))^{p-1}
\; &  \;
(Q^{--}(t,t'))^{p-1}
\end{array}
\right)
\label{saddleF}
\; .
\end{eqnarray}
We denote with a cross the standard operational product
\begin{eqnarray}
{\cal A} \otimes {\cal B} (t,t')
&=&
\left(
\begin{array}{c}
\int dt'' \, A^{+\gamma}(t,t'') B^{\gamma +}(t'',t') 
\;\;\;\;\;\;
\int dt'' \, A^{+\gamma}(t,t'') B^{\gamma -}(t'',t') 
\nonumber\\
\int dt'' \, A^{-\gamma}(t,t'') B^{\gamma +}(t'',t') 
\;\;\;\;\;\;
\int dt'' \, A^{-\gamma}(t,t'') B^{\gamma -}(t'',t') 
\end{array}
\right)
\; 
\end{eqnarray}
where a sum over $\gamma$ is assumed.
The saddle-point with respect to $\lambda^{\alpha\beta}(t,t')$ 
yields
\begin{equation}
{\cal L}(t,t') = \frac{\hbar}{i} {{\cal Q}^{-1}}(t,t') - {\cal O}p(t,t')
\; .
\label{saddlelambda}
\end{equation}
The matrix and time-operator inverse of  ${\cal Q}$ is denoted 
${\cal Q}^{-1}$. 
The saddle-point equation with respect to $Q^{\alpha\beta}(t,t')$ yields
\begin{equation}
{\cal L}(t,t') = -\frac{ip \, {\tilde J}^2}{2\hbar} \, F[{\cal Q}](t,t')
\; .
\label{saddleq}
\end{equation}
Equations~(\ref{saddlelambda}) and (\ref{saddleq}) imply, in a compact
matrix and time-operator  notation, 
\begin{equation}
\frac{i}{\hbar} \mbox{Op} \otimes {\cal Q}  
= I  - \frac{p \, {\tilde J}^2}{2\hbar^2} \, F[{\cal Q}] \otimes {\cal Q}
\; ,
\label{saddleQ}
\end{equation}
where $I$ is the identity: 
$I^{\alpha\beta}(t,t')= \delta^{\alpha\beta} \delta(t-t')$.
The saddle-point with respect to $z^\alpha$ yields
\begin{eqnarray}
\frac{i}{\hbar} &=& 
({\cal O}p + {\cal L})^{-1}_{++}(t,t) = \frac{i}{\hbar} Q^{++}(t,t)
\; ,
\label{z+}
\\
\frac{i}{\hbar} &=& 
({\cal O}p + {\cal L})^{-1}_{--}(t,t)= \frac{i}{\hbar} Q^{--}(t,t)
\label{z-}
\end{eqnarray}
and these equations lead, as expected, to the spherical constraint.

In the limit $N\to\infty$ one could also proceed as in
Refs.~[\raisebox{-.22cm}{\Large \cite{Sozi,Koge}}]
and write the full action $S_{\sc eff}$
in terms of a {\it single} variable. This is at the expense
of modifying the thermal kernel and the interaction term
in a self-consistent way, through the introduction of terms
arising from the non-linear interactions (the vertex
$\tilde D$ and the self-energy $\tilde \Sigma$, respectively). 

This procedure is not of particular usefulness for the analysis of 
the model we treat in this paper since the single variable
effective action is Gaussian. It does however 
become useful for dealing with quantum models whose single-spin 
effective action has higher order interaction  terms. An example 
is the quantum Sherrington-Kirkpatrick model. 
One could then envisage to derive an effective quantum Langevin equation
for the single variable\cite{Kleinert} and study this 
equation with an 
adequate numerical algorithm as the one developed by Eissfeller and 
Opper.\cite{Eiop}

\subsection{Dynamical equations for correlation and response}

The dynamic equations for the auto-correlation and response 
follow from the set of equations (\ref{saddleF})-(\ref{saddleQ})
and the definitions of the dynamic order parameters given in 
Eqs.~(\ref{Q++})-(\ref{Q--}). 
More precisely, the equation of motion for the response function follows from 
the subtraction of the $++$ and $+-$ components of Eq.~(\ref{saddleQ}):
\begin{eqnarray}
& & 
\left( m \partial^2_t + z^+(t) \right) R(t,t') + 
4 \int_{t'}^t dt'' \; \eta(t-t'') \, R(t'',t') 
\;\;\;\;\;\;\;\;\;\;\;
\nonumber\\
& & 
=
\delta(t-t') - \frac{p {\tilde J}^2}{2 i \hbar} \, 
\int_0^\infty dt'' 
\left[
(Q^{++}(t,t''))^{p-1} - (Q^{+-}(t,t'')^{p-1}) 
\right] \, 
R(t'',t')
\; ,
\label{eqR1}
\end{eqnarray} 
and the equation of motion for the auto-correlation function 
follows from the addition of the  $+-$ and 
$-+$ components of Eq.~(\ref{saddleQ}):
\begin{eqnarray}
& & 
\left( m \partial^2_t + \frac12 \, \left(z^+(t) + z^-(t) \right) \right) C(t,t') 
+ 
\frac{i}{2} \left(z^+(t) - z^-(t) \right) \hbar (R(t',t)-R(t,t') )
\;\;\;\;\;\;\;\;\;\;\;
\nonumber\\ 
& & 
-2\hbar \int_0^\infty dt'' \nu(t-t'') R(t',t'') +
4 \int_{0}^t dt'' \; \eta(t-t'') \, C(t'',t') 
\nonumber\\
& & 
=
-\frac{ p {\tilde J}^2}{2\hbar} \, 
\int_o^\infty dt'' \;
\mbox{Im} \left[
(Q^{++}(t,t''))^{p-1} Q^{+-}(t'',t') - (Q^{+-}(t,t''))^{p-1}Q^{--}(t'',t')
 \right]
\label{eqC1}
\; .
\end{eqnarray}
Written in this way, 
Eq. (\ref{eqC1}) is complex. Its imaginary part yields 
\begin{equation}
z(t) \equiv z^+(t)=z^-(t)
\; .
\end{equation}
Moreover, since the response is causal, products of advanced $R(t,t'')$ and 
retarded $R(t'',t')$ responses vanish identically for all $t,t''$: 
\begin{equation}
R(t,t'')R(t'',t) =0
\;\;\;\;\;\;\; \forall \; t,t''
\end{equation}
and one can show that for any integer $k >0$ and any constants
$c_1$, $c_2$
\begin{equation}
\left[ C(t,t') + c_1 R(t,t') + c_2 R(t',t) \right] ^k=
\left[ C(t,t') + c_1 R(t,t')  \right] ^k +
\left[ C(t,t') + c_2 R(t',t) \right] ^k
- C^k(t,t')
\; .
\end{equation}
Using this property one has
\begin{equation}
(Q^{++}(t,t''))^{p-1} - (Q^{+-}(t,t''))^{p-1} =
2i \mbox{Im} \left[ C(t,t'')-\frac{i \hbar}{2} R(t,t'') \right]^{p-1}
\end{equation}
and 
\begin{eqnarray}
& &\mbox{Im}\left[(Q^{++}(t,t''))^{p-1} Q^{+-}(t'',t') - 
(Q^{+-}(t,t''))^{p-1}Q^{--}(t'',t') \right]
= \nonumber\\
& & 2 C(t'',t') 
\mbox{Im} \left[ C(t,t'')-\frac{i \hbar}{2} R(t,t'') \right]^{p-1}
-\hbar R(t',t'') 
\mbox{Re}\left[ C(t,t'')-\frac{i\hbar}{2}( R(t,t'')+ R(t'',t)) \right]^{p-1}
\; .
\nonumber\\
\end{eqnarray}
We can identify the self-energy $\tilde \Sigma$ and the vertex $\tilde D$
as
\begin{eqnarray}
\Sigma(t,t')
+ 4 \eta(t-t') &\equiv& \tilde{\Sigma}(t,t') \equiv
-\frac{p \tilde{J}^2}{\hbar} 
\mbox{Im} \left[ C(t,t')-\frac{i \hbar}{2} R(t,t') \right]^{p-1}
\; ,
\label{sigmatilde}
\\
D(t,t')- 2 \hbar \nu(t-t') &\equiv&
\tilde{D}(t,t') \equiv
\frac{p \tilde{J}^2}{2}
\mbox{Re}\left[ C(t,t')-\frac{i\hbar}{2}( R(t,t')+ R(t',t)) \right]^{p-1}
\; .
\label{Dtilde}
\end{eqnarray}
For $t\geq t'$, they can be encoded in a single complex equation:
\begin{equation}
\tilde D(t,t') + \frac{i \hbar}{2} \, \tilde \Sigma(t,t') 
=
\frac{p {\tilde J}^2}2 \; \left( C(t,t') +  \frac{i \hbar}{2} \,R(t,t') 
\right)^{p-1}
\; .
\end{equation}
Note that the total self-energy $\Sigma$ and vertex $D$ are real
and have two contributions of different origin: one arises from the interaction of the 
system and the bath ($\eta$ and $\nu$) and one is caused by the 
non-linearities stemming from the average over disorder (that we 
called $\tilde \Sigma$ and $\tilde D$ in Eqs.~(\ref{sigmatilde}) and 
(\ref{Dtilde})). 
If $p=2$, $\tilde \Sigma$ and $\tilde D$ are
identical to the classical ones. Instead, if $p \geq 3$, the non-linear terms 
acquire an explicit dependence upon $\hbar$.

The dynamic equations can then be written in a  compact form 
\begin{eqnarray}
(m\partial^2_t + z(t) )R(t,t') 
&=& \delta(t-t') + \int_0^\infty dt'' \, \Sigma(t,t'') R(t'',t')
\; ,
\label{schwingerR}
\\
(m\partial^2_t + z(t)) C(t,t') 
&=& \int_{0}^\infty dt'' \, \Sigma(t,t'') C(t'',t') + \int_{0}^{t'} dt'' \, 
D(t,t'') R(t',t'')
\; ,
\label{schwingerC}
\end{eqnarray}
that we call later the $R$-eq. and $C$-eq., respectively.
It is important to realize that the self-energy $\tilde \Sigma(t,t')$
is proportional to the response function $R(t,t')$, which
in turns implies
\begin{equation}
\tilde \Sigma(t,t')= \Sigma(t,t')=0 \;\;\;\;\;\;\;\;\; {\mbox{for}} \;\;\;\;
t<t'
\; .
\end{equation}
This means that the upper limit of integration in Eqs.~(\ref{schwingerR}) and
(\ref{schwingerC}) is $t$, which renders the equations explicitly
causal.
There are no more independent equations for $R$ and $C$. 
The other two equations that can be obtained 
from Eq.~(\ref{saddleQ}) are the 
equation for $R(t',t)$, that is equivalent to the $R$-eq above,
and one equation that identically 
cancels by virtue of the identity (\ref{identityQo}). 

In their integrated form as Schwinger-Dyson equations the dynamic equations read:
\begin{eqnarray}
R(t,t') &=& G_o(t,t') + \int_{t'}^t dt'' \int_{t'}^{t''} dt''' \;
G_o(t,t'') \, \Sigma(t'',t''') \, R(t''',t')
\; ,
\label{intschwingerR}
\nonumber\\
C(t,t') &=& \int_{0}^t dt'' \int_{0}^{t'} dt''' \;
R(t,t'') \, D(t'',t''') \, R(t',t''')
\label{intschwingerC}
\; ,
\end{eqnarray}
with the propagator given by
\begin{equation}
G_o^{-1}(t,t') \equiv m\partial^2_{t} + z(t)
\; .
\label{freepropagator}
\end{equation}

Real and imaginary parts of 
Eqs.~(\ref{z+}) and (\ref{z-}) combined with  
Eq.~(\ref{saddlelambda}) imply the equal-times conditions 
\begin{eqnarray}
C(t,t) &=& 1
\; , 
\nonumber\\
R(t,t) &=&0
\; .
\end{eqnarray}
In addition, from Eq. (\ref{eqR1}) one obtains that the first derivative of the 
response function is discontinuous at equal times:
\begin{eqnarray}
\lim_{t'\to t^-} \partial_t R(t,t') &=& \frac{1}{m}  
\; ,
\nonumber\\
\lim_{t'\to t^+} \partial_t R(t,t') &=& 0 
\; ,
\end{eqnarray}
while from Eq. (\ref{eqC1}) one obtains that the 
correlation is continuous:
\begin{equation}
\lim_{t'\to t^-} \partial_t C(t,t') = \lim_{t'\to t^+} \partial_t C(t,t') = 0 
\; .
\end{equation}

The equation for $z(t) = z^+(t)=z^-(t)$ can be derived
from the Schwinger-Dyson equation (\ref{intschwingerC}), 
by imposing the spherical constraint through the evaluation at $t=t'$. 
Multiplying operationally by $G_o^{-1}$ one obtains
\begin{eqnarray}
z(t) 
&=&
\int_0^t dt'' \left[
\Sigma(t,t'') C(t,t'') + D(t,t'') R(t,t'') 
\right] 
\nonumber\\
& & 
+ m \int_0^t dt'' \int_0^t  dt''' \,  
 (\partial_t R(t,t'') ) \,D(t'',t''')\, (\partial_t R(t,t''') )
\; .
\end{eqnarray}
The last term is a consequence of having a kinetic term with 
second derivatives. Had we chosen a first derivative term 
in the kinetic energy the last term would have not appeared.
It can be easily identified with minus the second-derivative of the 
correlation at equal times by taking the limit $t'\to t^-$ in 
Eq.~(\ref{schwingerC}). Thus
\begin{equation}
z(t) 
=
\int_0^t dt'' \left[
\Sigma(t,t'') C(t,t'') + D(t,t'') R(t,t'') 
\right] 
-
m
\left.
\frac{\partial^2 }{\partial t^2} C(t,t') 
\right|_{t'\to t^-}
\; .
\label{zeq}
\end{equation}

In conclusion, Eqs.~(\ref{schwingerR}), (\ref{schwingerC}) and (\ref{zeq}) 
are the complete set of equations that determines the dynamics of the 
system.
In the following we make a time reparametrization
\begin{equation}
\hat t = M \gamma_o t
\end{equation}
that implies
\begin{eqnarray}
\hat C = C 
\;\;\;\;\;\;\;\;\;\;\;\;\;\;
\hat R = \frac1{M \gamma_o } R
\end{eqnarray}
and  
\begin{eqnarray}
\hat m = (M \gamma_o)^2 m
\; ,
\;\;\;\;\;\;\;\;\;\;
\hat \hbar =  {M \gamma_o} \hbar
\; ,
\;\;\;\;\;\;\;\;\;\;
\hat{\tilde J} = {\tilde J} 
\; ,
\;\;\;\;\;\;\;\;\;\;
\hat \beta = \beta
\;\;\;\;\;\;\;\;\;\;
\hat \Lambda = \frac{\Lambda}{M \gamma_o}
\; .
\end{eqnarray}
The system of units we use is 
\begin{eqnarray}
[\hat C] &=& [\hat R] = [\hat \hbar] =1
\; ,
\nonumber\\
\left[\hat m \right] 
&=& 
[\hat \beta] =[ 1/\hat{\tilde J}]  = [ 1/\hat{\Lambda}] = [\hat{t}] 
\; .
\end{eqnarray}
Hereafter we write the equation in reparametrized variables and 
drop all hats.

\section{Solutions to the mode coupling equations}
\label{scenario}

As already mentioned, the set of integro-differential equations
(\ref{schwingerR}), (\ref{schwingerC}) and (\ref{zeq}) is causal -- 
as in the classical case -- and one
can attempt a numerical solution by a simple algorithm.
The correlation $C(t,t')$ and response
$R(t,t')$ are {\it two time quantities}, that is, they depend on 
$t$ (which physically corresponds to the time of observation) and
$t'$ (which corresponds to the age of the system). 
In a normal situation, one  expects that
after an equilibration time
$t_{\sc eq}$ (in general model dependent) all
two-time functions will depend only on the time
difference $t-t'$. In other words, all two-time functions 
become time-translation invariant (TTI). Under these 
circumstances, to which we
refer as equilibrium evolution,
response and correlation are not independent quantities but are
linked via the fluctuation-dissipation theorem (FDT).

Nevertheless, it is also known that certain systems never reach
(at least for experimentally accessible time-scales) an equilibrium
dynamics. For arbitrary large $t'$ and in a certain
region of the parameter space, although one-time quantities, 
such as the energy-density, tend to an asymptotic limit, two-time quantities, 
such as the correlation and response,
depend both on $t$ and $t'$ 
in a non-trivial way. This is indeed
the typical situation for a glassy system.\cite{Struick,Nagel}

The question we would like to explore  is
whether Eqs.~(\ref{schwingerR}), (\ref{schwingerC}) and (\ref{zeq}) encode
a non-equilibrium evolution as described above. We know that indeed
this is the case for classical disordered systems. As shown
in Ref.~[\raisebox{-.22cm}{\Large \cite{letter}}], this kind of equations 
describe, at least partially,  the non equilibrium dynamics of glassy systems, 
notably {\it aging} effects of spin-glasses.
The salient feature
of these solutions is that, below a  critical temperature
$T_d\equiv T_c(\hbar=0)$, and for large $t'$, there exist at least two time
scales: for $t-t'$ small with respect to a characteristic time
 ${\cal T}(t')$, that depends upon the age of the system,
the dynamics is similar to an equilibrium evolution
(stationary regime) but for $t-t'$ large with respect to ${\cal T}(t')$
the dynamics becomes extremely slow
with aging effects. In this regime, correlation and response are related in 
a manner that violates FDT.

\begin{figure}
 \centerline{\hbox{
   \epsfig{figure=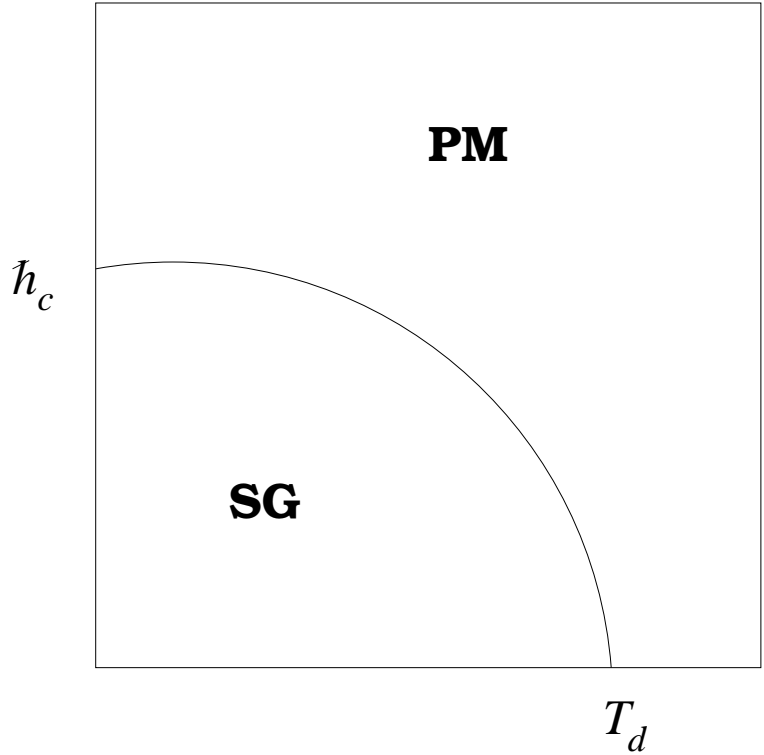,width=10cm}}
 }
\vspace{.5cm}
 \caption{The schematic phase diagram for the $p=3$ model.}
 \label{phasediagram}
 \end{figure}

In this paper we show that this also happens in the quantum problem
in a region of parameter space.
In what follows we concentrate on the case $p\geq 3$. As in the 
classical case, the 
$p=2$ model is connected to $O(N)$ models and has a different behavior.
In Fig.~\ref{phasediagram} we present 
a schematic phase diagram $(T,\hbar)$ for $p=3$.
As we show
below, the dynamics in the disordered (``paramagnetic')
and ordered (``spin-glass') phases are characterized by ``equilibrium'
and ``out-of-equilibrium'' effects, respectively. 
The dynamics for the quantum system in the glassy phase
has a similar pattern to the classical one, in the sense that
there are at least two
time-sectors in which the evolution is of a different nature. 
Inspired by the classical problem,\cite{Cuku} we
propose that also in the quantum case the weak-ergodicity breaking 
\cite{Bo,Cuku} and the weak long-term memory \cite{Cuku} 
properties hold. These two assumptions are based in part on the insight coming 
from the numerical solution of the full equations.

\subsection{The correlation function and the weak-ergodicity 
breaking scenario}

The {\it weak-ergodicity breaking scenario}
states that, for $t\geq t'$,  the correlation 
function decays in such a way that 
\begin{eqnarray}
 \lim_{t'\to\infty} C(t,t') 
&=&
q + C_{\sc st}(t-t')
\\
\lim_{t-t'\to \infty} C_{\sc st}(t-t') 
= 0 
\;\;\;\; &\Rightarrow & 
\;\;\;\; 
\lim_{t-t'\to \infty} \lim_{t'\to\infty} C(t,t') 
= q
\label{limitqq}
\\
\lim_{t \to\infty} C(t,t') &=& 0
\label{limit0}
\; .
\end{eqnarray}
In these equations $q$ is the Edwards-Anderson order parameter that 
characterizes the spin-glass phase.
This means that for large $t\geq t'$ such that $t-t'$ is small with respect to ${\cal T}(t')$,
the correlation function first decays from 1 to $q$ in a time-translational
invariant manner. ${\cal T}(t')$ is a growing function of $t'$ 
whose precise form depends on the model. The correlation 
goes further below $q$ to eventually reach $0$ in a manner that depends both
upon $t$ and $t'$ (the aging effect). This behavior suggests
the presence of at least two time-sectors in which the dynamics is
stationary and non-stationary, respectively.

 We then write the correlation as the 
sum of a stationary
and an aging contribution:
\begin{eqnarray}
C(t,t') &=& C_{\sc st}(t-t') +C_{\sc ag}(t,t')
\; .
\end{eqnarray} 
The matching conditions at equal times 
between $C_{\sc st}$ and $C_{\sc ag}$ are
\begin{eqnarray}
C(t,t)=1 \, \;\;\;\;\; &\Rightarrow&  \;\;\;\;\;  C_{\sc st}(0) +C_{\sc ag}(t,t) 
=1 
\end{eqnarray}
with
\begin{eqnarray}
C_{\sc st}(0)=1-q 
\;\;\;\;\;\;\;\;\;\;\;\;\;\;
C_{\sc ag}(t,t) = q
\; .
\end{eqnarray}
Together with 
Eq. (\ref{limitqq})  
they ensure that in the two-time sector in which
$C_{\sc st}$ decays from $1-q$ to $0$, $C_{\sc ag}$ is just a constant $q$. 
Instead, in the time-sector in which $C_{\sc ag}$ decays from
$q$ to $0$, $C_{\sc st}$ vanishes identically.

The name weak-ergodicity
breaking \cite{Bo,Cuku} reflects  the fact that for short time-differences the 
system
behaves as if it would be trapped in some region of phase space of ``size'' $q$ --
suggesting ergodicity breaking. However, it is always
able to escape this region in a time-scale ${\cal T}(t')$ 
that depends upon its age $t'$. Hence, trapping is gradual and 
ergodicity breaking is {\it weak}.

We recall that in the classical and purely relaxational case
(Langevin dynamics) the correlation functions are monotonic with
respect to both times $t$ and $t'$. 
However, in the presence of inertial terms there can be oscillations
and the decay can be non-monotonic. This will depend upon the relative
value of the mass $m$ with respect to the other parameters in the problem.
For the parameters we study in this paper,
the oscillations appear only in the stationary regime, the aging dynamics
having a monotonic decay towards zero. This is relevant since it
will allow us to use the general properties of monotonic correlation
functions proven in Ref.~[\raisebox{-.22cm}{\Large \cite{Cuku2}}] 
to find the two-time scaling of 
$C_{\sc ag}(t,t')$.

\subsection{The response function and the weak long-term memory scenario}

Regarding the response function, we propose
\begin{equation}
R(t,t') = R_{\sc st}(t-t') + R_{\sc ag}(t,t')
\end{equation}
with
\begin{eqnarray}
R_{\sc st}(t-t') &\equiv& \lim_{t'\to\infty} R(t,t')
\; .
\end{eqnarray}
When a system is in equilibrium, the response is simply 
related to the correlation via FDT. 
Here we need to extend this relation to a non-equilibrium quantum situation.
 As in the classical case we assume that the dynamics in the stationary 
regime is like an equilibrium dynamics and satisfies FDT. We then assume
that $R_{\sc st}(\tau)$ and  $C_{\sc st}(\tau)$
are related by
\begin{equation}
R_{\sc st}(\omega) = - \frac{2}{\hbar} \lim_{\epsilon\to  0^+} 
\int \frac{d\omega'}{2\pi} \; \frac{1}{\omega-\omega'+i \epsilon} \,
\tanh\left( \frac{\beta\hbar\omega'}{2}\right) \, C_{\sc st}(\omega)
\; 
\end{equation}
(see Appendix \ref{appFDT}).
In contrast, in the time-sector in which the dynamics is non-stationary and 
manifestly non-equilibrium, FDT will be modified. In Section \ref{generalFDT} 
we suggest 
an extension of FDT to quantum non-equilibrium 
systems  with slow dynamics, that we later check in the particular model
under study.

The {\it weak long-term memory scenario} states\cite{Cuku,Bocukume} 
that the system keeps a weak memory of 
what happened in the past. More precisely, the response function 
tends to zero when times get far apart,
\begin{equation}
\lim_{t\to \infty} R(t,t^*) =0 \;\;\;\;\;\;\;\;\;\;\;\;\;\forall \; 
\mbox{fixed} \; t^*
\; ,
\label{shortR}
\end{equation}
and its integral over a {\it finite} time-interval also vanishes
\begin{equation}
 \lim_{t\to\infty} \int_0^{t^*} dt'' \, R(t,t'') = 0
\;\;\;\;\;\;\;\;\;\;\;\;\;\forall \; \mbox{fixed} \; t^*
\label{finitetimes}
\end{equation}
but, its integral over an interval that grows with time gives a 
finite contribution:
\begin{equation}
\lim_{t-t^* \to\infty} \int_{t-t^*}^t dt'' \, R(t,t'') < \lim_{t\to\infty} 
\int_0^t dt'' \, R(t,t'')
\;\;\;\;\;\;\;\;\;\;\;\;\;\forall \; \mbox{fixed} \; t^*
\; .
\end{equation}
Note that the property (\ref{shortR}) implies
\begin{eqnarray}
\lim_{t-t'\to\infty} R_{\sc st}(t-t') &=& 0
\; .
\end{eqnarray}

An important quantity that reflects the decay of the response function 
is the two-time susceptibility defined as the integral of the response
over the time-interval $[t_w, t]$:
\begin{equation}
\chi(t,t_w) = f \int_{t_w}^t ds \, R(t,s)
\;
\end{equation}
($f$ is the strength of the applied perturbation, see Eq.~(\ref{respdef})).
In the classical case, a useful description of the 
departure from FDT is obtained by plotting $\chi(t,t_w)$ vs
$C(t,t_w)$ for several $t_w$ using $\tau\equiv t-t_w$ as a parameter. 
Since the classical FDT implies 
$\chi(t,t_w)=1/T \left( C(t,t_w) - C(t,0) \right)$, 
violations of the theorem manifest as departures from a straight line plot.
In the quantum case, the relation between 
$R$ and $C$ is more involved and {\it a priori}  one does not expect 
these plots to be useful. As we show in Section \ref{generalFDT}, 
it turns out that even 
for a quantum system these parametric plots are relevant.

\section{Dynamics in the paramagnetic phase}
\label{equilibriumdynamics}

In the paramagnetic
phase  (see Fig.~\ref{phasediagram})
one expects that, after a short non-equilibrium transient, 
a time-translational invariant (TTI)  solution 
establishes. We show that a TTI ansatz implies 
that FDT holds between $R$ and C and between 
$\tilde \Sigma$ and $\tilde D$.
By solving numerically the full set of equations
(\ref{schwingerR}), (\ref{schwingerC}) and (\ref{zeq})
we show the correctness of this ansatz and later discuss
some properties of the solutions.

After the short initial transient, one-time quantities such as
$z(t)$ should  reach a limit $z_\infty$. We evaluate it as 
\begin{equation}
z_\infty \equiv \lim_{t\to\infty} z(t) =
\int_{-\infty}^\infty \frac{d\omega}{2\pi} \; \left[ \Sigma(\omega) C(\omega) + 
D(\omega)
R(\omega) \right] - m \left. \partial^2_{t} C(t-t') 
\right|_{t'\to t^{-}\to\infty}
\; .
\end{equation}
The Fourier-transform of 
Eqs.~(\ref{schwingerR}) and (\ref{schwingerC}) yields
{\it the quantum mode-coupling equations}:
\begin{eqnarray}
R(\omega) &=& \frac{1}{-m \omega^2 + z_\infty + 
4 \eta(\omega) - \tilde \Sigma(\omega)}
\; ,
\label{111}
\\
C(\omega) &=& (2 \hbar \nu(\omega) + \tilde D(\omega) ) |R(\omega)|^2
\; .
\label{222}
\end{eqnarray}
Thus,  
\begin{eqnarray}
\mbox{Im} \, R(\omega) &=&- \mbox{Im} ( \tilde \Sigma(\omega) - 4  \eta(\omega) 
) |
R(\omega)|^2
\; ,
\label{ImRomega}
\\
\mbox{Re} \, R(\omega) &=&  
\left(-m \omega^2 + z_\infty - \mbox{Re }
( \tilde \Sigma(\omega) - 4  \eta(\omega))  \right)   |R(\omega)|^2
\; .
\label{ReRomega}
\end{eqnarray}


\begin{figure}
\centerline{\hbox{
\epsfig{figure=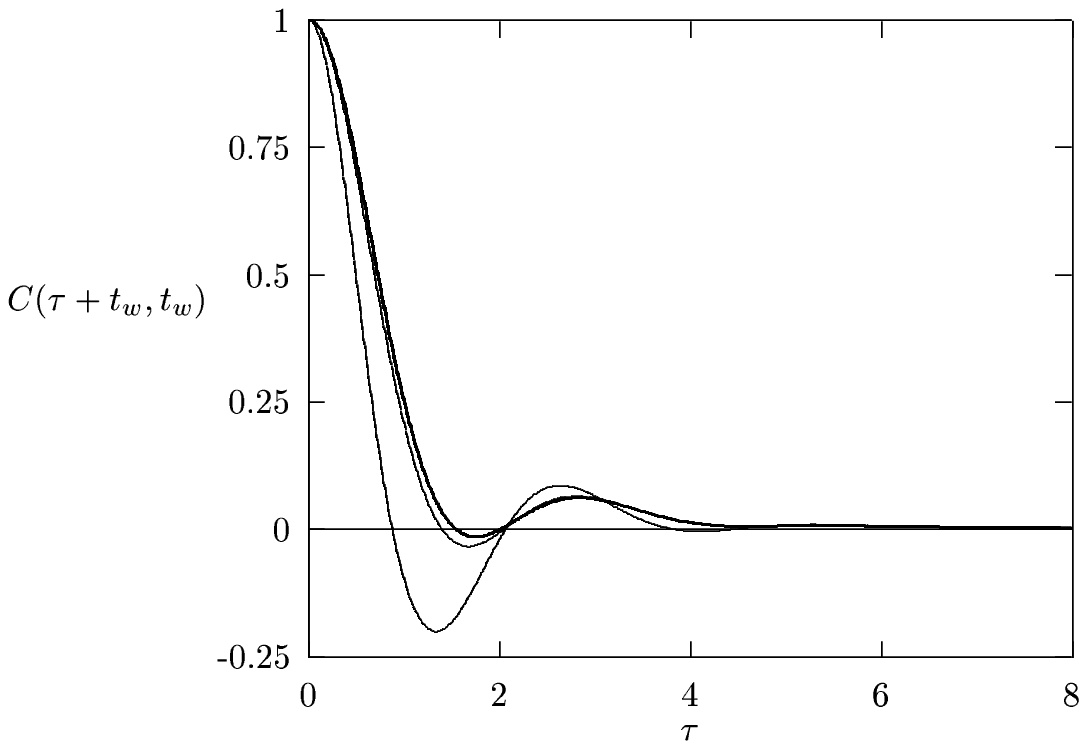,width=13cm}}
}
\vspace{-.5cm}
\caption{The auto-correlation $C(\tau+t_w,t_w)$
as a function of the time-difference $\tau$ for $t_w=2,4,8,16$
for the {\it classical} model in the paramagnetic phase, $T=3$ and $\hbar=0$.
The curves for $t_w=8$ and $t_w=16$ are superposed.
As in all following figures, $p=3$, $\Lambda=5$, $\tilde J=1$ and $m=1$.}
\label{figHT2}
\vspace{-.1cm}
\centerline{\hbox{
\epsfig{figure=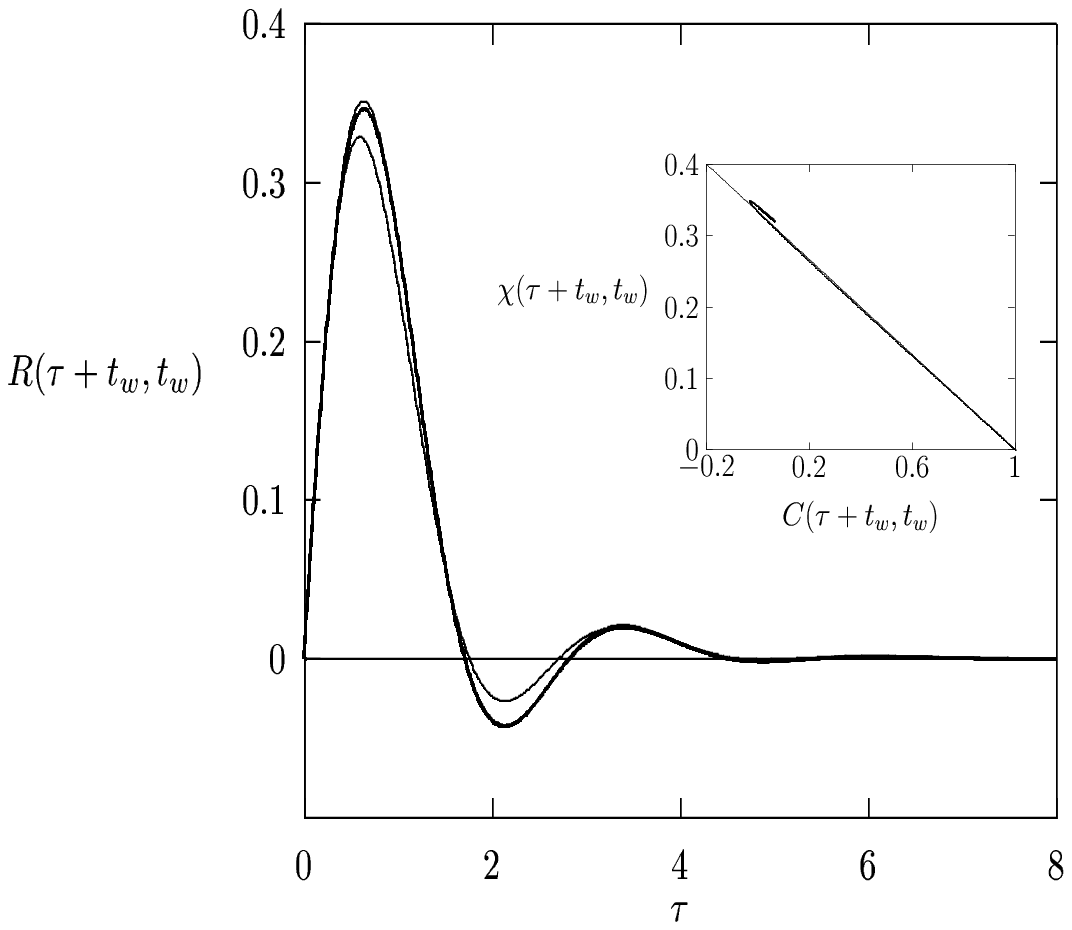,width=12cm}}
}
\caption{The same parameters as in Fig.~\ref{figHT2}.
The response function $R(\tau+t_w,t_w)$ as a 
function of the time-difference $\tau$ for 
$t_w=2,4,8, 16$. TTI establishes for $t_w \geq 8$. 
In the inset, the integrated response $\chi(\tau+t_w,t_w)$ 
vs. $C(\tau+t_w,t_w)$ in a parametric plot for waiting time
$t_w=4$ and  $\tau$ in $[0,28]$.
The classical FDT prediction is represented by the straight line 
of slope $-1/T=-1/3$; the $\chi$ vs. $C$ curve coincides 
with it showing that the classical FDT holds.
}
\label{figHT3}
\end{figure}

\begin{figure}
\centerline{\hbox{
\epsfig{figure=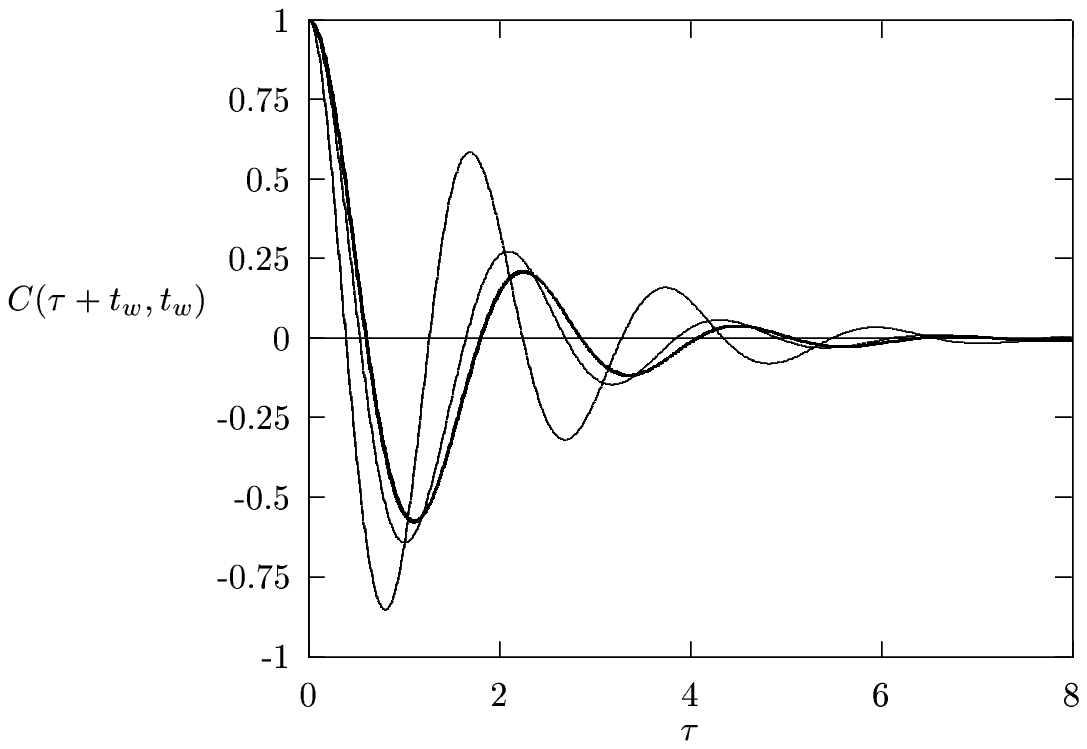,width=13cm}}
}
\caption{The auto-correlation $C(\tau+t_w,t_w)$ as a function of
the time-difference $\tau$ for $t_w=2,4,8,16$
for the {\it quantum} model in the paramagnetic phase, $T=0$ and $\hbar=6$. 
The curves for $t_w=8$ and $t_w=16$
fall on top of each other demonstrating TTI.}
\label{figHhbar3-1}
\centerline{\hbox{
\epsfig{figure=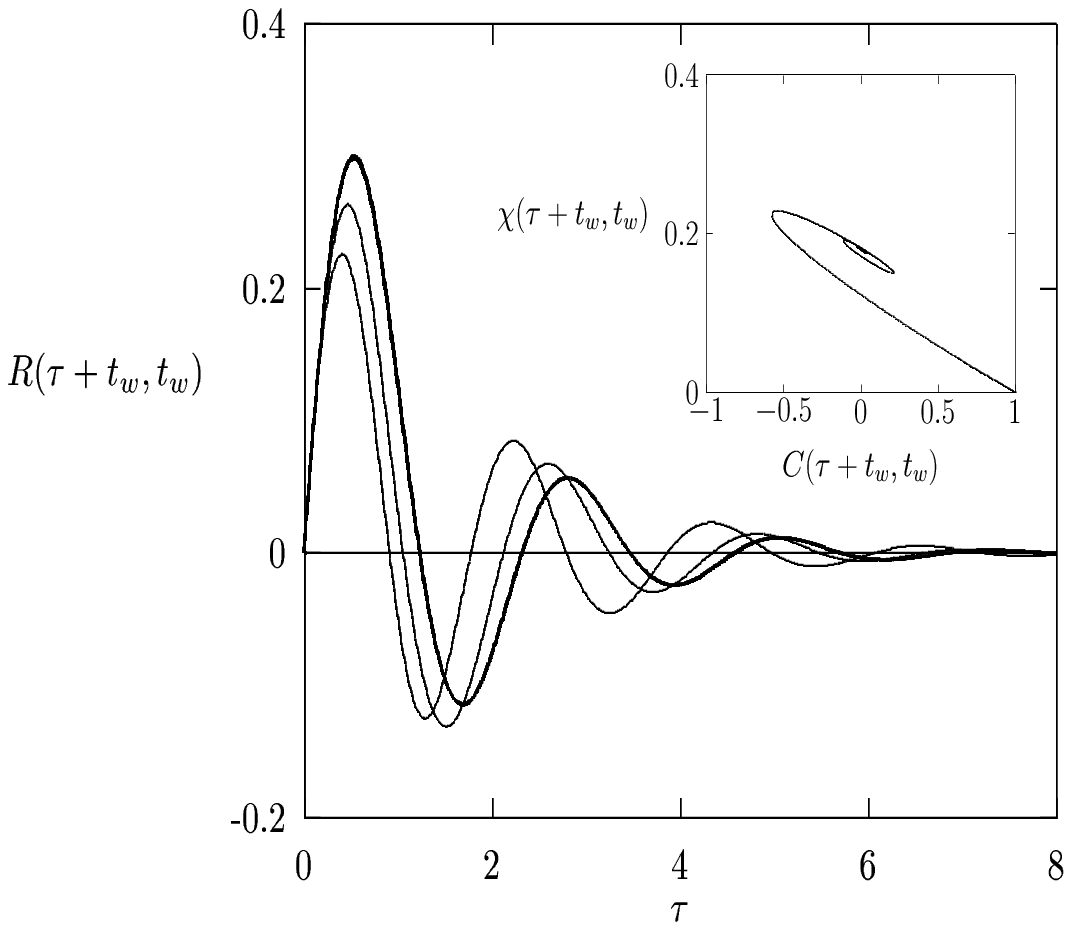,width=11cm}}
}
\caption{The response 
$R(\tau+t_w,t_w)$ as a function of the 
time-difference $\tau$ for the same parameters as in 
Fig.~\ref{figHhbar3-1}.
The waiting times are $t_w=2,4,8,16$ and 
TTI holds for $t_w\geq 8$. 
In the inset the integrated response $\chi(\tau+t_w,t_w)$ vs. 
the auto-correlation $C(\tau+t_w,t_w)$ in a parametric plot
for $t_w=8$. Since the system is in the 
paramagnetic phase with pure quantum fluctuations, 
the $\chi$ vs. $C$ plot is not expected to give 
us further information (see, however, Section~\ref{generalFDT}).}
\label{figHhbar3}
\end{figure}

If we now assume that $\tilde \Sigma$ and $\tilde D$ 
are related by FDT, it is easy to 
check that Eqs.~(\ref{111}), (\ref{222}) and (\ref{ImRomega})  
imply that $R$ and $C$ are also related by FDT.
The proof is completed by showing that FDT between $R$ 
and $C$ implies 
FDT between $\tilde \Sigma$ and $\tilde D$. This is easier to do in 
the time-domain, by verifying that
Eq. (\ref{FDTtemporal}) holds between $\tilde \Sigma$ and $\tilde D$ 
if it does between $R$ and $C$.

A solution of this type exists only in a restricted region of 
the phase diagram. In order to demonstrate the existence of a 
dynamic transition,  we solve the {\it full}
dynamic equations numerically and find that at a
critical line, both the TTI and FDT assumptions  break down.

For the purpose of solving Eqs.
(\ref{schwingerR}), (\ref{schwingerC}) and (\ref{zeq})
numerically, we proceed as in the purely relaxational 
classical case.\cite{Cuku,Frme}
Notice though that 
the presence of  non local kernels $\eta$ and $\nu$,  that
resemble the delta-function and appear convoluted with
the correlation and response, renders the numerical solution harder.
The larger the cut-off $\Lambda$, 
the smaller the iteration  step $\delta$ we need
to compute these integrals with a good precision. 
We found an acceptable data collapse for $\delta=0.0025,0.005,0.01$
if $\Lambda=5$. Note that the values of $\delta$ typically
used in the purely relaxational case are at least one order of magnitude
larger. This is the reason why the time-window we explore here is
rather narrow though it suffices to show the trend of the solution.

\begin{figure}
\centerline{\hbox{
\epsfig{figure=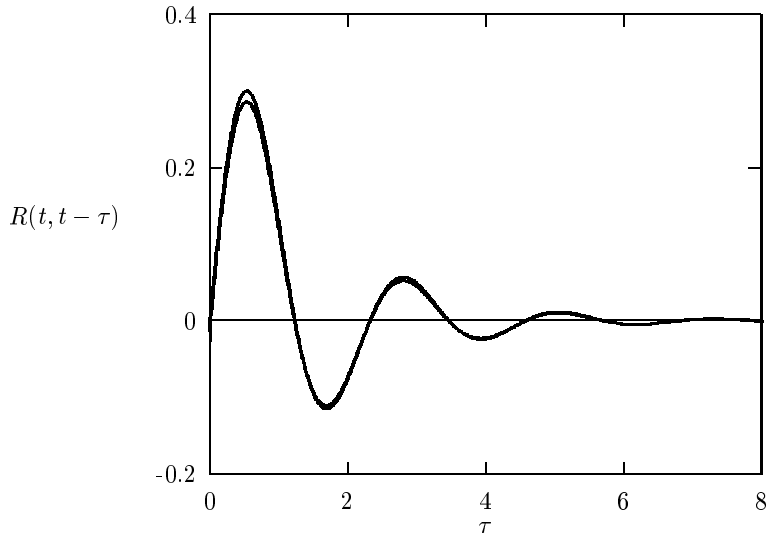,width=12cm}}
}
\caption{Check of FDT for $T=0$ and $\hbar=6$ (the paramagnetic phase).
We compare the response $R(t,t-\tau)$ for total time fixed $t=32$ and 
$\tau\in[0,t]$ obtained from the iteration of 
the dynamic equations, and the response  
computed from Eq.~(\ref{FDTFourier}) 
that assumes FDT. The correlation values used in Eq.~(\ref{FDTFourier}) are 
also obtained from the numerical algorithm. 
}
\label{RfouL5hbar6T0}
\end{figure}

\begin{figure}
\centerline{\hbox{
\epsfig{figure=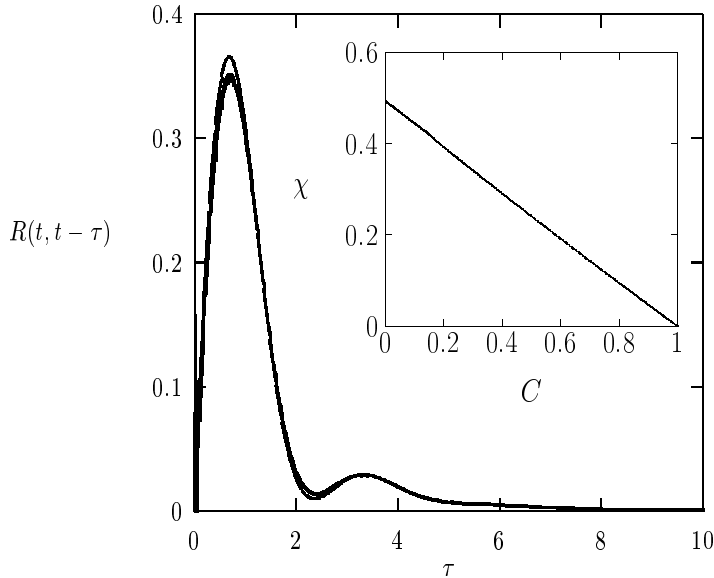,width=11cm}}
}
\caption{The same comparison as in Fig.~\ref{RfouL5hbar6T0} 
for $T=2$ and $\hbar=1$ (the paramagnetic phase).
In the inset, the 
$\chi$ vs $C$ plot for $t_w=16$. The straight line of slope $-1/2$ 
is the expectation for a classical 
FDT at this temperature. For $t_w$ larger than a characteristic time,
the FDT becomes the classical one, 
as expected in a high-temperature situation.
}
\label{mixedhb1T2}
\end{figure}

Even after rescaling time and all parameters in the problem,
there are still many free parameters left:
$(\tilde J, p, m, \Lambda,\hbar,T)$.
We here concentrate on the dependence upon only a small fraction of this set.
We fix hereafter $p=3$, $\tilde J=1$, $m=1$ and 
$\Lambda=5$
and study the behavior upon the remaining parameters $T,\hbar$. 
We discuss the data for the correlation, response and integrated response
for three points in the paramagnetic phase:

\begin{itemize}
\item
$\bbox{T=3>T_d,\hbar=0}$. This corresponds to 
the classical problem with  colored noise and inertia.
The classical critical temperature for the purely relaxational
case is $T_d \sim 0.6$. In Fig.~\ref{figHT2} we show the correlation
$C(\tau+t_w,t_w)$ vs. the time-difference $\tau$ for $t_w=2,4,8,16$.
In Fig.~\ref{figHT3} we plot the
response $R(\tau+t_w,t_w)$ vs $\tau$ for the same waiting times.
In both cases the decay is
very fast, typical of upper critical dynamics, and 
TTI quickly sets in (cfr. the curves for $t_w=8$ and $t_w=16$ that are indistinguishable in 
both plots). 
In the inset, we plot the integrated response $\chi(\tau+t_w,t_w)$ 
vs. $C(\tau+t_w,t_w)$ for $t_w=4$ using $\tau$ as a parameter.
The $\chi$ vs. $C$ plot very soon becomes a straight line of slope $-1/T=-1/3$ 
showing that the classical FDT is satisfied.

\item
$\bbox{T=0,\hbar=6>\hbar_c}$. This corresponds to 
a strong  quantum regime without thermal noise.
The data for correlations and 
response in Figs.~\ref{figHhbar3-1} and \ref{figHhbar3} 
are qualitatively similar to the 
ones for the classical problem at high temperatures in the sense 
that an equilibrium regime is 
quickly attained. The decay is fast 
and TTI holds. In the inset of Fig.~\ref{figHhbar3} we display the 
$\chi$ vs. $C$ plot. In this strong quantum case at $T=0$, the curve 
severely deviates from the classical 
expectation of a straight line of infinite slope. 
The quantum fluctuation-dissipation theorem is 
verified in Fig.~\ref{RfouL5hbar6T0} 
by comparing
the response function obtained numerically with 
the convolution of the kernel $\tanh(\beta\hbar\omega/2)$
with the numerical correlation function, see 
Eq.~(\ref{FDTFourier}). These two curves should be equal in 
equilibrium, as imposed by FDT, and we here check that this is indeed the 
case in the paramagnetic phase.

\item 
$\bbox{\bbox{T=2,\hbar=1> \hbar_c(T=2)}}$.
This is the quantum problem with both thermal and quantum fluctuations in the 
paramagnetic phase. 
In Fig.~\ref{mixedhb1T2} we check FDT for these parameters and observe, 
in the $\chi$ vs. $C$ curve displayed in the inset, that the FDT becomes the 
classical.
\end{itemize}

We continue the study of the paramagnetic phase by analyzing
the variation with $T$ and $\hbar$, close to the transition line,
of the out of phase susceptibility
\begin{equation}
\chi''(\omega) = \mbox{Im} R(\omega)
\; .
\end{equation}
This is 
the quantity that is most commonly measured experimentally. 
For instance, in the context of the study of quantum phase transitions, 
the dynamics of the randomly diluted, dipolar coupled, Ising magnet 
LiHo$_x$Y$_{1-x}$F$_4$ was studied by measuring,\cite{Wuetal} 
among other quantities, $\chi''(\omega)$.

\begin{figure}
\centerline{\hbox{
\epsfig{figure=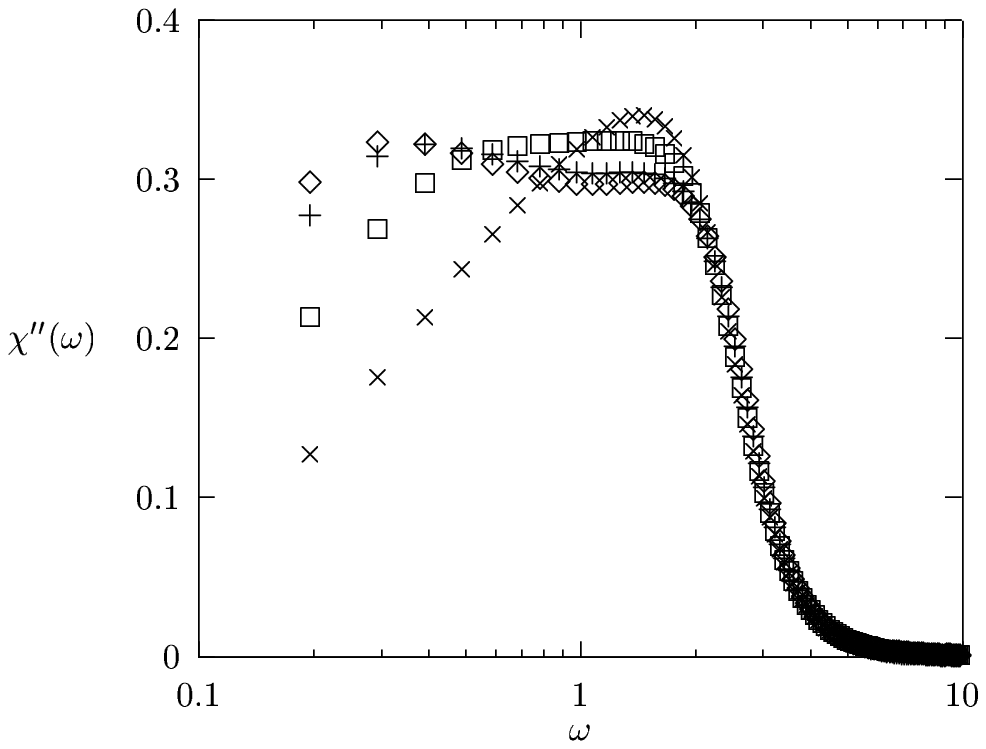,width=11.2cm}}
}
\vspace{-.3cm}
\caption{$\chi''(\omega)$ vs. $\omega$ for $T=1.3$ and $\hbar=0$ ($\Diamond$), 
$\hbar=1$ ($+$), $\hbar=2$ ($\Box$), $\hbar=3$ ($\times$).}
\label{chiT13hbar}
\centerline{\hbox{
\epsfig{figure=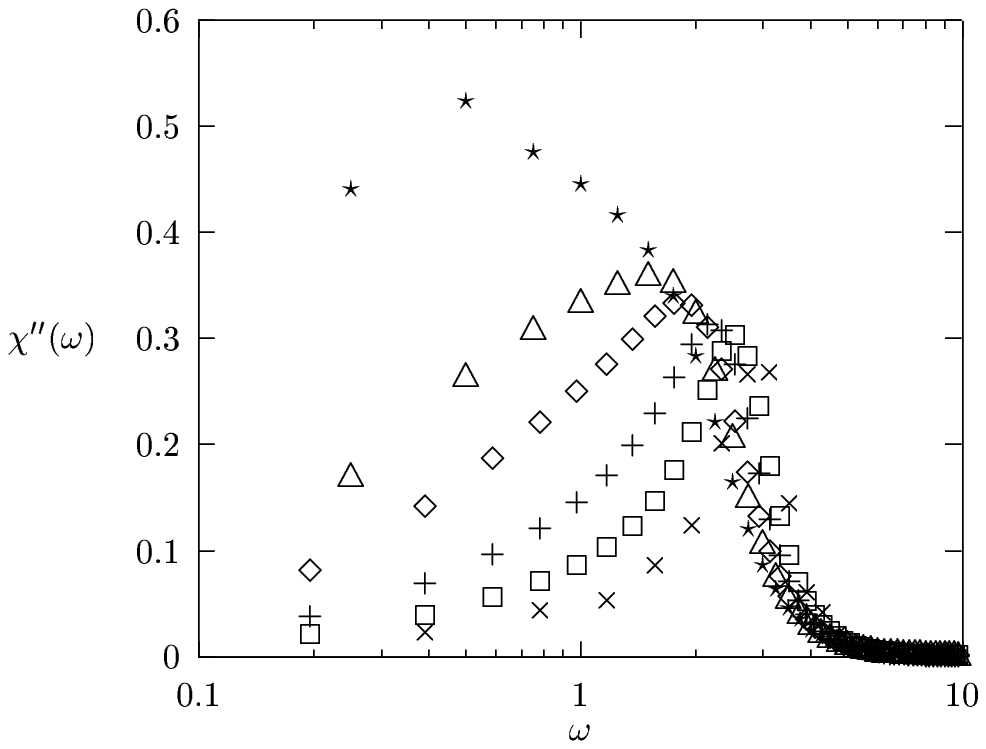,width=12.5cm}}
}
\caption{$\chi''(\omega)$ vs.
$\omega$ for $T=0.1$ and $\hbar=3.5$ ($\star$), $\hbar=4$ ($\triangle$),
$\hbar=4.5$ ($\Diamond$), $\hbar=5$ ($+$), 
$\hbar=5.5$ ($\Box$), $\hbar=6$ ($\times$).}
\label{chiT01hbar}
\end{figure}

We follow two routes to the transition: 

\begin{itemize}
\item
$\bbox{T=1.3 > T_d}$, $\bbox{\hbar=3,2,1,0}$.
One approaches the classical model by decreasing the strength 
of quantum fluctuation at fixed temperature $T=1.3 \sim 2 T_d$. 
In Fig.~\ref{chiT13hbar} we plot $\chi''(\omega)$ vs. $\omega$. 
In the curves for small $\hbar$ we observe a local
minimum  that is associated to the appearance of a 
plateau in the correlation function, at the value $q$,  when approaching $T_d$.
This is typical of ``discontinuous transitions'' as the one 
in the classical $p$-spin model\cite{Crhoso,Cuku} (and in 
super-cooled-liquids\cite{Go}). We here 
note that this feature survives the presence of small 
quantum fluctuations and progressively disappears when 
$\hbar$ increases. For $\hbar=3$, the minimum has been 
substituted by a single peak.

\item
$\bbox{T=0.1 < T_d}$, $\bbox{\hbar=6,5.5,5,4.5,4,3.5}$.
One approaches the critical line $(T_c,\hbar_c)$ from above by
modifying the strength of the quantum fluctuations while keeping
the temperature fixed $T=0.1$. 
Figure \ref{chiT01hbar} displays
$\chi''(\omega)$ vs. $\omega$ for this series of $\hbar$.  (We cannot get
closer to the transition line because the dynamics gets
so slow that the stationary limit cannot be reached.)  
The curves have a distinct peak that moves towards
decreasing frequencies for decreasing $\hbar$. 
The spectral width of the bell-shaped curves also 
increases when approaching the critical line. 
The low frequency part of the curves is
importantly suppressed with increasing $\hbar$.
All these features show 
that quantum fluctuations radically affect the relaxation. 

Contrary to Fig.~\ref{chiT13hbar}, 
there is no local minimum in any of the curves.
This is in accordance with a decreasing and eventually 
vanishing plateau value $q$ when the parameters approach the quantum 
critical point. We shall further discuss
the behavior of the model close to the transition line  in 
Section \ref{criticalline}.  
\end{itemize}

These results demonstrate the existence of a cross-over
in the disordered phase, separating a region where the scaling
laws are controlled by the quantum critical point $(0,\hbar_c)$ from another
region with different scaling laws controlled by the classical 
critical point $(T_d,0)$. 

Before concluding our analysis of the paramagnetic phase, 
let us compare our results with the situation encountered
for the system\cite{Wuetal} LiHo$_x$Y$_{1-x}$F$_4$, 
which is believed to be an experimental
realization of the transverse Ising model.\footnote{
Note however that experiments signal
a rather different dependence of the dynamics 
on the temperature history of the sample in the case of 
classical orientational glasses and classical spin-glasses
with short-range interactions.\cite{levelut,Nagel} 
This might give a warning on choosing the transverse Ising model to 
describe the dynamics of quantum dipolar glasses. 
}
Naturally, we do not expect to find the same behavior in both models
as we know that in the classical limit they are different:\cite{Bocukume}
the classical transition being first order for the $p$-spin model 
and second order for the Ising case. Our analysis indicates that
the dynamic transition of the $p$-spin model remains 
first order along the critical line but becomes second order at
the quantum critical point. 
This is the mirror situation to what has been observed 
in the dipolar glass. In Ref.~[\raisebox{-.22cm}{\Large \cite{Wuetal}}], 
it has been suggested 
that the quantum spin-glass transition may be first order.
Figure  \ref{chiT13hbar} displays the effect of quantum fluctuations 
close to the classical critical point and looks rather similar to 
Fig.~2 in Ref.~[\raisebox{-.22cm}{\Large \cite{Wuetal}}] which instead  
shows the dynamics close to the quantum critical point. 
Figure \ref{chiT01hbar} looks similar to Fig.~4 in 
Ref.~[\raisebox{-.22cm}{\Large \cite{Wuetal}}],
where the dynamics close to the classical critical point is presented.
This comparison indicates that this model belongs to a different 
class from the system LiHo$_x$Y$_{1-x}$F$_4$
and probably  from the transverse field  Ising model.\cite{SKpara,Rogr}

\section{Dynamics in the glassy phase}
\label{Nonequilibrium}

In this section we focus on the non-equilibrium dynamics
observed in the glassy phase. In order to justify 
the scenario discussed in Section \ref{scenario}, 
we first present the results
from the numerical solution of the full equations. Next, we 
analyze these equations in the stationary and 
aging regimes and obtain a self-consistent relation 
that determines the Edwards-Anderson parameter. 

\subsection{Numerical results}

\begin{figure}
\centerline{\hbox{
\epsfig{figure=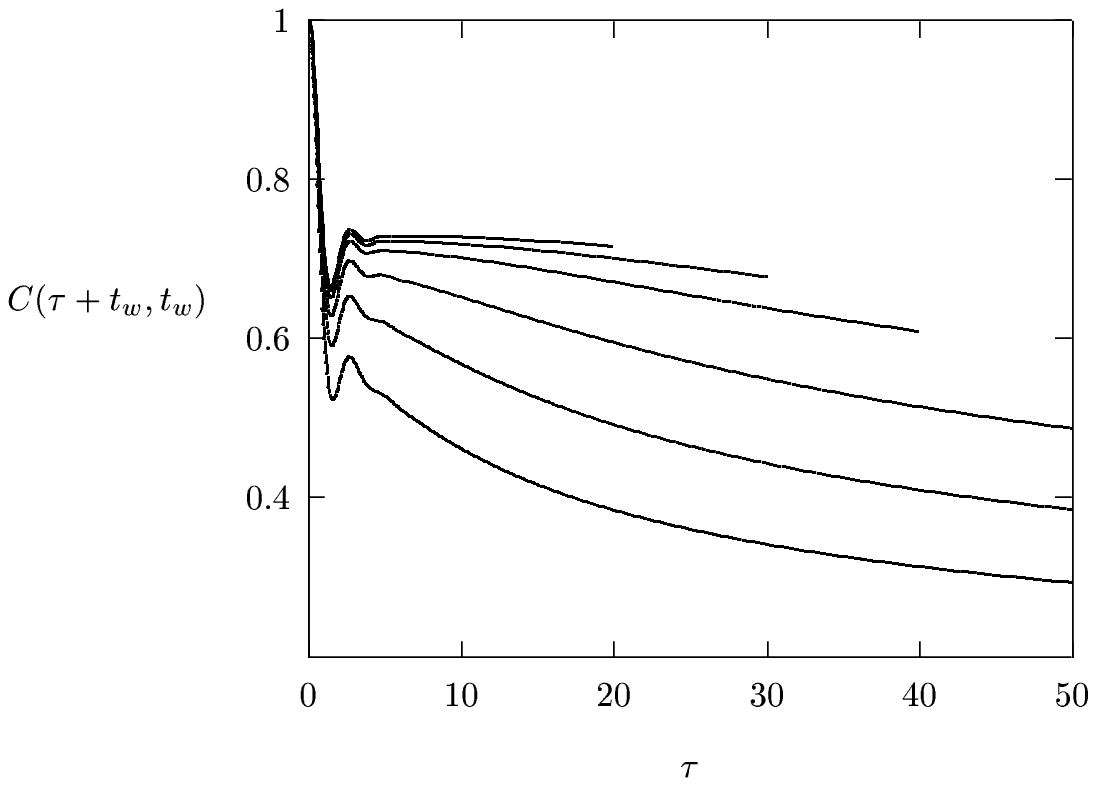,width=13cm}}
}
\caption{$C(\tau+t_w,t_w)$ vs. $\tau$ for 
$T=0$, $\hbar=1$ and from bottom to top  
$t_w=2.5,5,10,20,30,40$. 
The weak ergodicity breaking 
scenario and aging  are explicit. 
}
\label{cT0hb1}
\centerline{\hbox{
\epsfig{figure=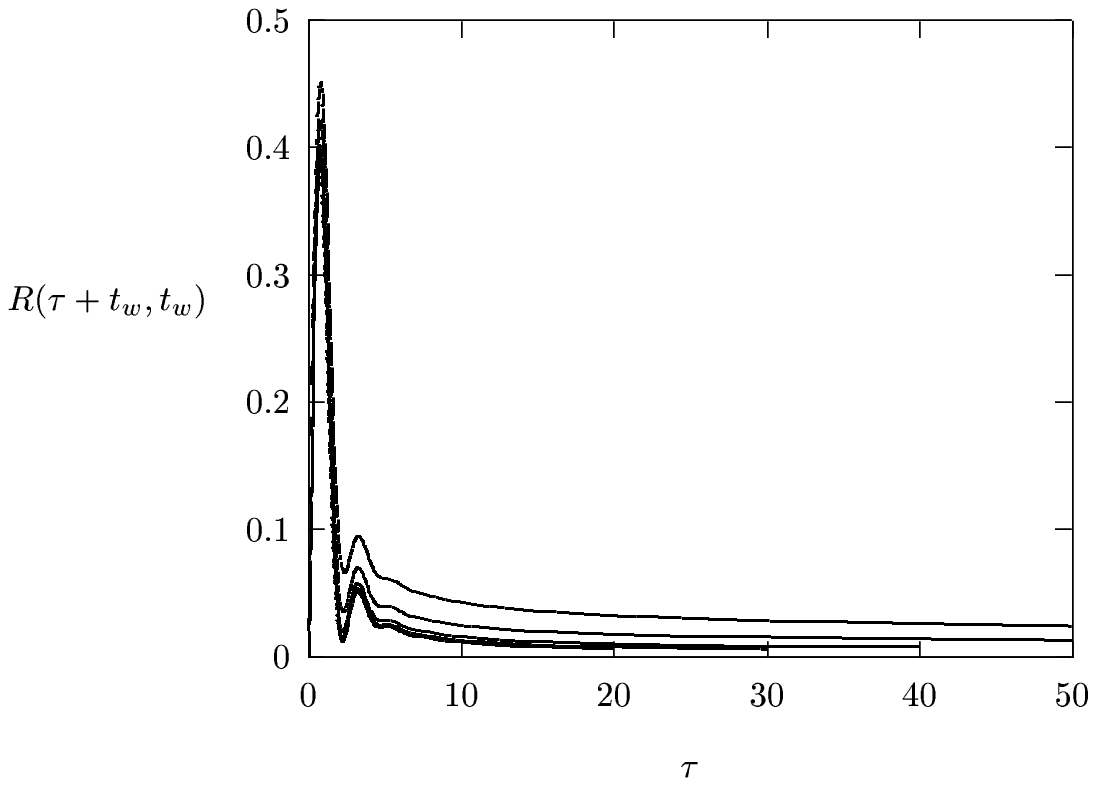,width=12.5cm}}
}
\vspace{.25cm}
\caption{$R(\tau+t_w,t_w)$ vs. $\tau$ for 
$T=0$, $\hbar=1$ and from top to bottom
$t_w=5,10,20,30,40$. The weak long-term memory 
scenario is explicit.
}
\label{rT0hb1}
\end{figure}

\begin{figure}
\centerline{\hbox{
\epsfig{figure=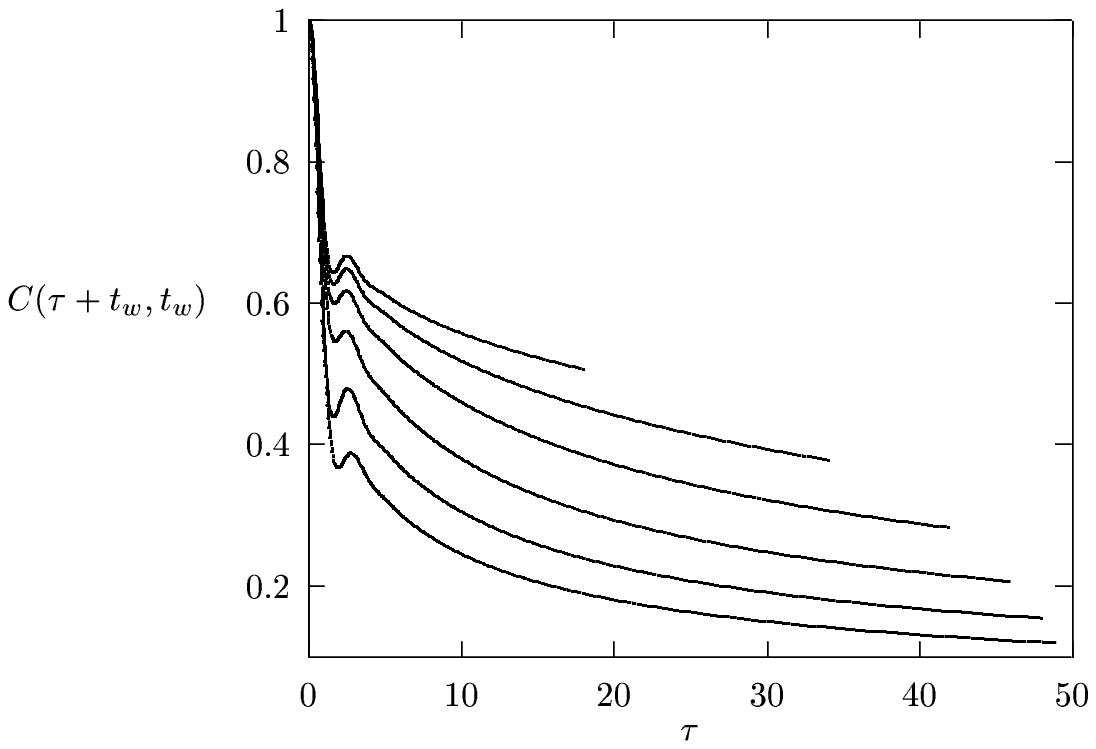,width=13.5cm}}
}
\caption{$C(\tau+t_w,t_w)$ vs. $\tau$ for 
$T=0.5$, $\hbar=1$ and from bottom to top  
$t_w=4,8,16,32$. 
The Edwards-Anderson parameter is slightly smaller 
and the decay in the aging regime is faster
than in  Fig.~\ref{cT0hb1}.
}
\label{cT05hb1}
\end{figure}

The numerical solution shows a very different behavior below the critical 
line $\hbar_c(T_c)$. In order to illustrate the slow dynamics, 
the breakdown of TTI 
and the weak-memory of the system
we choose the following parameters:
\begin{itemize}

\item
$\bbox{T<T_d, \hbar=0}$.
This is the classical, glassy model studied in 
Ref.~[\raisebox{-.22cm}{\Large \cite{Cuku}}]
but with the addition of inertia. 
The effect of inertia amounts to the presence of oscillations 
around $q$ but it does not change the qualitative behavior 
of the model. For example, as we shown analytically below, 
the value of $q$ is not modified by inertia if $m$ is small.

\item
$\bbox{T=0}$, $\bbox{\hbar=1}$.
This corresponds to pure quantum fluctuations.
Figure~\ref{cT0hb1} shows the correlation $C(\tau+t_w,t_w)$ vs. $\tau$ 
for the waiting times $t_w=0.5,1,2,4,8,16$. 
The curves have a stationary and an aging regime
separated by the Edwards-Anderson parameter $q\sim0.72$, 
see Eq.~(\ref{limitqq}). 
The correlations decay to zero at far apart times, as in 
Eq.~(\ref{limit0}). This demonstrates the weak ergodicity breaking 
scenario.

In Fig.~\ref{rT0hb1} we plot the response $R(\tau+t_w,t_w)$ vs. $\tau$
for the same waiting times. 
One observes a small time-difference regime, $\tau \leq 5$, 
where TTI rapidly establishes
and a large time-difference  regime, $\tau > 5$, 
where the response shows a decaying queue 
with a $t_w$ dependence.
The integration of this queue
over a time-interval scaling with $t_w$ will tell us about the 
violations of FDT in the quantum case (see Fig.~\ref{ML5hb01T0} below).
It also shows that the system has a weak long-term memory as described in 
Section~\ref{scenario}.

\item
$\bbox{T=0.5}$, $\bbox{\hbar=1}$.
The qualitative behavior is similar to that obtained for 
$T=0$ and $\hbar=1$ apart from the fact that thermal 
fluctuations slightly decrease the value of the Edwards-Anderson
parameter and accelerate significantly the decay in the aging regime,
as demonstrated in Fig.~\ref{cT05hb1}.

\end{itemize}

\subsection{The Lagrange multiplier $\bbox{z(t)}$: 
a one-time quantity that reaches a limit}

The Lagrange multiplier $z(t)$ is a one-time quantity, i.e. 
it depends only upon the 
total time $t$. In this formalism we assume that all one-time quantities reach 
a well-defined limit asymptotically. 
In Appendix \ref{integrals} 
we describe how we obtain the following equation 
for the asymptotic limit of $z(t)$:
\begin{eqnarray}
z_\infty&=&
A_\infty 
+ 
q \int_0^\infty d\tau' \;  \Sigma_{\sc st}(\tau') + 
\tilde D_q
\int_0^\infty d\tau' R_{\sc st}(\tau')
\nonumber\\
& & 
+ \int_0^\infty d\tau' \left[ \; \Sigma_{\sc st}(\tau') C_{\sc st}(\tau') + 
D_{\sc st}(\tau') R_{\sc st}(\tau') \; 
\right]
- m \left. \partial^2_\tau C_{\sc st}(\tau)\right|_{\tau\to 0}
\; .
\end{eqnarray}
Let us discuss each term in this expression.
We called $A_\infty$ the aging contribution:
\begin{eqnarray}
A_\infty 
&=& 
\lim_{t\to\infty} \int_0^t dt'' 
\left[ \; \tilde \Sigma_{\sc ag}(t,t'') C_{\sc ag}(t,t'') + 
\tilde D_{\sc ag}(t,t'') R_{\sc ag}(t,t'') \; 
\right]
\; . 
\end{eqnarray}
The kernels $\nu$ and $\eta$, that are related to the bath, do not contribute to $A_\infty$. 
This is so because we have assumed that for $\tau > {\cal T}(t')$
the kernels have already decayed to zero in such a way that they 
do not contribute to these integrals. 
More precisely, we are neglecting terms of the form
\begin{equation}
\lim_{t\to\infty} 
\int_0^t dt'' \, A(t-t'') B(t,t'')
\end{equation}
where $A$ is either $\nu$ or $\eta$ and $B$ is either 
$C_{\sc ag}$ or $R_{\sc ag}$.  

The second and third terms come from the constant (non-zero) limit of
the first decay of the correlation
$q\equiv \lim_{t-t'\to\infty} \lim_{t'\to\infty} C(t,t') $ and
equivalently of the vertex
\begin{equation}
\tilde D_q \equiv \lim_{t-t'\to\infty}\lim_{t'\to\infty}\tilde D(t,t')
\label{limitvertexFDT} 
\; .
\end{equation}
In the model under study 
\begin{equation}
\tilde D_q=\frac{{\tilde J}^2\, p }{2} \, q^{p-1}
\label{Dqstat}
\end{equation}
if we use $\lim_{\tau\to\infty} R_{\sc st}(\tau) \ll  q$, 
a property of the weak long-term memory scenario. 

\subsection{Dynamical equations in the stationary regime}
\label{stationaryregime}

If $(t,t')$ are such that $C(t,t') > q$, 
the discussion in Section \ref{scenario} 
implies $C(t,t')=q+C_{\sc st}(t-t')$ and 
$R(t-t')=R_{\sc st}(t-t')$. 
The Schwinger-Dyson equation for $R$ in this time sector reads
\begin{equation}
\left( m \partial^2_\tau + z_\infty \right) R_{\sc st}(\tau) 
=
\delta(\tau) + \int_0^\tau d\tau' \, \Sigma_{\sc st}(\tau-\tau') 
R_{\sc st}(\tau') 
\end{equation}
and it keeps the same form as in the high-temperature regime, apart 
from the fact that the constant $z_\infty$ has contributions 
from the aging regime.

The Schwinger-Dyson equation for $C$ reads
\begin{eqnarray}
\left( m \partial^2_\tau + z_\infty \right) (q + C_{\sc st}(\tau)) 
&=&
A_\infty 
+ q \int_0^\infty d\tau' \, \Sigma_{\sc st}(\tau') 
+ \tilde D_q
 \int_0^\infty d\tau' \,  R_{\sc st}(\tau')
\nonumber\\
& & 
+ 
\int_{-\infty}^\infty d\tau' \left[ \Sigma_{\sc st}(\tau+\tau') 
C_{\sc st}(\tau') + D_{\sc st}(\tau+\tau') R_{\sc st}(\tau') \right]
\; .
\label{Cst}
\end{eqnarray}
One can now Fourier-transform both equations
\begin{eqnarray}
R_{\sc st}(\omega) 
&=&  
\frac{1}{-m \omega^2 + z_\infty -  \Sigma_{\sc st}(\omega)}
\label{Rstat}
\; ,
\nonumber\\
\left(- m \omega^2 +z_\infty \right) C_{\sc st}(\omega) 
+ z_\infty q \delta(\omega) 
&=&
\left(
A_\infty +  q \Sigma_{\sc st}(\omega) + 
 \tilde D_q
R_{\sc st}(\omega)
\right)
\delta(\omega) 
\nonumber\\
& & 
+ \Sigma_{\sc st}(\omega) C_{\sc st}(\omega) + 
D_{\sc st}(\omega) R_{\sc st}(-\omega) 
\; .
\end{eqnarray}
The formal solution to the equation for $C_{\sc st}$ is
\begin{eqnarray}
C_{\sc st}(\omega)
&=& 
\left(-z_\infty q + A_\infty +  q \Sigma_{\sc st}(\omega)
+  \tilde D_q R_{\sc st}(\omega) \right)
\delta(\omega) R_{\sc st}(\omega)
+
D_{\sc st}(\omega) |R_{\sc st}(\omega)|^2 
\; .
\end{eqnarray}
The first term on the right-hand-side 
 has an imaginary and a real part.
The imaginary part vanishes identically since, due to FDT, 
both ${\mbox{Im}} R_{\sc st}(\omega)$ and  
${\mbox{Im}} \Sigma_{\sc st}(\omega)$ are 
proportional to $\tanh\left(\beta\hbar\omega/2\right)$
which is zero at $\omega=0$.
Concerning the real part of this first term, as we have assumed that $C_{\sc st}(\tau)$ goes to zero for $\tau\to\infty$, we need to impose 
the self-consistent condition
\begin{equation}
-z_\infty q + A_\infty +  q \Sigma_{\sc st}(\omega=0)+
 \tilde D_q R_{\sc st}(\omega=0)=0
\; .
\label{cond2}
\end{equation}
This is the condition that fixes the Edwards-Anderson parameter. 
We shall find it again in the next section as the matching condition 
between the stationary and aging regimes. 

The final equation for $C_{\sc st}(\omega)$ is
\begin{equation}
C_{\sc st}(\omega) = D_{\sc st}(\omega) |R_{\sc st}(\omega)|^2
\label{Cstat}
\; .
\end{equation}
One can check that these calculations are consistent with the results from 
$z_\infty$.
Actually, the integrals in $z$-eq. involving the stationary parts can be 
evaluated
with the help of the equations for 
$R_{\sc st}$ and $C_{\sc st}$, Eqs.~(\ref{Rstat}) and (\ref{Cstat}), 
and yield once again Eq.~(\ref{cond2}).  

Similarly to the high-temperature case one can now show that FDT for 
$\tilde \Sigma_{\sc st}$ and $\tilde D_{\sc st}$ implies FDT for 
$R_{\sc st}$ and $C_{\sc st}$. 
The remainder of the proof, i.e. to show that  FDT between $R_{\sc st}$ 
and $C_{\sc st}$ 
implies FDT between $\tilde \Sigma_{\sc st}$ and $\tilde D_{\sc st}$ 
depends only upon the 
from of $\tilde \Sigma_{\sc st}$ and $\tilde D_{\sc st}$ as functions of 
$R_{\sc st}$ and $C_{\sc st}$ and is not modified from 
the one discussed in Section~\ref{equilibriumdynamics}.
   
\subsection{Dynamical equations in the aging regime}     

If we now choose the times $t,t'$ to be well-separated so as to have 
$C(t,t') = C_{\sc ag}(t,t')\leq q$ and $R(t,t') = R_{\sc ag}(t,t')$,
the weak-ergodicity breaking and weak long-term memory 
hypotheses allow us to throw the second time
derivatives on the left-hand-side. We assume that their contribution is 
much weaker than the one of each of the integral terms on the right-hand-side. 
This is an assumption 
that we have to verify at the end of the calculation, once the solution for 
$C_{\sc ag}$ and $R_{\sc ag}$ is known. It corresponds to the over-damped limit:
\begin{eqnarray}
m \partial^2_t C_{\sc ag} \ll {\mbox{Terms in the}} \;\; {\sc rhs}_{\sc c}(t,t')
\; ,
\nonumber\\
m \partial^2_t R_{\sc ag} \ll {\mbox{Terms in the}}\;\; {\sc rhs}_{\sc r}(t,t')
\; .
\end{eqnarray}

Using the approximation
described in Appendix \ref{integrals},
the $R$-eq in the aging regime becomes
\begin{eqnarray}
z_\infty R_{\sc ag}(t,t') 
&=&
\tilde \Sigma_{\sc ag}(t,t')  \int_0^{\infty} d\tau' R_{\sc st}(\tau') 
+
R_{\sc ag}(t,t') \int_0^{\infty} d\tau' \, \Sigma_{\sc st}(\tau')
\nonumber\\ 
& & 
+
\int_{t'}^t dt'' \, \tilde \Sigma_{\sc ag}(t,t'') R_{\sc ag}(t'',t')
\; 
\end{eqnarray} 
and we call it the $R_{\sc ag}$-eq.
Similarly, the $C$-eq becomes
\begin{eqnarray}
z_\infty C_{\sc ag}(t,t')  
&=&
C_{\sc ag}(t,t') \int_0^{\infty} d\tau' \Sigma_{\sc st}(\tau') 
+ 
\tilde D_{\sc ag}(t,t') \int_0^{\infty} d\tau' R_{\sc st}(\tau') 
\nonumber\\
& & 
+
\int_0^t dt'' \; \tilde \Sigma_{\sc ag}(t,t'') C_{\sc ag}(t'',t') +
\int_0^{t'} dt'' \; \tilde D_{\sc ag}(t,t'') R_{\sc ag}(t',t'') 
\; 
\end{eqnarray}
and we call it the $C_{\sc ag}$-eq. 
In all integrals we approximated
$\Sigma_{\sc ag}(t,t') \sim \tilde \Sigma_{\sc ag}(t,t')$ and
$D_{\sc ag}(t,t') \sim \tilde D_{\sc ag}(t,t')$ 
since $\eta(t-t')$ and $\nu(t-t')$ are
very rapidly decreasing functions.

\subsection{The Edwards-Anderson parameter}

The Edwards-Anderson parameter $q$ is determined self-consistently from 
the matching of $\lim_{t\to\infty} C_{\sc ag}(t,t)= \lim_{t-t'\to\infty} 
\lim_{t'\to\infty} C(t,t')=q$. 
Taking the limit $t'\to t^-$ in the $R_{\sc ag}$-eq and $C_{\sc ag}$-eq one 
obtains
\begin{eqnarray}
z_\infty R_{\sc ag}(t,t) 
&=& 
\tilde  \Sigma_{\sc ag}(t,t) \int_0^\infty d\tau' \; R_{\sc st}(\tau')
+
R_{\sc ag}(t,t) \int_0^\infty d\tau' \; \Sigma_{\sc st}(\tau')
\; ,
\label{hola}
\\
z_\infty q 
&=&
A_\infty +
q \int_0^\infty d\tau' \; \Sigma_{\sc st}(\tau') +
\tilde D_{\sc ag}(t,t) \int_0^\infty d\tau' \; R_{\sc st}(\tau')
\; .
\label{Cq}
\end{eqnarray}

The first equation admits the solution $R_{\sc ag}(t,t)=0$ since 
$\tilde \Sigma_{\sc ag}(t,t)$ is proportional to $R_{\sc ag}(t,t)$
-- see Eq.~(\ref{sigmatilde}). 
This corresponds to the high-temperature solution 
where there is no aging regime. 
Here we concentrate on the other possibility, that is to say when
\begin{eqnarray}
z_\infty 
&=& 
\frac{\tilde \Sigma_{\sc ag}(t,t)}{R_{\sc ag}(t,t)} 
\int_0^\infty d\tau' \; R_{\sc st}(\tau') + 
\int_0^\infty d\tau' \; \Sigma_{\sc st}(\tau')
\; .
\end{eqnarray}
In actual fact, we can further approximate this equation 
by using one of the assumptions already used all over this section: that the 
response becomes smaller and smaller as time passes  
-- though its integral over 
an infinite interval gives a finite contribution. If 
we neglect all terms
that are proportional to $R_{\sc ag}(t,t)$ with respect to terms that are 
proportional 
to $q$, only the first term in the power expansions of
 $\tilde \Sigma$ and $\tilde D$ 
survive:
\begin{equation}
\left(\tilde \Sigma/R \right)_q \equiv 
\lim_{t\to\infty} 
\frac{\tilde \Sigma_{\sc ag}(t,t)}{R_{\sc ag}(t,t)}
\;\;\;\;\;\;\;\;\;\;\;\;\;\;\;\;\;\;\;
\tilde D_q \equiv \lim_{t\to\infty} \tilde D_{\sc ag}(t,t) 
\end{equation}
that in our model become
\begin{equation}
\left(\tilde \Sigma/R \right)_q
=
\frac{{\tilde J}^2 \, p(p-1)}{2} \; q^{p-2} 
\;\;\;\;\;\;\;\;\;\;\;\;\;\;\;\;\;\;\;
\tilde D_q 
= 
\frac{{\tilde J}^2 \, p}{2} \; q^{p-1}
\; ,
\end{equation}
in accord with the large $\tau$ limit of the stationary  values 
(\ref{Dqstat}).
Equations (\ref{hola}) and (\ref{Cq}) become
\begin{eqnarray}
z_\infty 
&=& 
\left(\tilde \Sigma/R \right)_q 
\int_0^\infty d\tau' \; R_{\sc st}(\tau') + \int_0^\infty d\tau' \; \Sigma_{\sc 
st}(\tau')
\; ,
\label{secondq}
\\
z_\infty q 
&=&
A_\infty +
q \int_0^\infty d\tau' \; \Sigma_{\sc st}(\tau') +
\tilde D_q  \, \int_0^\infty d\tau' \; R_{\sc st}(\tau')
\; .
\label{firstq}
\end{eqnarray} 
The second equation 
is the same as the one arising from the end of the 
stationary regime, Eq.~(\ref{cond2}).

We now study these equations with the help of Eqs.~(\ref{Rstat}) and 
(\ref{Cstat}) for the stationary functions. Using 
\begin{equation}
\int_0^\infty d\tau \; R_{\sc st}(\tau) 
= 
R_{\sc st}(\omega=0) = \frac{1}{z_\infty - \Sigma_{\sc st}(\omega=0)}
\label{secondqq}
\; ,
\end{equation}
one has
\begin{eqnarray}
1 &=& \left( \tilde \Sigma/R\right)_q \, R_{\sc st}^2(\omega=0)
\; .
\end{eqnarray}
We remind that the factor $R_{\sc st}^2(\omega=0)$ can be written
in terms of the stationary correlation function using FDT; therefore
this is a closed equation for the correlation that gives the Edwards-Anderson
parameter $q$.
In the case of the $p$-spin model it reads
\begin{eqnarray}
1 &=& \frac{{\tilde J}^2 p(p-1)}{2} q^{p-2} 
\left(\frac{1}{\hbar} \; 
P \int_{-\infty}^\infty 
\frac{d\omega'}{\omega'} \, 
\tanh\left(\frac{\beta\hbar\omega'}{2}\right) \, 
C_{\sc st}(\omega') \right)^2
\; .
\end{eqnarray}
In the classical case, the integral can be readily computed
and the final equation for $q$ is
\begin{equation}
\frac{ {\tilde J}^2 \, p(p-1)}{2} \; q^{p-2} (1-q)^2 =
 T^2
\; ,
\end{equation}
that coincides with the result for the purely relaxational dynamics.\cite{Cuku} 
For $p\geq 3$ fixed, $q$ is a function of temperature; it equals one at $T=0$ 
and tends to $q_d\equiv q(T_d) > 0$ at the dynamic critical temperature 
$T_d\equiv T_c(\hbar=0)$.   

\begin{figure}
\centerline{\hbox{
\epsfig{figure=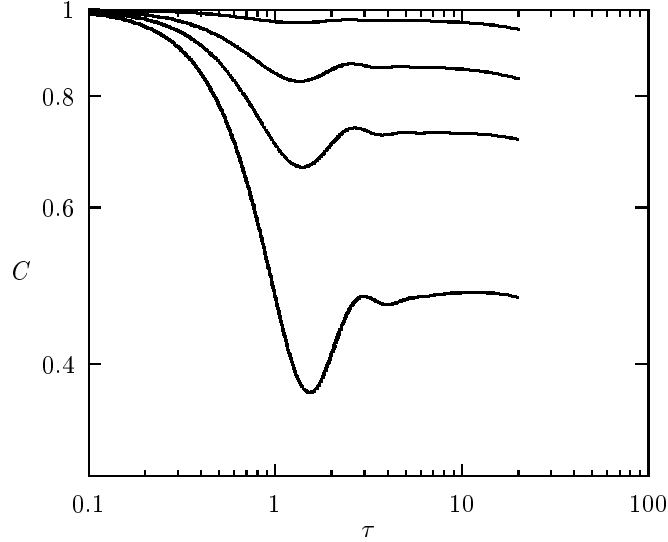,width=11cm}}
}
\caption{The $\hbar$ dependence of the Edwards-Anderson parameter $q$
that is read from the plateau in the log-log plot of the auto-correlation 
$C(\tau+t_w,t_w)$ vs. $\tau$ for $t_w=32$. From top to 
bottom $\hbar=0.1,0.5,1,2$ and the temperature is always $T=0$.
}
\label{plotq}
\end{figure}

If $\hbar\neq 0$, the Edwards-Anderson parameter $q$ depends upon $\hbar$ and $T$, 
$q(T,\hbar)$. It decreases when thermal and/or quantum fluctuations increase. One observes 
$q(0,0)=1$. At the critical line $(\hbar_c,T_c)$, $q(T_c,\hbar_c)\neq 0$
as long as $T\neq 0$. At the 
{\it quantum critical point} the nature of the transition changes 
and $q(\hbar_c,0)=0$.
We shall explain this result after 
introducing the modification of the quantum FDT needed to 
complete the solution.  Figure~\ref{plotq}  shows
the tendency of $q(T=0,\hbar)$ for $\hbar=0.1,0.5,1.2$.

\section{Fluctuation-dissipation relations out of equilibrium}
\label{generalFDT}

One of the important characteristics of the dynamic evolution
in equilibrium is the fluctuation-dissipation theorem theorem 
(FDT) that establishes a model independent relationship
(in the sense that only involves $T$ and $\hbar$ but no other
parameter in the model)  
between correlation and response functions. As recalled in
Appendix \ref{appFDT}, this result can  be easily
obtained, for the quantum and classical cases,
under the hypothesis of equilibrium.  In general,
 such a relation is not expected  to hold in a non equilibrium
situation and we can then say that FDT is ``violated" 
(of course what is violated
is not the theorem but the hypothesis under which
the result is derived). It is nevertheless surprising
that in the glassy phase of  certain classical models, the violation
of FDT \cite{Cuku} takes such a simple form that it is
possible to formulate a generalized version of this relation through
the introduction of an effective temperature $T_{\sc eff}$. \cite{Cukupe}
In this way the FDT takes the standard equilibrium 
form but with a model dependent effective temperature 
which is different from the bath temperature $T$
(in general $T_{\sc eff}$ is higher or equal than $T$ and it can take 
different values in different time-sectors).
We would like to explore in this section how these ideas
carry over to the quantum case.

\subsection{General considerations}

The standard quantum fluctuation-dissipation theorem  states
\begin{equation}
R(t-t') = \frac{2 i}{\hbar} \; \theta(t-t')
\int_{-\infty}^\infty \frac{d\omega}{2 \pi}  \; 
\exp(-i \omega(t-t')) \tanh\left( \frac{\beta \hbar \omega}{2} \right) \,
C(\omega)
\label{FDT1}
\end{equation}
with 
\begin{equation}
C(\omega)\equiv 2 {\mbox{Re}}
 \int_0^\infty d\tau \exp(i \omega \tau) 
C(\tau)
\label{FDT2}
\; .
\end{equation}
Guided by  the classical case, we introduce the effective 
temperature\cite{Cukupe}
\begin{equation}
T_{\sc eff}(t,t') \equiv \frac{T}{X(t,t')}
\end{equation}
and
we propose, as a generalization of FDT to the out of equilibrium case, 
the following relation 
\begin{equation}
R(t,t') = \frac{2 i}{\hbar} \; \theta(t-t')
\int_{-\infty}^\infty \frac{d\omega}{2 \pi}  \; 
\exp(-i \omega(t-t')) \tanh\left( \frac{X(t,t') \beta \hbar \omega}{2} \right) 
\, C(t,\omega)
\label{XFDT1}
\end{equation} 
with 
\begin{equation}
C(t,\omega)\equiv 2 {\mbox{Re}} \int_0^t ds \exp(i \omega (t-s)) C(t,s)
\label{XFDT2}
\; 
\end{equation}
and $X(t,t')$ a function of both times $t$ and $t'$. 
Evidently, 
Eqs.~(\ref{XFDT1}) and (\ref{XFDT2})
reduce to Eqs.~(\ref{FDT1}) and (\ref{FDT2}) when $t \to \infty$ if the 
evolution 
is TTI and the factor $X$ is set to one. 

Let us assume, for reasons that are intimately related to the time-separation
discussed in Section \ref{scenario}, that 
$X(t,t')$ is characterized by 
\begin{equation}
X(t,t')=
\left\{
\begin{array}{ll}
1
\;\;\;\;\;\;\;\;\;\; &{\mbox{if}} \;\;\;\;\;\;\;\;\;\; 
t-t' \leq {\cal T}(t') \;\;\; \Rightarrow \;\;\; C(t,t') \;\; > \;\; q
\nonumber\\
X_{\sc ag}(T,\hbar)
\;\;\;\;\;\;\;\;\;\; &{\mbox{if}} \;\;\;\;\;\;\;\;\;\; 
t-t' >  {\cal T}(t') \;\;\; \Rightarrow \;\;\; C(t,t')  \;\; 
\leq \;\; q 
\end{array}
\right.
\; 
\label{Xtwoscales}
\end{equation}
For times in the stationary 
regime $t-t' \leq {\cal T}(t')$ this implies: 
\begin{equation}
R_{\sc st}(t-t') = \frac{2 i}{\hbar} \; \theta(t-t')
\int_{-\infty}^\infty \frac{d\omega}{2 \pi}  \; 
\exp(-i \omega(t-t')) \tanh\left( \frac{\beta \hbar \omega}{2} \right) \,
C_{\sc st}(\omega)
\label{FDT3}
\end{equation}
with 
\begin{equation}
C_{\sc st}(\omega)\equiv 2 {\mbox{Re}}
 \int_0^\infty d\tau \exp(i \omega \tau) 
C_{\sc st}(\tau)
\; .
\label{FDT4}
\end{equation}
If we  suppose that (even at $T=0$) $\beta X_{\sc ag}(T,\hbar)$ is 
finite in the glassy  phase, it is then clear
that in the aging sector   
the Fourier integral is dominated by $\omega \sim 0$.
After using
\begin{equation}
\tanh \left( \frac{X_{\sc ag}\beta \omega}{2} \right) 
\sim \frac{X_{\sc ag}(T,\hbar) \beta \omega}{2}
\end{equation}
we obtain
\begin{equation}
R_{\sc ag}(t,t')= \beta X_{\sc ag}(T,\hbar) \; 
\frac{\partial}{\partial \, t'} C_{\sc ag}(t,t')
\;.
\label{FDT6}
\end{equation}

For a standard quantum system in equilibrium at finite 
temperature, there is 
a characteristic time, that depends upon temperature 
and the strength of quantum fluctuations, beyond which 
the evolution becomes classical. Concerning FDT, this means 
that the classical relation establishes with $X(t,t')=1$.
This is clear, for instance, for the model under consideration
at finite temperature in the paramagnetic phase (see the 
inset in Fig.~\ref{figHT3}).  
If our assumptions are correct, 
the situation in the glassy phase is different.
For small times compared with ${\cal T}(t_w)$, 
now a {\it waiting-time dependent} characteristic time,
correlation and response are related by the standard (quantum)
FDT. Instead, for time differences larger than ${\cal T}(t_w)$,
they are related in the same way as in the classical
case {\it but} with a $T$ and $\hbar$ dependent effective temperature
that is different from the one of the environment.

In general, the proposal given in Eq.~(\ref{FDT6}) will 
allow us to reduce the $C_{\sc ag}$-eq. and $R_{\sc ag}$-eq equations
to single equation for $C_{\sc ag}$. It is simple to check that, 
at least for the model under consideration, this equation 
has the same structure as the one for the classical problem.
All dependence on $\hbar$ enters through $q$, $X_{\sc ag}$ and 
$R_{\sc st}$. Hence, once 
the explicit form of $X_{\sc ag}$ is known, the aging equations 
are solved by the same ansatz as in the classical case:
\begin{eqnarray}
C_{\sc ag}(t,t') &=& q \jmath^{-1}\left( \frac{h(t')}{h(t)} \right)
\; ,
\label{agingC}
\\
R_{\sc ag}(t,t') &=& \frac{X_{\sc ag}}{T} \, q \partial_{t'} 
\jmath^{-1}\left( \frac{h(t')}{h(t)}\right)
\; .
\label{agingR}
\end{eqnarray}
Note that the general arguments introduced in 
Ref.~[\raisebox{-.22cm}{\Large \cite{Cuku2}}]
to characterize
the behavior  of {\it monotonic} correlation functions apply to 
the correlations in the aging regime and justify the solution 
(\ref{agingC})-(\ref{agingR}). 
Even if the inertia might make the correlation
oscillate, it does only in the stationary regime
since it oscillates around $q$, and it becomes monotonic in the aging 

In Eq.~(\ref{Xtwoscales}), we assumed that $X(t,t')$ can take only two
values in view of the model we study. However, 
in the classical case, a more general situation
arises in models with multiple time-scales, for which the factor measuring 
the violation of FDT is a non-constant function of the correlation in the 
aging regime.\cite{Cuku2,Frme} We expect the same structure to carry
through to quantum problems whose classical counterparts are of the 
multi-scale type.

\subsection{Analysis of the $\bbox{p\geq 3}$ model}
\label{criticalline}

We check, both analytically and numerically, the
viability of our assumptions for the $p$-spin model.
The factor $X_{\sc ag}$ measuring 
the violation of FDT
is determined by Eqs.~(\ref{firstq}) and (\ref{secondqq})  
\begin{equation}
0 = A_\infty - \frac{q}{R_{\sc st}(\omega=0)} + 
\tilde D_q \, R_{\sc st}(\omega=0)
\; .
\end{equation}
Using Eq.~(\ref{FDT6}) and the equivalent relation between
$\tilde \Sigma_{\sc ag}$ and $\tilde D_{\sc ag}$, we obtain
\begin{eqnarray}
A_\infty
&=&
\frac{X_{\sc ag}}{T} \, \lim_{t\to\infty} 
\left( \tilde D_{\sc ag}(t,t) C_{\sc ag}(t,t) \right)
=
\frac{X_{\sc ag}}{T} \, q \tilde D_q 
\; 
\end{eqnarray}
and
\begin{equation}
\frac{X_{\sc ag}}{T} \, 
=
\frac{(p-2)}{q} R_{\sc st}(\omega=0)
\,
= \sqrt{ \frac{2 (p-2)^2}{p(p-1) } } \; q^{-p/2}
\; .
\label{Xfinal}
\end{equation}
In the classical limit $X_{\sc ag}=(p-2)(1-q)/q$ and the result in 
Ref.~[\raisebox{-.22cm}{\Large \cite{Cuku}}]
is recovered.
Note that both in the classical and quantum case, $X=0$ if $p=2$.
Since the case $p=2$ is formally connected to ferromagnetic domain 
growth (in the mean-field approximation) there is still 
no-memory in the quantum domain growth. 

The classical critical point $(T_d,\hbar=0)$ is characterized by\cite{Cuku}
\begin{equation}
X_{\sc ag}(T_d,0) = 1 \;\;\;\;\;\;\;\; q(T_d,0) \neq 0
\; .
\end{equation}
Following the critical line towards the quantum critical 
point $(T=0,\hbar_c)$, one obtains
\begin{equation}
X_{\sc ag}(T_c,\hbar_c) = 1 \;\;\;\;\;\;\;\; 
q(T_c,\hbar_c)=\left( \frac{2 (p-2)^2}{\tilde J^2 p(p-1)} \right)^{1/p} \; T_c^{2/p}
\; 
\end{equation}
and a finite zero-frequency stationary susceptibility 
$R_{\sc st}(\omega=0) < +\infty$.
At the quantum critical point, $q$ vanishes and 
the zero-frequency stationary susceptibility diverges:
\begin{equation}
q(T_c\sim 0,\hbar_c) \propto T_c^{2/p} \to 0 \;\;\;\;\;\;\;\; \Rightarrow
\;\;\;\;\;\;\;\; R_{\sc st}(\omega=0) \propto T_c^{(2-p)/p} \to \infty
\; .
\end{equation}
(It is interesting to notice that the case $p=2$ is clearly different.)
On the other hand, if approaching the quantum critical 
point from below along the axis $(T=0,\hbar)$ 
\begin{equation}
T_{\sc eff}(T=0,\hbar) \sim (\hbar_c-\hbar)^{\alpha} 
\end{equation}
then
\begin{equation}
q\sim (\hbar_c-\hbar)^{2\alpha/p}
\;\;\;\;\;\;\;\; 
\Rightarrow
\;\;\;\;\;\;\;\;
R_{\sc st}(\omega=0) \sim (\hbar_c-\hbar)^{\alpha (1-p/2)}
\; .
\end{equation}

\begin{figure}
\centerline{\hbox{
\epsfig{figure=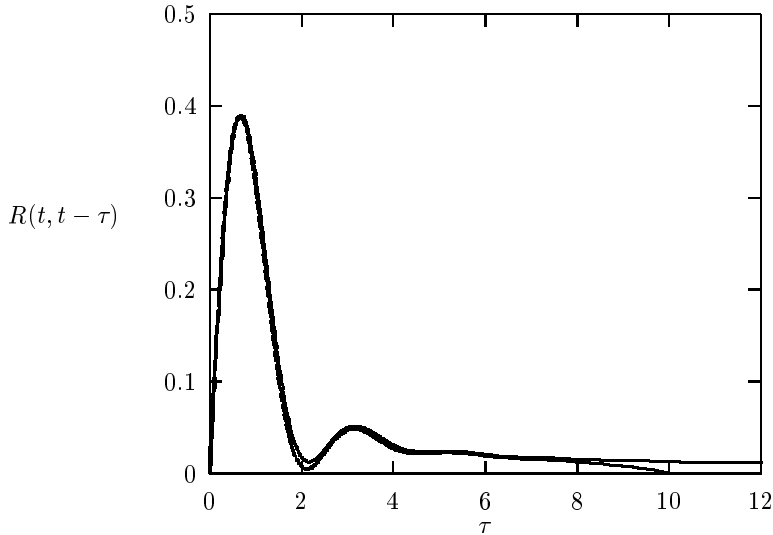,width=12cm}}
}
\caption{Check of the quantum FDT for the quantum model in the glassy 
phase  when times take values in the stationary regime $C\geq q$. 
The parameters  are $T=0$ and $\hbar=1.$ Thin curve: $R(t,t-\tau)$
from the numerical algorithm with fixed $t=32$;
bold curve: $R(t,t-\tau)$ from Eqs.~(\ref{FDT3}) and (\ref{FDT4})
using $C_{\sc st}(\tau)=C(t,t-\tau) - q$ with $C(t,t-\tau)$
from the numerical algorithm and $q\sim 0.7$. The curves coincide for $\tau < 7$.}
\label{FDThbarFourier}
\end{figure}

\begin{figure}
\centerline{\hbox{
\epsfig{figure=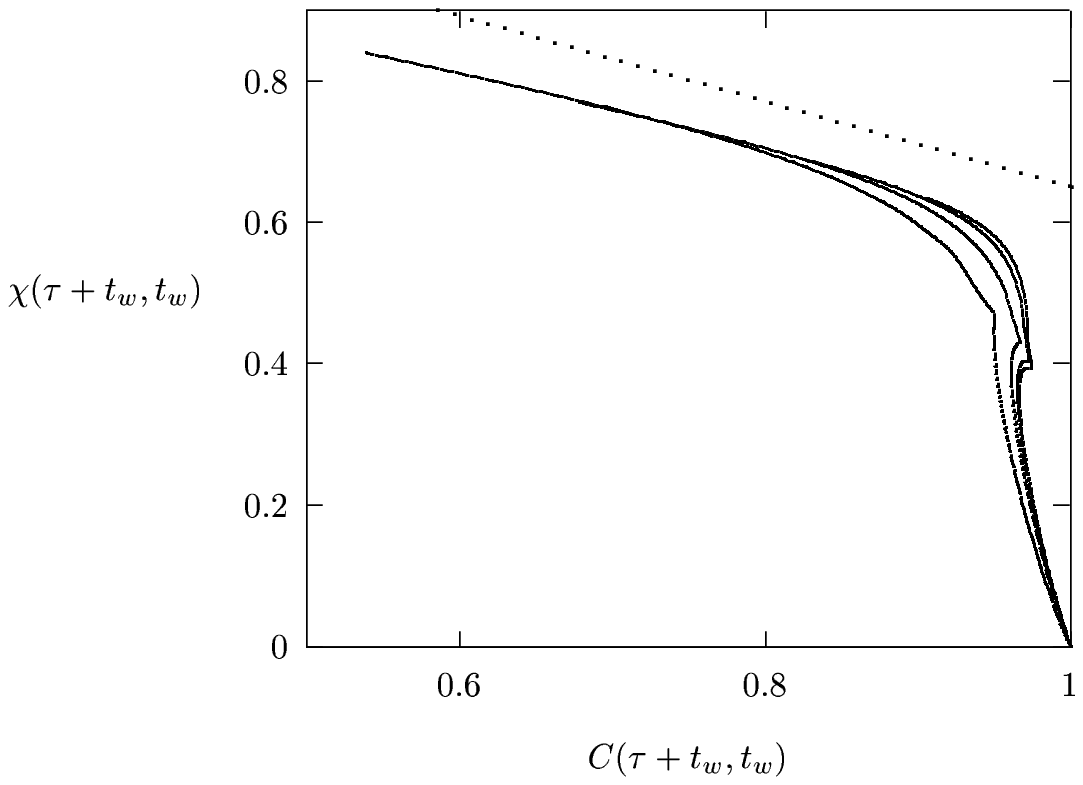,width=12cm}}
}
\caption{The $\chi$ vs $C$ plot 
for $T=0$ and $\hbar=0.1$. From bottom to top, 
different curves are associated to $t_w=5,10,20,30$.
One observes how a constant asymptotic limit is approached
for increasing $t_w$. The dots are a straight line
of slope $-1/T_{\sc eff}= -X_{\sc ag}/T=-0.6$ (the analytic result).  
}
\label{chiC1}
\centerline{\hbox{
\epsfig{figure=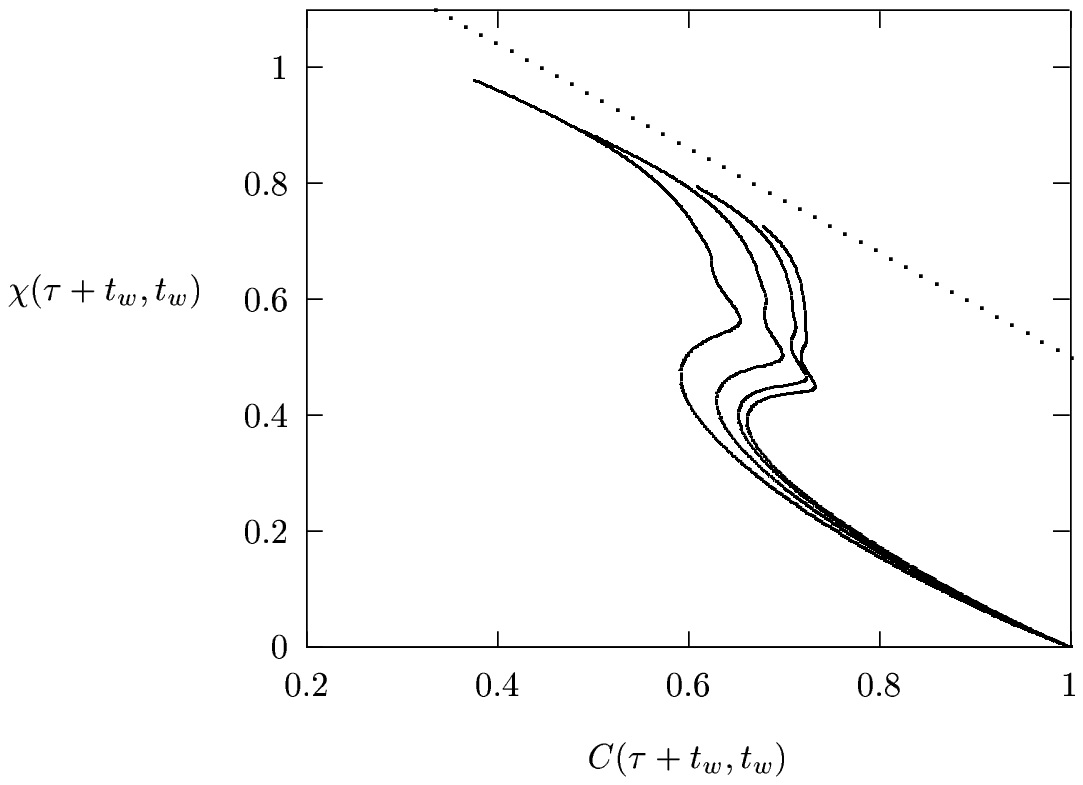,width=13cm}}
}
\caption{The $\chi$ vs $C$ plot 
for $T=0$ and $\hbar=1$.
The waiting times are $t_w=5,10,20,30$ and the 
dotted straight line has a slope $-1/T_{\sc eff}=-X_{\sc ag}/T=-0.9$ 
(the analytic result).
}
\label{chiC2}
\label{ML5hb01T0}
\end{figure}     

We next check numerically the relation between response and correlation
in the glassy phase,
${\bbox{T=0, \hbar=0.1 < \hbar_c}}$
corresponding  to pure quantum fluctuations. 

First,  we use  Fourier analysis to check the 
quantum FDT  for short time differences.
In Fig.~\ref{FDThbarFourier} we compare two different ways of obtaining
$R(t,t-\tau)$ for total time $t$ fixed and equal to $32$.
The thin curve is the result from the direct numerical solution of
the dynamic equations and shows a fast evolution for short 
time differences and a tail that gives rise to
the weak memory of the system. The bold curve is obtained from Eqs.~(\ref{FDT3})
and (\ref{FDT4})  using $C_{\sc st}(t,t-\tau)=C(t,t-\tau)-q$ 
with $C(t,t-\tau)$ obtained numerically.
We have estimated 
the Edwards-Anderson parameter as $q\sim 0.7$. One observes that 
for small time-differences the two curves coincide 
showing that FDT holds in this time-sector. 

Second, we check the generalized FDT in the aging regime.
For this purpose,
it is convenient to use the $\chi$ vs. $C$ plots defined in 
Section \ref{scenario}.
In Fig.~\ref{chiC1}  we 
observe, as predicted by Eqs.~(\ref{FDT3}), (\ref{FDT4}) and (\ref{FDT6}),
two  behaviors according to the relative value of 
$C$ and $q$.  The portion of the graph that is significant
for the aging regime  corresponds
to  $C< q$. 
As claimed in Eq.~(\ref{FDT6}) 
the $\chi$ vs $C$ curve approaches a time-independent asymptotic limit 
that is  a straight line of slope 
$-X_{\sc ag}/T$. In this case, $T_{\sc eff}=1.6$ and 
this shows that even a $T=0$, $T_{\sc eff} \neq 0$.

The effective temperature depends  on
$T$ and $\hbar$ in a non trivial way. 
In  Figure \ref{chiC2} we show the $\chi$ vs. 
$C$ plot for $\hbar$ =1 and $T=0$. In this case $T_{\sc eff}=1.1$.

\begin{figure}
 \centerline{\hbox{
   \epsfig{figure=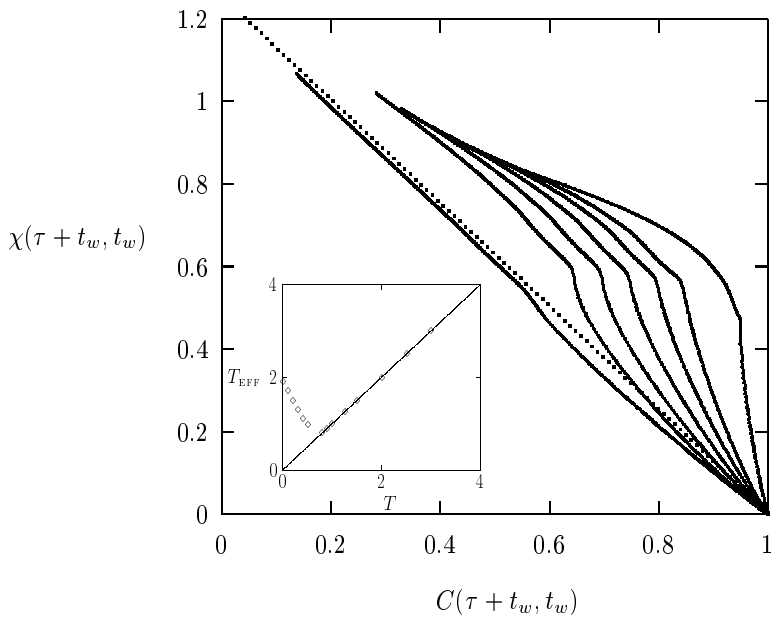,width=11.5cm}}
 }
\vspace{.5cm}
 \caption{The $\chi$ vs. $C$ curves for the quantum model at 
$\hbar=0.1$ and different temperatures. From top to bottom
$T=0.1, 0.2, 0.3, 0.4, 0.5, 0.8$. In
the inset, the estimated effective temperature $T_{\sc eff}=T/X_{\sc ag}$
as a function of temperature for the same $\hbar$. See the text for a discussion.}
 \label{teffmixed}
 \end{figure}

In Fig.~\ref{teffmixed} we work at fixed 
$\hbar=0.1$ and study the variation of $T_{\sc eff}$ with T. From the $\chi$ vs $C$ plots
we determine how $T_{\sc eff}$ varies with $T$ by measuring the slopes
of the different curves and by using Eq.~(\ref{Xfinal}).
The six curves above the dotted straight line correspond to 
temperatures below $T_c$. From top to bottom $T=0.1, 0.2,0.3,0.4,
0.5$. One notices
that $T_{\sc eff}>T$ in all these curves. The curve below the dotted straight 
line corresponds to $T=0.8>T_c$. It has a slope $-1/T$
(indicated by the dots) and $T_{\sc eff}=T$.
In the inset we plot the temperature dependence of the effective temperature. 
$T_{\sc eff}$ is linear in $T$ when the 
bath temperature is above $T_c$ but severely deviates from 
the linear behavior for temperatures below the critical line.
Two curious features are that, in the glassy phase, 
$T_{\sc eff}$ increases when either $T$ or $\hbar$ decrease. 
The surprising increase of $T_{\sc eff}$ as $T$ decreases is also obtained 
in the classical limit.\cite{Cuku}  

\section{Summary and conclusions} 
\label{conclusions}

In this paper we presented a formalism 
that allows us to analytically study 
the  behavior of quantum  glassy models
within the typical  situation encountered in experiments, 
i.e. out of equilibrium. 

\vspace{.5cm}

Faced to the question as to how relevant quantum fluctuations
are for the asymptotic real-time dynamics, one would have naively answered
that they are irrelevant since quantum mechanics
is not expected to play a relevant role in the
large time sector, meaning large $t_w$ and for all $t\geq t_w$.
We show in this paper that this would have been a wrong conclusion
and that the dynamic behavior is much richer:
 
 Quantum effects, together with temperature,
determine {\it where} in parameter space 
a glassy phase exists. We show that sufficiently
strong quantum fluctuations destroy the glassy phase
at arbitrary low temperatures driving the system towards
a paramagnetic phase. 
We predict then the existence of a dynamic quantum phase transition. 
This has  no analogue in the 
classical model where glassiness is always present at $T=0$.

Quantum fluctuations also dictate 
{\it when}, in the glassy phase, non-equilibrium effects manifest.
For any waiting-time $t_w$ after preparation, there exists a 
characteristic time ${\cal T}(t_w)$ that determines the end of a 
stationary regime and the entrance into the aging regime. 
This means that for time differences $t-t_w < {\cal T}(t_w)$ the 
dynamics is stationary while for $t-t_w \geq {\cal T}(t_w)$ 
two-point functions have explicit waiting-time dependences.
The characteristic time ${\cal T}(t_w)$ depends on temperature, the strength of quantum
fluctuations and the waiting-time $t_w$. 

The dynamics in the paramagnetic phase satisfies the 
fluctuation-dissipation theorem. In the glassy phase, 
we found that the quantum FDT is satisfied in the stationary regime
and that a generalization is needed in the aging regime.
This generalization defines a two-time function $X(t,t')$ that 
is consistent with the 
definition of an effective temperature $T_{\sc eff}(t,t') = T/X(t,t')$
and that we
later characterized, based on the numerical evidence for
this model. 
When $T_{\sc eff}$ is different from zero, the modification of the quantum 
FDT takes a similar form as the classical one but  with a 
$T$ and $\hbar$ dependent effective temperature. 

The characteristic time ${\cal T}(t_w)$ is reminiscent of a 
decoherence time in the sense that it separates  the stationary 
regime in which the influence of quantum effects is very strong
and explicit from the aging regime in which the influence of quantum effects 
is implicit, appearing through, e.g.,
the $\hbar$ dependence of $T_{\sc eff}$.

\vspace{.5cm}

One of the virtues of the approach we have followed is that it
is close in spirit to the dynamic approach to classical systems. 
This encourages us to attempt to extend many of the recent 
advances in the understanding of classical glassy dynamics 
to the quantum problem. We mention below some of the questions 
that merit further attention. 

The model here studied has a modification of the quantum 
FDT given by a piecewise
function $X=1$ if $C>q$ (${\cal T}(t_w) > t-t_w$) and a constant 
$X_{\sc ag}(T,\hbar)< 1$ if $C \leq q$ (${\cal T}(t_w) \leq t-t_w$).
It is known that in classical systems a more general situation can 
arise. Another family of models exists 
with  the factor measuring the FDT violation given by  $X=1$ if 
$C>q$  and a non-constant function of the 
auto-correlation
$X_{\sc ag}(C)< 1$ if $C \leq q$.
We expect the arguments put forward in Section \ref{generalFDT} 
to apply to this kind of system with minor modifications.
It would be of particular interest to see how this structure appears 
in a concrete model as, for instance, the problem of a particle 
in a long-range correlated random potential.

In the classical case, an intriguing connection between the factor that 
measures the FDT violation\cite{So,Cuku,Cuku2,Frme,Frmepape}  
in the non-equilibrium approach $X_{\sc ag}(C)$
and the Parisi function $x(q)$, $0\leq q < q_{EA}$
of the replica analysis in equilibrium, 
has been noticed and checked in several occasions. The structure of both 
functions is the same and, on top of this,  their explicit values coincide 
if an argument of ``marginality''
is used to determine $x(q)$ in the equilibrium case.
In Ref.~[\raisebox{-.22cm}{\Large \cite{Giledou}}], several quantum models related, 
but not equal, to the one 
considered here were studied with a Gaussian variational method 
and replica theory in the imaginary time Matsubara representation. 
In order to obtain a physical 
value for the conductivity of the model, Giamarchi 
and Le Doussal used the marginality condition to fix the replica breaking 
point parameter $x=x(q)$, $q< q_{EA}$.
A careful comparison of the dynamic $X_{\sc ag}$
with the static $x$ is now in order. Nevertheless, this cannot be 
readily done by comparing the result in 
Ref.~[\raisebox{-.22cm}{\Large \cite{Giledou}}] with ours since 
the way in which both systems are coupled to the environment are not
necessarily equivalent. 
Thence, if a formal relation between FDT-violations and the replica
$x(q)$ 
exists at the quantum level is an interesting question that 
deserves further study.

The last remark we wish to make 
about the FDT violations is that in the classical 
case it has been related to the production of entropy during 
evolution.\cite{Cudeku,Lale}  It would be desirable to see
whether such a connection also holds at the quantum level.

Concerning the formalism here developed, 
it can be extended in several directions. 
On the one hand, one would like to study models with a spatial structure
such as the one associated to the motion of a {\it manifold} of internal 
dimension $d$
in a random quenched potential. This quantum problem has many 
interesting applications
and, in the case of an embedding space with a large dimensionality 
can be treated, for example, within a dynamic Gaussian 
variational approach.\cite{Frme,Gile}
On the other hand, it would be extremely interesting to study the opposite
limiting case of low dimensions and analyze, for example, the real-time 
dynamics of the Ising chain in a transverse field. 
In addition, one could investigate the effect induced on the dynamics by 
different environments like considering sub or super Ohmic
baths, non-linear couplings and even baths of a completely  different nature
as those generated by spin variables.\cite{Stamp}

The Martin-Siggia Rose generating functional for classical stochastic
processes takes a particularly simple form if
the time-space is enlarged to a superspace by introducing a couple of 
Grassmann coordinates, and the variables of the problem are encoded in 
a superfield.\cite{Zinn} This can also be done for systems with disorder.\cite{Ku}
The dynamic equations are then written in terms of a single super-correlator
that encodes the auto-correlation and response. The equilibrium 
theorems, viz. causality, invariance under time-translations and 
the fluctuation-dissipation theorem, follow from the symmetries 
of the MSR effective action written in this way.\cite{Ku} 
The supersymmetric formulation of the classical dynamics has
been helpful to find a solution of mean-field like models with an internal 
dimension like the problem of the manifold in a random media but, 
most importantly, it has highlighted the formal connection between 
the out of equilibrium formalism and the replica approach to the statics. 
It would be very interesting to identify a supersymmetric structure 
in the quantum problem, probably at the level of the Keldysh-Schwinger 
generating functional. This could provide a means to establish
a relation between the quantum dynamic formalism here presented and 
the quantum replica approach.

\vspace{.5cm}

We believe that an analytical study of aging effects along the lines
presented in this paper will be relevant in the understanding of 
realistic glassy systems.

Indeed, even at very low temperatures, where quantum
effects are important, a glass is a non-equilibrium system
that evolves over many orders of magnitude in time.\cite{librolowT}
Aging effects, demonstrating this non-equilibrium evolution,
have been observed with, for example, burning hole experiments 
in organic glasses (see Ref.~[\raisebox{-.22cm}{\Large \cite{Fayer}}] 
and references therein).
In the context of magnetic disordered systems with explicit 
quenched disorder like LiHo$_x$Y$_{1-x}$F$_4$, 
a great effort has been devoted to the study of the quantum 
phase transition and the dynamics close to the quantum critical
point. It would be interesting to perform dynamic measurements in the glassy
phase of these compounds to search for aging effects and 
check if a scenario as the one here discussed with, for example, a two step 
relaxation of correlation functions and FDT violations is also present.  

Another area of research where 
the relevance of quantum effects in the very low temperature
dynamics is currently under study are 
systems of  magnetic nano-particles
such as $\gamma-$Fe$_2$O$_3$  particles (maghemite)
embedded in a silica matrix.
The search for a non-stationary evolution in
samples at small concentrations ($\sim 0.3\%$) 
has given a negative result. \cite{Eric} 
However, in experiments performed at higher temperatures (where 
quantum effects are not expected to play an important role)
with samples with a larger concentration ($\sim 17\%$)  much of 
the phenomenology of spin-glasses has been recovered.\cite{Nordblad} 
It is then licit to ask whether there is a region in the
temperature-concentration parameter-space 
where the systems exhibit aging driven by quantum fluctuations. 
In the area of magnetic nano-particles, a notion of 
an effective temperature greater than the temperature
of the environment was advocated in Ref.[\raisebox{-.22cm}{\Large \cite{Chgu}}]
and it would be interesting to explore the connection with 
effective temperature defined in this paper. 

We have chosen to comment on three kind of materials where effects like the ones 
obtained in this article might be observed. Surely enough, many other glassy systems
with or without explicit disorder might exhibit a similar phenomenology.
 
\vspace{1cm}

\centerline{{\bf Acknowledgments}}
\vspace{.5cm}

We want to especially thank J. Kurchan for very useful discussions and 
constant encouragement during the preparation of this work. We also wish to 
thank T. Evans, A. Georges, T. Giamarchi, D. Grempel, P. Le Doussal,
R. Revers, M. Rozenberg and E. Vincent for useful discussions. 
At the early stages of this work we have also benefited from discussions with 
L. Bettencourt and H. de Vega.   
The authors want to thank the International Center for Theoretical Physics, 
Trieste, Italy for hospitality during the ``Summer College in Condensed Matter
on Statistical physics of frustrated systems''. 
L.F.C. wishes to thank the Institute for Theoretical Physics, 
University of California at Santa Barbara for hospitality
during the work-shop ``Jamming and rheology'' partially supported by the
grant No PHY94-07194. G.L is financially 
supported by the EC grant N ERBFMBICT 961226.

\newpage

\appendix

\section{The Schwinger-Keldysh path-integral}
\setcounter{equation}{0}
\renewcommand{\theequation}{A.\arabic{equation}}
\label{appKeldysh}
\vspace{.5cm}

The generating functional (\ref{generating}) is more 
conveniently expressed with the 
help of path-integrals:
\begin{eqnarray}
{\cal Z}[{\bbox \xi}^+, {\bbox \xi}^-]  &=&
\int d{\bbox \phi}    
\; 
\langle {\bbox \phi}  | T^* \exp\left( -\frac{i}{\hbar} 
\int_{0}^\infty dt \, {\bbox \xi}^-(t) {\bbox \phi}(t) \right)
 T \exp\left( \frac{i}{\hbar} 
\int_{0}^\infty dt \, {\bbox \xi}^+(t) {\bbox \phi}(t) \right) | 
  \hat \rho(0) | {\bbox \phi} \rangle
\nonumber\\
&=&
\int d{\bbox \phi}d{\bbox \phi}_1 d{\bbox \phi}_2    
\; 
\langle {\bbox \phi}  | 
T^* \exp\left( -\frac{i}{\hbar} \int_{0}^\infty dt \, {\bbox \xi}^-(t) {\bbox \phi}(t) \right)
|{\bbox \phi}_1 \rangle 
\nonumber\\
& & 
\;\;\;\;\;\;\;\;\;\;\;\;\;\;\;\;\;
\langle {\bbox \phi}_1 |
 T \exp\left( \frac{i}{\hbar} \int_{0}^\infty dt \, {\bbox \xi}^+(t) {\bbox \phi}(t) \right)  
| {\bbox \phi}_2 \rangle  
\langle {\bbox \phi}_2  | \hat \rho(0) | {\bbox \phi }\rangle
\; .
\end{eqnarray}
The two matrix elements are now expressed as usual Feynman path-integrals
\begin{eqnarray}
{\cal Z}[{\bbox \xi}^+, {\bbox \xi}^-]  
&=&  
\int d{\bbox \phi}    d{\bbox \phi}_1    d{\bbox \phi}_2     
\int_{{\bbox \phi}^+(0)={\bbox \phi}_2}^{{\bbox \phi}^+(\infty)=  
{\bbox \phi}_1 } 
 {\cal D}{\bbox \phi}^+  
\int_{{\bbox \phi}^-(0)={\bbox \phi}}^{{\bbox \phi}^-(\infty)= 
{\bbox \phi}_1} 
{\cal D}{\bbox \phi}^-  
\nonumber\\
& &
\exp\left[ \frac{i}{\hbar} \, \left( S[{\bbox \phi^+] - S [\bbox \phi^-}]
+\int dt \,  {\bbox \xi}^+ {\bbox \phi}^+
- \int dt \, {\bbox \xi}^- {\bbox \phi}^-
\right) \right] 
\;\;
\langle {\bbox \phi}_2  | \hat \rho(0) |  {\bbox \phi}  \rangle
\; .
\end{eqnarray}
The first integrals $\int d {\bbox \phi}$, etc. are standard 
while we denote $\int {\cal D} {\bbox \phi}^+$ the path-integrals.
The doubling of degrees of freedom $({\bbox \phi}^+,{\bbox \phi}^-)$ 
is a consequence of having two matrix elements, one for each source
$({\bbox \xi}^+,{\bbox \xi}^-)$.
Formally this expression is rewritten as
\begin{eqnarray}
{\cal Z}[{\bbox \xi}^+, {\bbox \xi}^-]  
&=&  
\int {\cal D}{\bbox \phi}^+  
\int {\cal D}{\bbox \phi}^-  
\exp\left[ \frac{i}{\hbar} \, \left( S[{\bbox \phi^+] - S [\bbox \phi^-}]
+ \int dt \,  {\bbox \xi}^+ {\bbox \phi}^+
- \int dt \, {\bbox \xi}^- {\bbox \phi}^-
\right) \right] 
\nonumber\\
& & 
\;\;\;\;\;\;\;\;\;\;\;
\times 
\langle {\bbox \phi}^+ | \hat \rho(0) |  {\bbox \phi}^- \rangle
\; .
\end{eqnarray}

\section{Quantum fluctuation-dissipation theorem}
\setcounter{equation}{0}
\renewcommand{\theequation}{B.\arabic{equation}}
\label{appFDT}
\vspace{.5cm}

In this appendix we recall the quantum fluctuation-dissipation theorem. 
Proofs and descriptions of this theorem  can be found in several 
textbooks.\cite{Kubo,Parisi} 
We express it in the time-domain in a 
form  that we use in Section \ref{equilibriumdynamics} to show that 
the dynamic equations 
are compatible with a TTI-FDT ansatz. We also write it in 
a mixed time-Fourier notation
that gives us insight as to how extend it to the case of glassy 
non-equilibrium dynamics.

If at time $t'$ the system is characterized by a density functional 
$\rho(t')$,  the two-time correlation functions read 
\begin{equation}
C_{AB}(t,t') \equiv \langle A(t) B(t') \rangle = \frac1Z \mbox{Tr} \left[ A(t) 
B(t') \rho(t') \right]
\label{corr}
\end{equation}
where the time-dependent operators, in the Heisenberg representation, are 
defined as
\begin{eqnarray}
O(t) \equiv \exp\left( \frac{iHt}{\hbar} \right) O(0) \exp\left( 
-\frac{iHt}{\hbar} \right) 
\; .
\end{eqnarray}
The trace is defined in the usual way, 
$\mbox{Tr} [ \, \bullet \, ] \equiv \sum_\alpha \langle \psi_\alpha | \bullet | 
\psi_\alpha \rangle$, 
with $\{\psi_\alpha\}$ an orthonormal basis in Fock space.
The normalization is given by $Z\equiv \mbox{Tr} \left[ \rho(t') \right]$.

Since operators do not commute, 
the  quantum two-time auto-correlation functions are not symmetric in times
\begin{equation}
C_{AA}(t,t') = \langle A(t) A(t') \rangle \neq   \langle A(t') A(t) \rangle 
\; .
\end{equation}
One thus defines the symmetrized and antisymmetrized correlation functions:
\begin{eqnarray}
C_{\{A,B\}}(t,t') &=& \frac12 \, \langle A(t) B(t') + B(t') A(t) \rangle
\; ,
\nonumber\\
C_{[A,B]}(t,t') &=&  \frac12 \,  \langle A(t) B(t') - B(t') A(t) \rangle
\; ,
\end{eqnarray}
respectively.

In linear response theory
$R_{AB}(t,t')$ (see Eq.~(\ref{respdef}) for the definition)
and the correlation $C_{[A,B]}(t,t')$
are related by the Kubo formula
\begin{equation}
R_{AB}(t,t') = \frac{i}{\hbar} \; \theta(t-t') \, 
\langle \, [ A(t), B(t') ] \, \rangle = \frac{2 i}{\hbar} \; \theta(t-t') \, 
C_{[A,B]}(t,t')
\label{FDTasym}
\; .
\end{equation}

If the system has reached equilibrium with a heat-bath at temperature $T$ at 
time $t'$, the density functional $\rho(t')$ is 
just the Boltzmann factor $\exp(-\beta H)$. 
It is then immediate to show that,
in equilibrium, time-translation invariance holds 
\begin{equation}
C_{AB}(t,t') = C_{AB}(t-t') 
\; .
\end{equation}
In addition, Eq.~(\ref{corr}) with $\rho = \exp(-\beta H)$  imply the 
KMS condition
\begin{equation}
C_{AB}(t,t') = C_{BA}(t',t+i\beta \hbar) = C_{BA}(-t-i\beta \hbar, -t')
\; .
\end{equation}
Using now the KMS properties 
and assuming, for definiteness, that $t > 0 $ it is easy to 
verify the following equation
\begin{equation}
C_{\{A,B\}}(\tau) + \frac{i\hbar}{2} R_{AB}(\tau) =
C_{\{A,B\}}(\tau^*) - \frac{i\hbar}{2} R_{AB}(\tau^*)
\; ,
\label{FDTtemporal}
\end{equation}
where $\tau = t + i\beta \hbar/2$. This is a way to 
express FDT through an analytic continuation to complex times
that we use in Section~\ref{equilibriumdynamics} to show that a 
TTI and FDT solution 
holds in the paramagnetic phase.

In terms of the Fourier transformed $C_{AB}(\omega)$ defined by 
\begin{equation}
C_{AB}(t-t') = \int_{-\infty}^\infty \frac{d\omega}{2\pi} \, \exp(-i \omega 
(t-t')) \, C_{AB}(\omega)
\end{equation}
the KMS relations read
\begin{equation}
C_{AB}(\omega) = \exp(\beta \hbar \omega) C_{BA}(-\omega) 
\end{equation}
and lead to 
the following relation between Fourier transforms
of the correlation functions:
\begin{eqnarray}
2 C_{[A,B]}(\omega) &=& \left(1 - \exp(-\beta \hbar \omega) \right) \, 
C_{AB}(\omega) 
\; ,
\nonumber\\
2 C_{\{A,B \}}(\omega) &=& \left(1 + \exp(-\beta \hbar \omega) \right) \, 
C_{AB}(\omega)
\; , 
\nonumber\\
C_{[A,B]}(\omega) &=& \tanh\left( \frac{\beta \hbar \omega}{2} \right) \, 
C_{\{A,B \}}(\omega)
\; .
\end{eqnarray}
Back in Eq. (\ref{FDTasym}) this implies FDT
\begin{eqnarray}
R_{AB}(t-t') &=& \frac{i}{\hbar} \, \theta(t-t') \, 
\int_{-\infty}^\infty  \frac{d\omega}{\pi} \;  \exp(-i \omega (t-t'))  \; 
\tanh\left( \frac{\beta \hbar \omega}{2} \right) \, C_{\{A,B \}}(\omega)
\label{FDTFourier}
\end{eqnarray}
Using 
\begin{equation}
\int_0^\infty dt \, \exp(i \omega t)
=
\lim_{\epsilon\to 0^+} \frac{i}{\omega+i\epsilon} =
\pi \delta(\omega) + i \frac{P}{\omega}
\end{equation}
one has
\begin{eqnarray}
R_{AB}(\omega) &=& - \frac{1}{\hbar} \,
\lim_{\epsilon\to 0^+} \int_{-\infty}^\infty \frac{d\omega'}{\pi } \; 
\frac{1}{\omega-\omega'+i\epsilon} \; 
\tanh\left( \frac{\beta \hbar \omega'}{2} \right)  \, 
C_{\{A,B \}}(\omega')
\end{eqnarray}
from which we obtain the real and imaginary relations
\begin{eqnarray}
\mbox{Im} R_{AB}(\omega) &=& 
\frac{1}{\hbar} \, \tanh\left( \frac{\beta \hbar \omega}{2} \right) C_{\{A,B 
\}}(\omega')
\; ,
\nonumber\\
\mbox{Re} R_{AB}(\omega) &=& -\frac{1}{\hbar} \, P \int_{-\infty}^\infty 
\frac{d\omega'}{\pi} \; 
\frac{1}{\omega-\omega'} \; 
\tanh\left( \frac{\beta \hbar \omega'}{2} \right)  \, 
C_{\{A,B \}}(\omega')
\; .
\end{eqnarray}
If 
$\beta\hbar\omega/2  \ll 1$,
$\tanh(\beta\hbar\omega/2 ) \sim \beta\hbar\omega/2 $ and 
Eq.~(\ref{FDTFourier}) becomes the classical FDT:
\begin{equation}
R_{AB}(\tau) = - \frac{1}{T} \frac{dC_{AB}(\tau)}{d\tau}
\end{equation}
with $\tau=t-t'$.

In the zero temperature limit, and for $A=B$, FDT reads
\begin{eqnarray}
R(\omega) &=& -\frac{2}{\hbar} \, 
\lim_{\epsilon\to 0^+}
\int_{-\infty}^\infty \frac{d\omega'}{2\pi}
\; 
\frac{1}{\omega-\omega'+i \epsilon} \, \mbox{sign}(\omega') \, C(\omega')
\; ,
\nonumber\\
R(t)&=&
\frac{2}{\hbar} \, \theta(t) \int_0^\infty \frac{d\omega}{\pi} \;
\sin(\omega t) \, C(\omega)
\; .
\end{eqnarray}
A finite integrated response
\begin{equation}
\int_0^\infty dt R(t)  = R(\omega=0) = \frac{2}{\hbar} 
\int_0^\infty \frac{d\omega'}{\pi} \;
\frac{C(\omega')}{\omega'}
\end{equation}
implies
$C(\omega=0) =0$.
In the glassy model we study in the paper, below $\hbar_c(T_c)$, 
the response and correlation turn out to be 
the sum of two contributions. One is a stationary part that satisfies FDT.
Since one expects it to yield a finite contribution to the susceptibility
$\chi(\tau+t_w,t_w)$, one must have:
\begin{equation}
\int_0^\infty d\tau \, R_{\sc st}(\tau) 
\;\;\;\;\;\; \mbox{Finite}
\;\;\;\;\;\;\;\;\;\;
\Rightarrow
\;\;\;\;\;\;\;\;\;\;
\int_0^\infty d\tau \, C_{\sc st}(\tau) =0
\; .
\end{equation}
Thus, the stationary part of the correlation function
oscillates around zero or, in other words, the full
correlation oscillates around $q$ at zero temperature. 
This is observed in the numerical solution to the dynamic
equations.

\section{The classical limit}
\setcounter{equation}{0}
\renewcommand{\theequation}{C.\arabic{equation}}
\label{classicallimit}
\vspace{.5cm}

The classical limit can be checked at every stage of the calculation. 
We shall here show how one can recover (i) the Martin-Siggia-Rose (MSR) action 
from the Keldysh-Shwinger (KS) action (\ref{action1}); (ii) the equations 
of motion, for the classical correlation and response from the
equations of motion for their quantum partners.

This Appendix is intended for readers already familiar with the MSR formalism.
We refer to Ref.~[\raisebox{-.22cm}{\Large \cite{MSR}}] for a detailed description 
of the method.

Let us identify, in the limit $\beta \hbar \to 0$, the quantum linear 
combinations on the left with the 
classical variables on the  right
\begin{eqnarray}
\frac{{\bbox \sigma}^+(t) - {\bbox \sigma}^-(t)}{\hbar} \to \hat {\bbox s}(t)  
\; ,
\;\;\;\;\;\;\;\;\;\;\;\;\;  
{\bbox \sigma}^+(t) \to  {\bbox s}(t) + \frac{ \hbar}{2}  \hat {\bbox s}(t)
\; ,
\nonumber\\
\frac{{\bbox \sigma}^+(t) + {\bbox \sigma}^-(t)}{2}  \to  {\bbox s}(t)
\; ,
\;\;\;\;\;\;\;\;\;\;\;\;\;  
{\bbox \sigma}^-(t)  \to   {\bbox s}(t) - \frac{\hbar}{2}  \hat {\bbox s}(t)
\; .
\label{identifications}
\end{eqnarray}

In the classical limit $ \hbar \to 0$, and when the cut-off tends to infinity, 
the kernels $\eta$ and
$\nu$ become
\begin{eqnarray}
4 \eta(t-t') &=& 4 M \theta(t-t') \, \gamma_o \delta'(t-t') 
\; ,
\nonumber\\
2 \hbar \nu(t-t') &=& 4 M \gamma_o k_B T \delta(t-t')
\; .
\end{eqnarray} 
With these identifications, we establish a connection 
between the KS and the MSR  actions.

The friction and thermal terms become
\begin{eqnarray}
\lim_{\hbar\to 0}  \frac{i}{\hbar} S_{\sc t} 
&=&
-2 M \gamma_o \int_0^\infty dt \; i \hat {\bbox s}(t) \frac{d}{dt} {\bbox s}(t)   
+ 2 k_B T M \gamma_o \int_0^\infty  dt\; (i \hat {\bbox s}(t))^2
\; .
\end{eqnarray}
By calling $\Gamma_o^{-1} \equiv 2 M \gamma_o$ we recover two 
terms of the MSR action.

The kinetic and constraint terms reduce to 
\begin{eqnarray}
\lim_{\hbar\to 0} \, \frac{i}{\hbar}S_{\sc 0} &=&
\int_0^\infty  dt \; 
\left( -i \hat {\bbox s}(t) m \partial^2_t {\bbox s}(t)  
                             - \frac{ 1}{2}\,  i \hat {\bbox s}(t) 
(z^+(t)+z^-(t)) {\bbox s}(t)
\right)
\nonumber\\
& & 
+ \frac{i}{2 \hbar} 
 \int_0^\infty  dt \,
 {\bbox s}(t) \left(z^+(t) -z^-(t) \right) {\bbox s}(t)
+ \frac{i}{\hbar}  \frac{N}{2} \int_0^\infty  dt \,
 \left( z^+(t) - z^-(t) \right)
\; .
\end{eqnarray}
By calling $z^+(t)-z^-(t) = z_o(t)$ and $z^+(t)+z^-(t) = z(t)$
the last two terms impose the spherical constraint and are
usually included in the path-integral measure, while the second term 
is the one left in the MSR action:
\begin{equation}
\lim_{\hbar\to 0} \frac{i}{\hbar}S_{\sc 0}
=
-\int_0^\infty dt \; 
i \hat {\bbox s}(t) (m \partial^2_t + z(t))  {\bbox s}(t)  
\; .
\end{equation}

Finally, the terms that depend upon disorder become
\begin{equation}
\lim_{\hbar\to 0} 
- \frac{i}{\hbar} \int_0^\infty dt \; 
V[{\bbox \sigma^+}, J] - V[{\bbox \sigma^-}, J]  
= 
- \int_0^\infty dt \;
i\hat{\bbox s} \, \frac{\delta V[{\bbox s}, J] }{\delta {\bbox s}} 
\; .
\end{equation}

Putting everything together, in the large friction limit
$\Gamma_o^{-1} \equiv 2 M \gamma_o \gg 1$ such that 
one neglects the second derivative terms in the kinetic part of the action, 
one recovers the usual MSR action for the Langevin dynamics of the classical 
model
in a thermal bath at temperature $T$:
\begin{equation}
\lim_{\hbar\to 0} \frac{i}{\hbar} \, S_{KS}
=
-\int dt \; 
\left[
  i \hat {\bbox s}(t) \left( \Gamma_o^{-1} \frac{d}{dt} + z(t) \right)  {\bbox 
s}(t)   
- k_B T \Gamma_o^{-1} \; (i \hat {\bbox s}(t))^2
 +   \frac{\delta V[{\bbox s}, J] }{\delta {\bbox s}(t)} 
i \hat{\bbox s}(t)   \right]
\; .
\end{equation}

In order to study the classical limit of the dynamic 
equations (\ref{schwingerR}) and  (\ref{schwingerC})
we recognize the classical limit of the 
correlation and response as follows:
\begin{eqnarray}
C(t,t') &=& \frac12 \langle {\bbox \sigma}^+(t) {\bbox \sigma}^-(t') +  {\bbox 
\sigma}^-(t) {\bbox \sigma}^+(t') \rangle
\to 
\langle {\bbox s}(t) {\bbox s}(t') \rangle
\; ,
\nonumber\\
R(t,t') &=& \frac{i}{\hbar} \langle {\bbox \sigma}^+(t) {\bbox \sigma}^-(t') +  
{\bbox \sigma}^+(t) {\bbox \sigma}^-(t') \rangle
\to 
\langle {\bbox s}(t) i \hat {\bbox s}(t') \rangle
\; .
\end{eqnarray} 
In a similar way, it is now easy to verify that 
the classical dynamic equation
for the $p$-spin model follow from the $\hbar\to 0$ limit of 
Eqs.~(\ref{schwingerR}) and (\ref{schwingerC}).


\section{Treatment of integrals}
\setcounter{equation}{0}
\renewcommand{\theequation}{D.\arabic{equation}}
\label{integrals}
\vspace{.5cm}

Integrals over time windows spanning the interval $[0,t]$ appear
in the dynamic equations  (\ref{schwingerR}) and (\ref{schwingerC}). 
In the large times limit we approximate them 
in the way we here describe.

\subsection{First type of integral}

Integrals of the form
\begin{equation}
I_1(t) \equiv \int_0^t dt'' A(t,t'') B(t,t'')
\; 
\end{equation}
appear, for example, in $z(t)$. 
The idea is to 
separate the integration time-interval as
\begin{equation}
[0,t] = [0,\delta] \;\; U \;\; [\delta,t^-] \;\;  U \;\; [t^-,t]
\; .
\end{equation}
If $\delta$ is chosen to be a finite time,
all functions  can be approximated by $A(t,0)$ that vanishes
when $t\to \infty$. Since  the integration interval is finite, 
this term can be neglected.
In the second interval the functions vary in the aging regime and 
in the third interval they vary in the stationary regime. Thus
\begin{eqnarray}
I_1(t) 
&\sim&
\int _0^\delta dt'' A(t,t'') B(t,t'') 
+
\int_\delta^{t^-} dt'' A_{\sc ag}(t,t'') B_{\sc ag}(t,t'')
\nonumber\\
& & 
+
\int_{t^-}^t dt'' \, 
\left[ 
\left(
A_{\sc st}(t-t'')
+
\lim_{t-t''\to\infty} \lim_{t''\to\infty}  A(t,t'')  
\right)
\left(
B_{\sc st}(t-t'')
+
\lim_{t-t''\to\infty} \lim_{t''\to\infty}  B(t,t'') 
\right)
\right]
\; .
\nonumber\\
\end{eqnarray}
We assume that this separation is sharp and that we can 
neglect the corrections associated to mixing of the three
regimes. In the third term we replaced $A$ and $B$ 
in terms of $A_{\sc st}$, $B_{\sc st}$.

Since in all cases either $A$ or $B$ is proportional to
the response, 
we can neglect the first term using the
property (\ref{finitetimes}) of the weak long-term memory scenario.
We can then replace the lower limit of the second integral by $0$ and 
the upper limit of the second integral
by $t$. In addition, assuming that $B$ is proportional to 
the response, 
\begin{equation}
\lim_{t-t''\to\infty} \lim_{t''\to\infty}  B(t,t'') =0
\; .
\end{equation}
This yields
\begin{eqnarray}
 I_1(t)
&\sim&
\int_0^t dt'' A_{\sc ag}(t,t'') B_{\sc ag}(t,t'')
+
\left( \lim_{t-t''\to\infty} \lim_{t''\to\infty}  A(t,t'') \right)
\int_0^{t-t^-\to\infty} d\tau' \; B_{\sc st}(\tau') 
\nonumber\\
& & 
+
\int_0^{t-t^-\to\infty} d\tau' \; 
 A_{\sc st}(\tau') B_{\sc st}(\tau') 
\; .
\end{eqnarray}

\subsection{Second type of integral}

Another type of integrals is:
is
\begin{equation}
I_2(t,t') \equiv \int_{t'}^t dt'' A(t,t'') B(t'',t')
\; .
\end{equation}
In particular, if $B=1$, $A(t,t'')=R(t,t'')$ and $t'=0$, this integrals
yields the susceptibility. Let us assume that $t$ and $t'$ are 
far apart;
we start as for the first type of integrals, by dividing the 
time interval in three subintervals
\begin{equation}
[t',t] =[t',t'^+] \;\; U \;\; [t'^+,t^-] \;\; U \;\;[t^-,t]
\end{equation}
and by approximating the 
functions by their functional form inside each of the intervals:
\begin{eqnarray}
I_2(t,t') &\sim& 
\int_{t'}^{t'^+} dt''  A_{\sc ag}(t,t'') 
B(t''-t')
+ 
\int_{t'^+}^{t^-} dt''  A_{\sc ag}(t,t'') B_{\sc ag}(t'',t') 
\nonumber\\
& & 
+ 
\int_{t^-}^{t} dt'' A(t-t'') 
B_{\sc ag}(t'',t')\; .
\end{eqnarray}
In the first term $B(t-t'')$ can
be replaced by $B(t-t'')= 
B_{\sc st}(t-t'')+\lim_{t-t''\to\infty} \lim_{t''\to\infty}B(t,t'') $.
The same applies to $A$ in the last term.
All functions vary fast in the stationary regime but very slowly in the 
aging regime. The next assumption is that functions
in the aging regime that are convoluted with functions
in the stationary regime, can be considered to be 
constant and taken out of 
the integral. That is to say
\begin{eqnarray}
I_2(t,t') 
&\sim&
A_{\sc ag}(t,t') \int_{t'}^{{t'}^+} dt'' \, B(t''-t') 
+
\int_{t'}^t dt''  A_{\sc ag}(t,t'') B_{\sc ag}(t'',t')
\nonumber\\
& & 
+
B_{\sc ag}(t,t')
\int_{t^-}^t dt'' \, A(t-t'')
\nonumber\\
&\sim&
A_{\sc ag}(t,t') \int_0^\infty d\tau' \, B(\tau') 
+
\int_{t'}^t dt''  A_{\sc ag}(t,t'') B_{\sc ag}(t'',t')
\nonumber\\
& & 
+
B_{\sc ag}(t,t')
\int_0^\infty d\tau' \, A(\tau')
\; ,
\end{eqnarray}
where we used $t-t^-\to\infty$ and ${t'}^+-t'\to\infty$.
Typically, $A(\tau')$ and $B(\tau')$ are proportional to the response
function. Using FDT the integrals in the first and third term can then be 
performed in the classical limit or they can be expressed as functions
of the correlation in the quantum case.  
The second term instead depends exclusively on the 
aging dynamic sector.

All other integrals can be evaluated, in the large-time limit,
in a similar way.



\begin{thebibliography}{99}

\bibitem[a]{add1}
E-mail address: leticia@physique.ens.fr
\bibitem[b]{add2}
E-mail address: lozano@ipno.in2p3.fr
\bibitem[c]{add3}
Unit\'e propre du CNRS,  associ\'ee
\`a\ l'Ecole Normale Sup\'erieure et \`a\ l'Universit\'e de Paris Sud.
\bibitem[d]{add4}
 Unit\'e de Recherche des
Universit\'es Paris 11 et 6 associ\'ee au CNRS.

\bibitem{Struick} 
L. C. E. Struick, {\it Physical aging in amorphous systems and other 
materials} (Houston: Elsevier, 1978).
I. Hodge; Science {\bf 267}, 1945 (1996).

\bibitem{spin-glasses}
L. Lundgren, P. Svedlindh, P. Nordblad and O. Beckmann,
 Phys. Rev. Lett. {\bf 51}, 911 (1983).
E. Vincent, J. Hammann, M. Ocio, J.P. Bouchaud and  L. F.  Cugliandolo,
{\it Slow dynamics and aging} in Sitges Conference on Glassy Systems, 
M. Rub\'{\i} ed. (Springer-Verlag, 1997), cond-mat/9607224.

\bibitem{levelut}
F. Alberici, P. Doussineau and A. Levelut,
J. Phys. I France {\bf 7},  329 (1997); Europhys. Lett. {\bf 39}
329 (1997). F. Alberici-Kious, J-P Bouchaud, L. F. Cugliandolo, 
P. Doussineau and A. Levelut,
{\it Aging in} K$_{1-x}$Li$_x$ Ta$_3$: 
{\it a domain growth interpretation}, cond-mat/9805208.


\bibitem{Nagel} R. L. Leheny and S. Nagel; Phys. Rev. {\bf B57},
5154 (1998).

\bibitem{Bonn} D. Bonn, H. Tanaka, G. Wegdam, H. Kellay, and J. Meunier; 
{\it Crossover behavior in the aging of a colloidal glass}, 
preprint LPSENS/1997. 

\bibitem{Fihu} 
A. J. Bray and M. A. Moore, J. Phys. {\bf C17}, L463 (1984)
and in {\it Heidelberg Colloquium on Glassy
Dynamics}, Lecture Notes in Physics {\bf 275}, ed. J. L. van Hemmen and
I. Morgenstern (Springer-Verlag, Berlin).
G.J. Koper and  H.J. Hilhorst,
J. Phys. (France) {\bf 49}, 429 (1988).
D.S.  Fisher and D.A. Huse,
Phys. Rev. Lett. {\bf 56}, 1601 (1986);
Phys. Rev. {\bf B38}, 373 (1988).

\bibitem{Phasespace}
P. Sibani and K.H. Hoffmann,
Europhys. Lett. {\bf 4}, 967 (1987);
Phys. Rev. {\bf A38}, 4261 (1988);
Europhys. Lett. {\bf 16}, 423 (1991).

\bibitem{rusos}
V. Dotsenko, J. Phys. {\bf C18}, 6023 (1985).
V. Dotsenko, M. Feigel'man and L. B. Ioffe,
{\it Spin-glasses and related problems}, Soviet Scientific
Reviews {\bf 15} (Harwood 1990). 
M. V. Feigel'man and V. Vinokur,
J. Phys. I (France) {\bf 49}, 1731 (1988).

\bibitem{Bo}
J-P Bouchaud, J Phys. I (France) {\bf 2}, 1705 (1992).
J-P Bouchaud and D. S. Dean,
J. Phys. I (France) {\bf 5},  265 (1995).

\bibitem{Kula} J. Kurchan and L. Laloux, J. Phys. {\bf A29},
1929 (1996).

\bibitem{Cuku} L. F. Cugliandolo and J. Kurchan, 
Phys. Rev. Lett. {\bf 71}, 173  (1993);
Phil. Mag.  {\bf B71}, 501 (1995).

\bibitem{Cuku2} L. F. Cugliandolo and J. Kurchan, 
J. Phys. {\bf A27}, 5749 (1994).

\bibitem{Frme}
S. Franz and M. M\'ezard, Europhys. Lett. {\bf 26},  209 (1994);
Physica {\bf A209}, 1 (1994).
L. F. Cugliandolo and P. Le Doussal, Phys. Rev. {\bf E53}, 1525  (1996).

\bibitem{Cukule} L. F. Cugliandolo, J. Kurchan and P. Le Doussal, 
Phys. Rev. Lett. {\bf 76}, 2390 (1996).

\bibitem{Ri} H. Rieger, Annual Review of Computational Physics {\bf II},
D. Stauffer ed. (World Scientific, Singapore, 1995).

\bibitem{numerics} G. Parisi, Phys. Rev. Lett. {\bf 79}, 3660 (1997).
W. Kob and J-L Barrat,  Phys. Rev. Lett. {\bf 78}, 4581 (1997).
A. Barrat, Phys. Rev. {\bf E57}, 3629 (1998). 
H. Yoshino, {\it Aging Dynamics of an Elastic 
String Diffusing in a Disordered Media}, cond-mat/9802283.
J-L Barrat and W. Kob, 
{\it Fluctuation dissipation ratio in an aging Lennard-Jones glass},
cond-mat/9806027.

\bibitem{Bonn2} D. Bonn, H. Kellay and J. Meunier, in preparation.
T. Grigera and N. E. Israeloff, in preparation.

\bibitem{Kith} T. D. Kirkpatrick and D. Thirumalai, 
Phys. Rev. {\bf B36}, 5388 (1987). 
T. R. Kirkpatrick and P. Wolynes, Phys. Rev. {\bf A35}, 3072 (1987);
Phys. Rev. {\bf B36}, 8552 (1987).
T. R. Kirkpatrick, D. Thirumalai, P. G. Wolynes;
Phys. Rev. {\bf A40}, 1045 (1989).

\bibitem{Go} W. G\"otze, in {\it Liquids, freezing and
glass transition}, JP Hansen, D. Levesque, J. Zinn-Justin eds.
(North Holland, 1989).
W. G\"otze and  L. Sj\"ogren, Rep. Prog. Phys. {\bf 55}, 241  (1992).



\bibitem{Frhe}
S. Franz and J. Hertz, Phys. Rev. Lett. {\bf 74}, 2114 (1995).   
J-P Bouchaud, L. F.  Cugliandolo, J. Kurchan and M. M\'ezard,
Physica {\bf A226}, 243 (1996). 

\bibitem{Bocukume} J-P Bouchaud, L. F.  Cugliandolo, J. Kurchan and M. M\'ezard,
{\it Out of equilibrium dynamics in spin-glasses and other glassy systems}, 
in ``Spin-glasses and random fields', A. P. Young ed. (World Scientific, 
Singapore, 1997), cond-mat/9702070.

\bibitem{Gile} T. Giamarchi and P. Le Doussal, 
{\it Statics and dynamics of disordered
elastic systems}
in ``Spin-glasses and random fields', A. P. Young ed.
(World Scientific, Singapore, 1997) , cond-mat/9705096.


\bibitem{Wuetal} W. Wu, B. Ellmann, T. F. Rosenbaum, G. Aeppli and D. H. Reich, 
Phys. Rev. Lett. {\bf 67}, 2076 (1991). 
W. Wu, D. Bitko, T. F. Rosenbaum and G. Aeppli,  Phys. 
Rev. Lett. {\bf 71}, 1919 (1993). T. F. Rosenbaum, 
J. Phys. {\bf C8}, 9759 (1996).

\bibitem{ferroelectrics} 
E. Courtens, J. Phys.Lett. (Paris)  {\bf 43}, L199 (1982); Phys. Rev. Lett. 
{\bf 52}, 69 (1984). E. Matsushita, T. Matsubara, Prog. Theor. Phys. 
{\bf 71}, 235 (1984). 
R. Pirc, B. Tadic and R. Blinc, Z. Phys. {\bf B61}, 69 (1985);
Phys. Rev. {\bf B36}, 8607 (1987). 



\bibitem{quantumreplicas} 
P. Shukla and S. Singh, Phys. Rev. {\bf B23}, 4661 (1981).
A. J. Bray and M. A. Moore, {\bf 13} L665 (1985).
K. D. Usadel, Solid State Commun. {\bf 58}, 629 (1986).
T. Yamamoto and H. Ishii, J. Phys. {\bf C20}, 6053 (1987).
Y. Y. Godschmidt and P. Y. Lai, Phys. Rev. Lett. {\bf 64}, 2467 (1990).
Y. Y. Godschmidt, Phys. Rev. {\bf B41}, 4858.
V Dobrosavljevic and D Thirumalai, J. Phys. {\bf A23}, L767 (1990). 
L. De Cesare, K. Lubierska-Walasek, I. Rabuffo and K.  Walasek, Physica {\bf 
A214}, 499 (1995); J. Phys. {\bf A29}, 1605 (1996).
T. K. Kope\'c, Phys. Rev. {\bf B54}, 3367 (1996).

\bibitem{Giledou}
T. Giamarchi and P. Le Doussal, 
Phys. Rev. {\bf B53} 15206 (1996).

\bibitem{Niri} T. Nieuwenhuizen, Phys. Rev. Lett. {\bf 74}, 4289 (1995), 
{\it ibid} 4293. T. Nieuwenhuizen and F. Ritort, Physica {\bf A250}, 8 (1998).


\bibitem{Fisher} D. S. Fisher, Phys. Rev. Lett. {\bf 69}, 534 (1992); 
Phys. Rev. {\bf B51}, 6411 (1995).


\bibitem{SKpara} J. Miller and D. A. Huse, 
Phys. Rev. Lett. {\bf 70}, 3147 (1993). 

\bibitem{Rogr} D. Grempel and M. Rozenberg,
Phys. Rev. Lett. {\bf 80}, 389 (1998).
M. Rozenberg and D. Grempel,
{\it Dynamics of the Ising
Spin-Glass model in a transverse field}, cond-mat/9802106.

\bibitem{MC} H. Rieger and A. P. Young, {\it Quantum Spin-glasses}
(Springer-Verlag, Berlin, 1996), 
cond-mat/9607005.
R. N. Bhatt, {\it Quantum spin-glasses}, in ``Spin-glasses and random fields', 
A. P. Young ed. (World Scientific, Singapore, 1997).

\bibitem{Sachdev}
Y. Ye, N. Read and S. Sachdev, Phys. Rev. Lett. {\bf 70}, 4011 (1993).
N. Read, S. Sachdev and Y. Ye, Phys. Rev. {\bf B52}, 384 (1995). 

\bibitem{Igri}
F. Igloi and H. Rieger, Phys. Rev. Lett. {\bf 78}  2473 (1997),
cond-mat/9709260.
H. Rieger and F. Igloi, Europhys. Lett. {\bf 39}  135 (1997). 
F. Igloi, D. Karevski and H. Rieger; cond-mat/9707185. 

\bibitem{librolowT}
{\it Amorphous solids: low temperature properties}, W. A. Phillips ed.
(Springer-Verlag, 1981).

\bibitem{Fayer} L. R. Narasimhan, K. A. Littau, D. W. Pack, Y. S. Bai, A. 
Elsechner and M. D. Fayer, Chem. Rev. {\bf 90}, 439 (1990).
Y. S. Bai and M. D. Fayer, Phys. Rev. {\bf B39}, 11066 (1989). 


\bibitem{MSR} 
C. P. Martin, E. Siggia and H. A. Rose,
Phys. Rev. {\bf A8}, 423 (1973).
H. K. Janssen, Z. Phys. {\bf B23}, 377 (1976); 
{\it Dynamics of critical phenomena and related Topics},
Lecture Notes in Physics {\bf 104}, C. P. Enz ed., 
(Springer-Verlag, Berlin, 1979).

\bibitem{Langevin} The relevance of the quantum Langevin equation 
has been extensively treated in the literature. Some references are
G. W. Ford, M. Kac and M. Mazur, J. Math. Phys. {\bf 6}, 504 (1965).
G. W. Ford and M. Kac, J. Stat. Phys {\bf 46}, 803 (1987). 

\bibitem{Gardiner} C. W. Gardiner, 
{\it Quantum noise} (Springer-Verlag, Berlin, 1991).



\bibitem{closedpath} J. Schwinger, J. Math. Phys. {\bf 2}, 407 (1961).
L. V. Keldysh, Zh. Eksp. Teor. Fiz. {\bf 47}, 1515 (1964), Sov. Phys JETP {\bf 
20}, 235 (1965). 
P. Danielewicz, Ann. Phys. {\bf 152}, 239 (1984). 

\bibitem{Feve} R. P. Feynman and F. L. Vernon, Ann Phys. {\bf 24}, 118 (1963). 


\bibitem{pspinIsing} B. Derrida,
Phys. Rev. Lett.  {\bf 45}, 79 (1980);
Phys. Rev. {\bf B24}, 2613 (1981).
D. J. Gross and M. M\'ezard, Nucl. Phys. {\bf B240}, 431 (1984).

\bibitem{Sozi} H. Sompolinsky and A. Zippelius,
Phys. Rev. Lett. {\bf 47}, 359 (1981);
Phys. Rev. {\bf B25}, 6860 (1982).

\bibitem{So}
H. Sompolinsky, Phys. Rev. Lett. {\bf 47}, 935 (1981);
Phil. Mag. {\bf 50}, 285 (1984).

\bibitem{p2} J. M. Kosterlitz, D. J. Thouless and R. C. Jones, 
Phys. Rev. Lett. {\bf36}, 1217 (1976). 

\bibitem{Cide} S. Ciuchi and F. de Pasquale, Nucl. Phys. {\bf B300} [FS22], 
31 (1988).

\bibitem{Cude} L. F. Cugliandolo and D. S. Dean, 
J. Phys. {\bf A28},  4213 (1995); {\it ibid} L453 (1995). 
 
 
\bibitem{Crso} 
A. Crisanti and H.-J. Sommers, Z. Phys. {\bf B87}, 341 (1992).
A. Cavagna, I. Giardina and G. Parisi, J. Phys. {\bf A30}, 4449 (1997); 
{\it ibid} 7021. 

\bibitem{Crhoso} 
A. Crisanti, H. Horner and H.-J. Sommers,
Z. Phys. {\bf B92}, 257 (1993).

\bibitem{Hojayo} A. Houghton, S. Jain and A. P. Young, 
Phys. Rev. {\bf B28}, 2630 (1983).

\bibitem{Biyo} K. Binder and A. P. Young, Rev. Mod. Phys. {\bf 58}, 801 (1986). 

\bibitem{Mepavi} M. M\'ezard, G. Parisi and M. A. Virasoro, 
{\it Spin glass theory and beyond} (World Scientific, Singapore, 1987). 



\bibitem{letter} L. F. Cugliandolo and G. Lozano,
Phys. Rev. Lett. {\bf 80}, 4979 (1998), cond-mat/9712090.



\bibitem{chinos} G. Zhou, Z. Su, B Hao, Y. Lu, Phys. Rep. {\bf  118}, 1 (1985). 

\bibitem{alemanes} H. Grabert, P. Schramm and G-L Ingold, Phys. Rep. {\bf 168}, 
115 (1988). 

\bibitem{alemanes2} C. Greiner and S. Leupold, {\it Stochastic interpretation 
of Kadanoff-Baym equations and their relation to Langevin processes}, 
hep-ph/9802312.

\bibitem{Hupazh} B. L. Hu, J. P. Paz and Y. Zhang, Phys. Rev. {\bf D45}, 2843 
(1992), {\it ibid} {\bf D47}, 1576 (1993).

\bibitem{Huetal} E. Calzetta and B. L. Hu, Phys. Rev.{\bf  D35}, 495 (1987); 
{\it ibid} {\bf D37}, 2878 (1988).

\bibitem{deVega} D. Boyanovsky, H. J. de Vega and R. Holman, 
{\it Erice Lectures on inflationary reheating}, hep-ph/9701304.



\bibitem{Cale} A. Caldeira and A. Legget, Phys Rev. {\bf A31}, 1059 (1985). 

\bibitem{Leggetreview} A. J. Legget, S. Chakravarty, 
A. T. Dorsey, M. P. A. Fisher, A. Garg and W. Zwerger, 
Rev. Mod. Phys. {\bf 59}, 1 (1987).

\bibitem{cirano} C. De Dominicis, Phys. Rev. {\bf B18}, 4913 (1978).

\bibitem{Koge} A. Georges and G. Kotliar, 
Phys. Rev. {\bf B45}, 6479 (1992).
A. Georges, G. Kotliar, W. Krauth and M. Rozenberg, 
Rev. Mod. Phys. {\bf 68}, 13 (1996).

\bibitem{Kleinert} H. Kleinert and S. V. Shavanov, Phys. Lett. {\bf A200},
224 (1995). 

\bibitem{Eiop} H. Eissfeller and M. Opper, Phys. Rev. Lett. {\bf 68}, 2094 
(1992).

\bibitem{Cukupe} L.F. Cugliandolo, J.Kurchan and L.Peliti, 
Phys.Rev. {\bf E55},
3898 (1997).

\bibitem{Frmepape}  S. Franz, M. M\'ezard, G. Parisi and L. Peliti,
{\it Measuring equilibrium properties
in aging systems}, cond-mat/9803108.

\bibitem{Cudeku} L. F. Cugliandolo, D. S. Dean and J. Kurchan,  
Phys. Rev. Lett. {\bf 79}, 2168 (1997).

\bibitem{Lale} L. Laloux and P. Le Doussal,
Phys. Rev. {\bf E57}, 6296 (1998).

\bibitem{Stamp} M. Dub\'e  and P. C. E. Stamp; J. Low Temp. Phys. 
{\bf 110}, 779 (1998). 

\bibitem{Zinn} J.Zinn Justin, {\it Quantum Field Theory and Critical
Phenomena}, (Clarendon Press, Oxford, 1989).

\bibitem{Ku} J.Kurchan, J.Phys. I (France) {\bf 1},  1333 (1992).


\bibitem{Eric}
R. Sappey, E. Vincent, M. Ocio, J. Hammann,
J. Magn. Magn. Mat. {\bf 177}, 957 (1998). 
R. Sappey, E. Vincent, N. Hadacek, F. Chaput, J.P. Boilot, D. Zins,
Phys. Rev. {\bf B56}, 14551 (1997).

\bibitem{Nordblad} 
T. Jonsson, J. Mattsson, C. Djurberg, F. A. Kahn, P. Nordblad and
P. Svedlindh, Phys. Rev. Lett. {\bf 75}, 4138 (1995).
T. Jonsson, J. Mattsson, P. Nordblad, P. Svedlindh,
J. Magn. Magn. Mat. {\bf 168}, 269 (1997)

\bibitem{Chgu}
E. M. Chudnovsky and L. Gunther, Phys. Rev. Lett. {\bf 60}, 661 (1988).


\bibitem{Kubo}
R. Kubo, M. Toda and N. Hashitume, {\it Statistical Physics II: Non-equilibrium 
Statistical Mechanics}, Springer Verlag (1992).

\bibitem{Parisi} G. Parisi, {\it Statistical Field Theory}, Frontiers in 
Physics, Lecture Notes 
Series, Addison-Wesley (1988). 

\end{thebibliography}
\end{document}